\documentclass[aps,prx,reprint,superscriptaddress,showpacs,showkeys, nofootinbib]{revtex4-2}

\usepackage{bm}
\usepackage[T1]{fontenc}
\usepackage{lmodern}
\usepackage{verbatim}

\usepackage{ccaption}

\usepackage[english]{babel}

\usepackage[normalem]{ulem}

\usepackage{graphicx}
\graphicspath{{./}{./figures/}}

\usepackage{placeins}
\usepackage{float}

\usepackage{amsmath,amssymb}

\usepackage[utf8x]{inputenc}

\usepackage{textcomp,marvosym}

\usepackage{nameref,hyperref}

\usepackage[right]{lineno}

\usepackage{microtype}
\DisableLigatures[f]{encoding = *, family = * }

\usepackage{array}

\newcolumntype{+}{!{\vrule width 2pt}}

\newlength\savedwidth

\usepackage[aboveskip=1pt,labelfont=bf,labelsep=period,justification=raggedright,singlelinecheck=off]{caption}

\makeatletter
\renewcommand{\@biblabel}[1]{\quad#1.}
\makeatother

\usepackage{mdframed}
\newcounter{boxcounter}
\renewcommand{\theboxcounter}{Box \arabic{boxcounter}}

\definecolor{light-gray}{gray}{0.92}

\usepackage[dvipsnames]{xcolor}

\makeatletter
\DeclareRobustCommand{\texcomment}{%
  \begingroup\sethlcolor{Blue}\color{white}\hl{*}\endgroup
  \@ifnextchar\bgroup{}{\ }%
}
\makeatother

\DeclareRobustCommand{\edit}[1]{{\begingroup\color{black}{#1}\endgroup}}

\makeatletter
\def\l@subsection#1#2{}
\def\l@subsubsection#1#2{}
\makeatother

\begin{document}

\title{The Hands-On Growth Laws Theory Cookbook}

\author{Rossana Droghetti}
\email{rossana.droghetti@ifom.eu}
\affiliation{IFOM ETS - The AIRC Institute of Molecular Oncology, via Adamello 16, 20139 Milan, Italy}

\author{Mattia Corigliano}
\affiliation{IFOM ETS - The AIRC Institute of Molecular Oncology, via Adamello 16, 20139 Milan, Italy}

\author{Ludovico Calabrese}
\affiliation{Biozentrum, University of Basel and Swiss Institute of Bioinformatics, Basel, Switzerland}

\author{Philippe Fuchs}
\affiliation{Centre de Biologie Structurale (CBS), Univ Montpellier, CNRS, INSERM, Montpellier 34090, France}
\affiliation{Dipartimento di Fisica, Università degli Studi di Milano, via Celoria 16, 20133 Milan, Italy}

\author{Abhishek Vaidyanathan}
\affiliation{Humanitas University, 20072 Pieve Emanuele, Milan, Italy}
\affiliation{IRCCS Humanitas Research Hospital, Rozzano 20089, Milan, Italy}

\author{Johannes Keisers}
\affiliation{Centre de Biologie Structurale (CBS), Univ Montpellier, CNRS, INSERM, Montpellier 34090, France}

\author{Gabriele Micali$^\mathparagraph$}
\email{gabriele.micali@humanitasresearch.it}
\affiliation{IRCCS Humanitas Research Hospital, Rozzano 20089, Milan, Italy}

\author{Marco Cosentino Lagomarsino$^\mathparagraph$}
\email{marco.cosentino-lagomarsino@ifom.eu}
\affiliation{IFOM ETS - The AIRC Institute of Molecular Oncology, via Adamello 16, 20139 Milan, Italy}
\affiliation{Dipartimento di Fisica, Università degli Studi di Milano, via Celoria 16, 20133 Milan, Italy}
\affiliation{INFN sezione di Milano, via Celoria 16, Milan, Italy}

\author{Luca Ciandrini$^\mathparagraph$}
\email{luca.ciandrini@umontpellier.fr}
\affiliation{Centre de Biologie Structurale (CBS), Univ Montpellier, CNRS, INSERM, Montpellier 34090, France}
\affiliation{Institut Universitaire de France (IUF)}

\begingroup
\renewcommand\thefootnote{$\mathparagraph$}
\footnotetext{These authors contributed equally to this work.}
\endgroup

\begin{abstract}
    This tutorial covers  the emerging field of coarse-grained cellular growth modeling, and aims to bridge the gap between theoretical foundations and practical application. By adopting an original  ``cookbook'' approach, it is designed to offer a hands-on guide for constructing and analyzing different key aspects of cellular growth, focusing on available results for bacteria and beyond. The tutorial is structured as a series of step-by-step ``recipes'', and covers essential concepts, recent literature, and key challenges. It aims to empower a broad audience, from students to seasoned researchers, to replicate, extend, and innovate in this scientific area. Specifically, each section provides detailed, bare-bone models to start working in each area, from basic steady-state growth to variable environments and focusing on different key layers relevant to biosynthesis, transcription, translation, nutrient sensing and protein degradation, links between cell cycle and growth, ending with ecological insights.
\end{abstract}

\maketitle

\tableofcontents
\section{Introducing the Cookbook} \label{sec1:introduction}
 \vspace{-.3cm}

At the intersection of physics and biology, a well-known cultural gap exists, with each field approaching similar questions from distinct perspectives~\cite{knight_bridging_2002, bialek_perspectives_2017}. 
Physicists typically seek universal principles and common features, while biologists tend to ask specific detailed questions for a precise biological system or condition, exploring the nuances and complexities of life~\cite{anderson_more_1972}. 
As a consequence, biology often focuses on mechanisms that are specific to molecules or organisms, while physics-driven approaches strive to achieve simplified descriptions of living systems that unify multiple observations, aiming to identify some underlying principles that can explain the data.

Early microbiologists, such as Monod, Schaechter, Maaløe, Neidhardt, and Kjeldgaard (to name a few), arguably approached their work in a way more similar to physicists~\cite{jun_fundamental_2018}. They focused on identifying patterns and regularities in cellular phenomena like cell size, growth cycles, and macromolecular composition. Some of these remarkable observed regularities, today often referred to as ``growth laws'', describe how cells regulate their growth in relation to nutrient availability, biosynthesis rates, and environmental conditions.

\begin{figure*}[h!t]
    \centering
    \includegraphics[width=.8\textwidth]{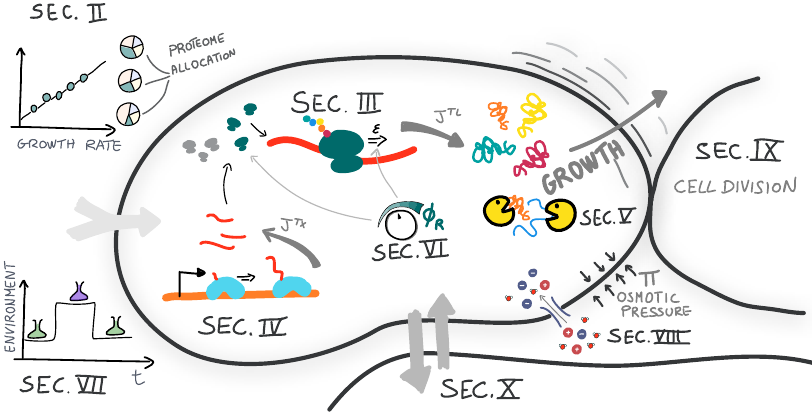} 
    \caption{ \edit{ \textbf{Overview of the topics covered in this tutorial.} The topics covered in this tutorial begin with the classical proteome allocation theory (Sec.~\ref{sec2:classical}), which introduces the phenomenology of growth laws. Building on this foundation, Sec.~\ref{sec3:translation} explores the dependence of the translation elongation rate and ribosomal activity on growth, providing a mechanistic interpretation of these trends based on ribosome recruitment. Section~\ref{sec4:transcription} further extends the framework by incorporating transcriptional dynamics and traffic theory, while Sec.~\ref{sec5:degradation} refines the model to account for slow-growth regimes by including protein degradation. 
    Subsequently, Sec.~\ref{sec6:mecha_regulation} examines the mechanistic regulation of ribogenesis in both prokaryotes and eukaryotes, offering a molecular perspective on the emergence of growth laws. Section~\ref{sec7:shifts} then illustrates how to model cellular behavior in fluctuating environments. 
    The final three sections address broader topics beyond the classical theory: density homeostasis and volume–mass coupling (Sec.~\ref{sec8:density}); the interplay between growth laws and cell-cycle progression (Sec.~\ref{sec9:cell_cycle}); and, finally, the insights gained from a community and ecological perspective (Sec.~\ref{sec10:communities}).}
    }
    \label{fig:graphical_abstract}
\end{figure*}

However, it is only thanks to a renewed focus in recent years~\cite{scott_interdependence_2010}, with advancements in both experimental techniques and mathematical models, that these principles have evolved into robust predictive theoretical frameworks~\cite{Scott2011,jun_fundamental_2018, Serbanescu2021, scott_shaping_2022}. 
In bacteria, this maturation has contributed towards encouraging biologists and physicists to embrace the underlying simplicity within complex biological systems, incorporating this theoretical language into everyday scientific practice. 

As a consequence, coarse-grained resource allocation and flux-balance models are widely explored today as powerful tools for dissecting the general mechanisms behind cell growth. These models primarily offer quantitative predictions of how cells allocate their resources for biosynthesis and growth~\cite{scott_shaping_2022}, while also beginning to provide insights into the ecological and evolutionary dynamics of populations~\cite{Deris2013, huang_evolution_2024, labourel_proteome_2024, pacciani-mori_constrained_2021}.

With an increasing body of experimental evidence, the theory surrounding cellular growth has grown more complex. This complexity often makes it difficult to untangle the assumptions and approximations that can lead to differing interpretations of the data, and ultimately to validate the models and test their predictions. 
While many recent articles and reviews have synthesized this extensive body of work (see e.g.~\cite{Scott2011,scott_shaping_2022, schaechter_brief_2015, jun_fundamental_2018, bruggeman_searching_2020, belliveau_fundamental_2021, chure_optimal_2023, golding_2024, Zim2025}),
highlighting both theoretical foundations and empirical validation of various models, we believe that a practical gap remains. 
There is a need for a comprehensive, hands-on guide that equips researchers with the practical tools and knowledge to build, analyze, and extend these models comparing scenarios and enhancing our power to falsify hypotheses and validate mechanisms. 

This tutorial aims to address that gap by adopting an original ``cookbook'' approach. It does not aim to be a comprehensive synthesis of the literature, but a step-by-step guide designed to serve both as an educational resource and a catalyst for innovation. Structured as a series of ``recipes'', it provides readers with a toolkit to navigate the field providing instructions on how to construct and analyze models in different contexts, starting from established frameworks. 
\edit{The topics illustrated in this tutorial spans from the ``classical'' protein allocation theory to their connection with cell size, cell cycle and bacteria communities, as shown by the graphical abstract, Fig.\ref{fig:graphical_abstract}. }

Each section is dedicated to a key component of the mathematical framework for understanding cellular growth dynamics, and structured as follows.
We begin by guiding the reader through the essential concepts and recent literature, identifying the state of the art, along with key open problems and challenges ({\it Main questions and motivation} subsections). We then present a detailed, bare-bones model, explaining the underlying assumptions, mathematical formulations, and computational techniques ({\it Ingredients and Recipes}). Finally, we discuss the forefront of each area, offering insights into the latest developments and potential avenues for future research ({\it Outlook}). When crucial, we will provide insights into specific experimental techniques and their hidden limitations, helping readers navigate the subtleties of data interpretation and connect theory with practice.

The cookbook structure is designed to be accessible to a broad audience, from students, to early-career researchers and experienced scientists who are new to this field. 
Our goal is to empower readers not only by providing the essential ingredients to replicate existing models but also to push beyond the current frontier by creating new models. By offering strategies to interpret experimental data, recognize hidden assumptions, and develop new theoretical approaches, this tutorial aims
to encourage creative exploration and innovation. 

\subsection*{A Brief Introduction to Growth laws} 

Before dealing with the detailed models and recipes provided by this cookbook, it is essential to lay some groundwork by introducing the key concept of ``growth laws''. These empirical relationships form the backbone of our understanding of cellular growth and resource allocation. This section briefly covers what growth laws are, why they are important, and the main open questions that continue to drive research in this field. This will provide the necessary context for the concepts presented in later sections\footnote{The {\it gourmand}, more expert readers, can safely skip this part.}. 

\noindent\textbf{What are the growth laws?} 
Growth laws can be defined as quantitative empirical relationships that link key physiological parameters, such as cell size, cell-cycle variables, cell biochemical composition, and nutrient and metabolic fluxes to the growth rate.

The roots of these findings trace back to the foundational period of microbiology \cite{schaechter_brief_2015, jun_fundamental_2018}. Monod’s pioneering work in the 1940s and 1950s formalized the relationship between cellular growth and environmental factors~\cite{monod_recherches_1942}. Later, Schaechter, Maaløe, and Kjeldgaard discovered robust relationships between cellular components and the physiological state of cells~\cite{Schaechter1958}, laying the foundation for what we now refer to as empirical growth laws. 
\edit{Their  work  demonstrated  that  fast-growing  cells  were  larger and  contained  higher  levels  of  key  macromolecules  like  RNA  
and  proteins  compared  to  slower-growing cells~\cite{Schaechter1958}.}
%

In the 1960s, Neidhardt and colleagues introduced for the first time what we call today the  growth-dependent partitioning of the proteome~\cite{scott_shaping_2022}, using 2D gel electrophoresis to separate and quantify proteins~\cite{neidhardt_effects_1963}. They observed that fast-growing cells exhibited a higher RNA/protein ratio, shedding light on the connection between cellular composition and growth rate~\cite{neidhardt_studies_1960}. 

More recently, with the advent of proteomics, researchers have been able to classify protein groups based on their collective response to environmental changes~\cite{hui_quantitative_2015, Mori2021}. Hwa and co-workers were the first in linking gene expression directly to cellular physiology, combining experimental evidence with a theoretical framework that offers deeper insights into how cells allocate their resources under different growth conditions~\cite{scott_shaping_2022, klumpp_growth_2009, klumpp_bacterial_2014}. 

The choice to use the growth rate as the independent variable is more important than it seems. Indeed, in this theory the growth rate can be effectively used as an order parameter to describe cellular physiology.\edit{ We will see that when environments differ only in sugar identity, growth rate alone predicts some aspects of the internal molecular composition --for example ribosomal concentration}. In this way, growth laws are effectively independent from the molecular and microscopic details of the external environment.

\noindent\textbf{Why are growth laws important?} Growth laws have been instrumental in advancing our understanding of cellular physiology~\cite{scott_shaping_2022, jun_fundamental_2018}. They have successfully linked cellular growth rates with resource allocation, capturing general physiological relationships without requiring a detailed understanding of the chemical composition of the growth medium~\cite{scott_interdependence_2010}. Moreover, they predict how cell physiology changes across nutrients, how cells survive antibiotics~\cite{scott_interdependence_2010, greulich_growth-dependent_2015, Dai2016}, and how cells respond to fluctuations in environmental conditions~\cite{Erickson2017, wu2023aa}.  
The body of theory developed in the recent years provides a unifying framework for interpreting proteomics data and metabolic fluxes, enhancing our understanding of how key physiological ``summary'' variables are linked to processes like nutrient sensing, transcription, translation, and the regulation of ribosomes, tRNAs, and amino acids, and how these elements are tightly co-regulated~\cite{basan_overflow_2015, hui_quantitative_2015, mori_functional_2023, scott_emergence_2014, Dai2016, hu_optimal_2021}.

While the existence of across-species growth laws is not guaranteed, their emergence in biological systems stems from important underlying principles. These principles include flux balance, which implies the efficient allocation of resources and metabolic pathways; autocatalysis, the self-sustaining feedback loops inherent to biological systems that drive exponential growth; and the necessity for cellular macromolecular composition to remain stable despite environmental fluctuations. These general constraints impose order on otherwise complex biological processes, allowing for predictive relationships to emerge. As a result, a physics-based approach is possible and fruitful, bridging concepts from molecular biology, physiology, and systems biology.

\noindent\textbf{Where is the frontier?} 
Growth laws serve as a powerful tool for both experimentalists and modelers, offering a general, coarse-grained parameter-poor theory for cellular physiology.
Initially developed for exponential growing conditions, today, the theoretical framework developed to describe empirical growth laws has been extended beyond steady exponential growth to cells undergoing transitions~\cite{Erickson2017, KoremKohanim2018, wu2023aa} as well as cells in in non-growing states~\cite{Bren2013}. 
Applications of these theories span multiple bacterial species and even extend to yeast~\cite{kafri_rethinking_2016, metzl-raz_principles_2017, xia_proteome_2022}, hinting that general quantitative principles of proteostasis may hold across kingdoms. 
Many open questions remain, e.g.: Which regulatory mechanisms determine responses in fluctuating or stressful environments? To what extent are growth laws conserved across organisms? Can we integrate growth laws for population averages with single-cell observations to uncover new complexities? Can the framework developed for describing physiology be extended to ecologically interacting organisms? What are the evolutionary drivers behind these laws, and what constraints do they impose?~\cite{Mori2024, Zhu2025} We believe these questions will drive many of the future advances in quantitative biology.
What role do they play in spatially structured populations and communities? What are the evolutionary drivers behind the emergence of the laws and what are the evolutionary constraints imposed by the laws? We believe these questions will drive many of the future advances in quantitative biology.

\section{Appetizer: Classic Proteome Allocation Theory}
\label{sec2:classical}
This section introduces the classical phenomenological theory that links cell physiology and gene expression across various growth conditions, laying the foundation for understanding growth laws. This section provides a broad overview for unexperienced readers, ensuring accessibility and clarity for those new to the field.

\subsection*{MAIN QUESTIONS AND MOTIVATIONS}
\noindent\textbf{Intracellular composition is growth dependent.} 
A central focus of this tutorial is proteome allocation: the distribution of cellular resources, specifically proteins, among essential functions such as biosynthesis, metabolism, and regulation. In particular, we will explore ribosome allocation 
--how much the cell invests in protein synthesis-- a major component of the proteome that adapts dynamically to varying environmental and physiological conditions~\cite{scott_interdependence_2010, hui_quantitative_2015}.

\begin{figure}[h!]
    \centering
    \includegraphics[width = 1\columnwidth]{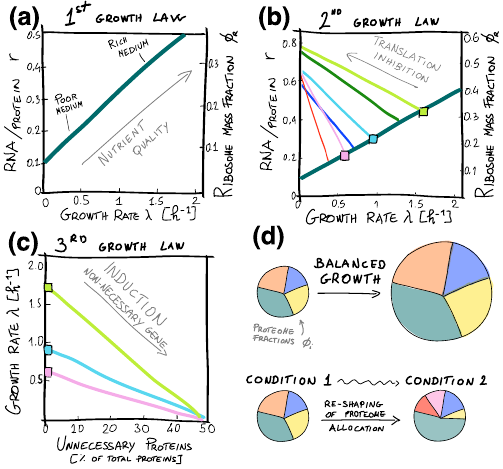}
    \caption{\textbf{Linear relationships between growth rate and ribosome allocation (RNA/protein ratio or $\phi_\text{R}$), observed under different conditions.} 
    (\textbf{a}) Variation in nutrient quality alters growth rate and ribosome allocation. 
    (\textbf{b}) Sub-lethal doses of the translation-inhibiting antibiotic chloramphenicol also affect ribosome allocation and growth rate. 
    (\textbf{c}) The burden of over-expressing  an unnecessary protein leads to reduced growth. The squares in panels b and c represent identical conditions (no induction of unnecessary proteins and no addition of chloramphenicol), enabling direct comparison of the plots. \edit{Panels (\textbf{a})-(\textbf{c}) are inspired by and adapted from~ Ref.~\cite{scott_interdependence_2010}.} 
    (\textbf{d}) While proteome sectors remain stable during exponential balanced growth, their relative allocation varies across different environmental conditions.
    }
    \label{fig1}
\end{figure}

A growth law describing ribosome allocation was first identified in the 1960s through systematic measurements of RNA and protein content in {\it Aerobacter aerogenes}~\cite{neidhardt_studies_1960}, and later confirmed in other bacteria~\cite{karpinets_rna_2006, scott_interdependence_2010} and eukaryotes~\cite{kafri_rethinking_2016, metzl-raz_principles_2017, xia_proteome_2022}. 
Since ribosomal RNA (rRNA) --the structural component of the ribosomes-- is far more aboundant than messenger (mRNA) and transfer (tRNA) RNAs, the RNA-to-protein ratio serves as a reliable proxy for the fraction of the proteome dedicated to ribosomes and, consequently, to protein synthesis~\cite{neidhardt_studies_1960, ecker_ribosome_1963}. 
These experiments, performed under growth conditions limited by the quality of the carbon source, revealed a striking linear relationship between the RNA-to-protein ratio and the growth rate~$(\lambda)$, see Fig.~\ref{fig1}(a). Thus, ribosome allocation and growth rate are directly proportional to each other. 

A deeper understanding of ribosome allocation and growth was achieved by Scott and coworkers~\cite{scott_interdependence_2010}, who not only confirmed the linear relationship between the ribosome mass fraction and steady-state growth rate under carbon-limited conditions, but also introduced the first quantitative model in which nutrient influx and protein synthesis are regulated by proteome sectors that follow the empirical growth laws. 
By studying cells exposed to sub-lethal doses of translation-inhibiting antibiotics, they corroborate~\cite{harvey_how_1980} a second linear relationship: ribosome allocation increases as growth rate decreases under translation inhibition, Fig.~\ref{fig1}(b). Finally, Scott and coworkers investigated the physiological burden caused by over-expressing unnecessary proteins. Their experiments quantified how this burden reduces the growth rate by diverting cellular resources away from essential biosynthetic processes, thereby lowering the ribosome allocation available for protein synthesis, Fig.~\ref{fig1}(c). 

Subsequent studies have shown that growth laws extend beyond ribosomal proteins. Many proteins can be grouped into co-regulated sectors, each exhibiting a linear dependence on the growth rate~\cite{hui_quantitative_2015, mori_functional_2023}. This suggests a coordinated regulation of distinct proteome sectors in response to various growth perturbations.

\vspace{0.75ex}
\noindent\textbf{Interdependence between gene expression and growth.}
Gene expression and cellular growth are tightly interconnected, with growth rate both reflecting and shaping the physiological state of the cell~\cite{klumpp_growth_2009, klumpp_bacterial_2014}. Global factors such as the abundance of active RNA polymerases, ribosomes, and transcription factors, scale with growth and directly influence transcription and translation rates~\cite{Dai2016, balakrishnan_principles_2022}. Consequently, gene expression cannot be studied in isolation from cellular growth, as both processes are inherently coupled. 
For instance, synthetic genetic circuits often behave unpredictably under varying growth conditions because changes in growth alter the availability of cellular resources~\cite{klumpp_growth_2009}. On the other hand, these circuits can also be designed to leverage growth-regulated dynamics, enabling more robust functionality and seamless integration with the host's physiology~\cite{ceroni_burden-driven_2018}.
Another interesting and non trivial behavior highlighted by this theory regards the regulation of constitutive genes~\cite{klumpp_growth_2009}. Constitutive genes are, by definition, not subject to any specific regulation, and therefore, one would expect these genes to always result in the same level of expression, regardless of the growing conditions. This turns out to be not true, as the regulation of other genes (regulated, like the ribosomal one) introduces competition for the gene expression machinery, resulting in a shrinkage of the resource available to the constitutive genes~\cite{scott_interdependence_2010, hui_quantitative_2015, Dai2016, Ripamonti2025}, producing an effective down-regulation of constitutive genes and proteins. It turns out there is a need of actively regulating the expression of genes, if the cell want to always keep  their expression constant.

Growth laws offer insights into how cells allocate resources to balance competing demands and adapt to environmental conditions. Yet, whether these strategies are optimized for efficiency across varying growth regimes remains an open question~\cite{scott_interdependence_2010, klumpp_bacterial_2014}.

\subsection*{INGREDIENTS AND RECIPES\footnote{The recipe introduced here builds upon ideas and models presented in Ref.~\cite{scott_interdependence_2010}. For each section, further relevant references are cited in `Main questions and motivations' and in `Outlook'.}}

\noindent\textbf{Balanced exponential growth.}
Growth laws are best understood under balanced exponential growth: a steady-state condition in which all intracellular components increase at the same constant rate~$\lambda$.
In such a scenario, the total cell culture's biomass~$M$ grows according 
to~$\dot{M} = \lambda M$ (where the notation~$\dot{x}$ stands for the time derivative of the quantity~$x$,~$\text{d}x/\text{dt}$). Crucially, balanced exponential growth implies that not just the overall biomass but all intracellular components grow at the same rate~$\lambda$, maintaining a fixed macromolecular composition and balanced biosynthetic fluxes (see~\ref{box1:exp_balanced}). \edit{Although special macromolecules such as the genome and cell-cycle regulators may follow distinct dynamics during the cell cycle, they must, on average, obey similar rules across a population --since newborn cells inherit, on average, the same amounts of these identity-- defining components~\cite{golding_2024}.}

Note that the concept of balanced growth (first introduced by Campbell~\cite{campbell_synchronization_1957}) extends beyond mere exponential growth. More widely, a key underlying concept is that unbalanced fluxes of different biosynthetic elements are not sustainable in the long term 
and subject to constant homeostatic regulation. This happens even when biosynthesis is out of steady state and the cell will eventually turn to a different macromolecular composition and growth state. 

\begin{mdframed}[backgroundcolor=light-gray]
\refstepcounter{boxcounter}
{\bf \large \theboxcounter. Balanced Exponential growth}\\
\label{box1:exp_balanced}
\noindent During steady-state balanced exponential growth, extensive quantities such as the total biomass of a cell culture $M$ and each molecular species $Y_\mathrm{i}$ grow effectively at the same pace, as follows
\begin{equation*}
\begin{aligned}
    \dot{M} = \lambda M \quad &
    \dot{Y_\text{i}} = \lambda Y_\text{i} \, ,
\end{aligned}
\end{equation*}
Assuming constant dry-mass density~$\rho_\text{DM}$,~$M = \rho_\text{DM} V$, consequently the total culture volume $V$ also grows exponentially with the same rate (we will discuss density homeostasis in Sec.~\ref{sec8:density}). 
In \emph{E.~coli}, the population-average dry-mass density is remarkably constant across conditions~\cite{kubitschek_independence_1984}.  

As a consequence, intensive properties such as concentrations, number and mass fractions, which are ratios of extensive quantities, remain constant during cell growth. For example, the concentration~$[P_\text{i}]$ of  protein type~$i$ remains constant (under the assumption of constant density), since
\begin{equation*}
	\cfrac{\text{d} [P_\text{i}]}{\text{dt}} = \cfrac{\dot P_\text{i}}{ V} - \cfrac{\dot V}{V} \cfrac{P_\text{i}}{V} = \lambda [P_\text{i}] - \lambda [P_\text{i}] = 0 \,.
\end{equation*}
Thus, during balanced exponential growth, the protein composition within the cell culture remains constant, although it is still influenced by external factors like growth media, temperature, and other experimental conditions that can alter the growth rate, Fig.~\ref{fig1}(d). Therefore, the proteome composition serves as a cellular ``fingerprint'', reflecting the environment. While it remains fixed under stable conditions, it adapts to different environments as cells adjust their molecular composition.
\end{mdframed}

\vspace{0.75ex}
\noindent\textbf{Proteome sectors.}
In the context of steady exponential growth, working with number or mass fractions of protein subsets is preferable to absolute values, as these fractions remain constant during balanced growth (\ref{box1:exp_balanced}). Notably, empirical observations show that these fractions can vary significantly across different conditions, especially when the growth rate is perturbed~\cite{scott_interdependence_2010, hui_quantitative_2015}, Fig.~\ref{fig1}(d).
Building on this property, we define ``proteome sectors'' as distinct groups of proteins whose proteome fractions exhibit a common response to specific \edit{limiting factors}~\cite{hui_quantitative_2015}. Proteins within a sector display consistent trends --such as increasing or decreasing expression levels-- with respect to changes in growth rate. 
Interestingly, this regulatory classification often aligns with functional classification, as proteins that are regulated together frequently perform related roles (e.g. ribosomes and translation affiliated proteins). Figure~\ref{fig1}(a) and (b) illustrates examples of this behavior.

Specifically, the fractional extent of a sector~$\phi_\text{i}$ is quantified by the ratio of the mass of the proteins of that sector to the total protein mass~$M_\mathrm{p}$~\cite{scott_interdependence_2010}:
\begin{equation}
   \phi_\text{i} = \frac{M_\text{i}}{M_p} . \label{eq:def_sector}
\end{equation}
By definition, all sectors must sum to 1, 
\begin{equation}
    \sum_\text{i} \phi_\text{i} = 1\,,
    \label{eq:sector_constraint}
\end{equation}
and this constraint needs to be satisfied both at steady state and in dynamic conditions. 
Clearly, it is possible to define various types of proteome sectors based on the specific questions addressed and the desired level of granularity. For example, one can define sectors for membrane proteins, transporters, metabolite production, etc.~\cite{Serbanescu2020new, belliveau_fundamental_2021, reyes-gonzalez_dynamic_2022}. However, as some sectors share the overall growth rate trend, the definition of additional sectors should be justified by the requirements of the model or the study (see for example the division protein sector introduced in Sec.~\ref{sec9:cell_cycle}).
Once the protein sectors are defined, the proteome composition of a cell is given by~$\bm{\phi}=\{\phi_\text{1},\ldots,\phi_\text{i},\ldots,\phi_\text{S} \}$, where~S is the total number of defined sectors. 
In growth law theories, proteome composition is a key determinant of the cell’s physiological state during steady-state exponential growth, reflecting how cellular resources are allocated and prioritized under given conditions.

\vspace{0.75ex}
\noindent\edit{\textbf{Proteome allocation in different growth conditions and perturbations.}}
Exponential growth occurs when bacteria grow steadily in a stable, non-fluctuating environment. However, the exact growth rate and proteome composition depend on the factors limiting growth, which may be external (e.g., nutrient availability, temperature) or internal (e.g., ribosome abundance or enzyme catalytic efficiency)~\cite{hui_quantitative_2015}.

\edit{Systematic changes in growth conditions that tune the same type of limiting factor reveal that} proteome allocation can be divided into three broad categories: one sector increases with the growth rate, another decreases, and a third remains constant~\cite{hui_quantitative_2015}. 
Experimental studies have demonstrated that these growth rate dependencies are linear across a wide range of growth rates, indicating a predictable and modular organization of the proteome. These three behaviors encompass all proteins, although two proteins may be up- or down-regulated at different rates.
This highlights the significance of proteome sectors both experimentally and conceptually. Biologically, it reflects the coordination in the expression levels of the vast majority of proteins. Conceptually, it provides a robust framework for modeling cellular processes, highlighting the utility of defining proteome sectors.

In summary, proteome sectors~$\phi_\text{i}$ are defined as groups of proteins whose levels change coherently (e.g., showing similar trends \textit{vs.} growth rate) in response to the same limitation~\cite{hui_quantitative_2015, Mori2021}. This coherence reflects shared regulatory pathways among sector members and suggests coordinated functions. It is important to note, however, that (i) different growth limitations may lead to distinct sector responses and (ii) by this definition, the assignment of proteins to sectors can vary depending on the class of limitations considered~\cite{hui_quantitative_2015}.\\

\vspace{0.75ex}
\noindent\textit{\edit{Steady growth }under carbon limitation.} 
We begin by considering carbon-limited growth (C-lim), where growth is restricted by the availability of a specific carbon substrate. In this context, the growth rate depends on the nutrient ``quality''; for instance, wild-type \textit{E.~coli} grows faster on glucose than on glycerol due to the higher quality of glucose as a substrate~\cite{scott_interdependence_2010}.
For now, we will stick to this simplified definition of the nutrient quality, but a recent work~\cite{Mukherjee2024_plasticity} highlights the limitation of this interpretation, as discussed later in Sec.~\ref{sec10:communities}.
Under these growth conditions, numerous studies~\cite{neidhardt_studies_1960, Dennis1974, scott_interdependence_2010, hui_quantitative_2015, Dai2016, Mori2021}, confirmed a linear relationship between~$\phi_\text{R}$, the R sector which includes ribosomes and ribosome-affiliated proteins, and the steady state growth rate~$\lambda$ in \textit{E. coli}, a relationship that holds across diverse bacterial species. This relationship, often referred to as the ``first growth law''~\cite{scott_interdependence_2010}, is expressed as:
\begin{equation}
\phi_\text{R} = \phi_\text{R}^\text{min} + \frac{\lambda}{\gamma_\text{0}}, \label{eq:GL_first}
\end{equation}
where~$\phi_\text{R}^\text{min}$ represents the minimum ribosomal investment at~$\lambda = 0$, and we will see that the slope~$1/\gamma_\text{0}$ is linked to the translation elongation rate~\cite{scott_interdependence_2010, Erickson2017}.
Fig.~\ref{fig1}(a) shows the first growth law, how $\phi_\text{R}$ increases linearly with the growth rate when cells are cultured in minimal media with different carbon sources or rich nutrients (e.g. casaminoacids).

Due to the limited cellular resources, if the fractional size of the ribosomal sector grows to sustain growth, then this must happen at the expense of some other portion of the proteome, given that the average cellular protein density remains constant~\cite{Erickson2017}. The sector whose size shrinks with increasing nutrient quality is generally called P  sector,~$\phi_\text{P}$, and it includes unregulated proteins, as well as proteins involved in nutrient uptake and metabolism. In nutrient-poor conditions, where growth is slow, cells prioritize resource acquisition and metabolism, increasing the P sector at the expense of the R sector~\cite{scott_interdependence_2010}.

The down-regulation of ribosomal proteins in favor of the catabolic ones in low-quality environments is called ``stringent response''. This conserved regulatory mechanism is also observed in budding yeast~\cite{warner_yeast_1978}, as discussed in Sec.~\ref{sec6:mecha_regulation}.\\

\vspace{0.75ex}
\noindent\textit{Growth under translation inhibition.}
Scott et {\it al.}~\cite{scott_interdependence_2010} investigated how the R sector responds to sub-lethal doses of the antibiotic chloramphenicol, which inhibits translation by targeting ribosomes. Chloramphenicol prevents the formation of peptide bonds between amino acids and interacts with the ribosome’s exit tunnel, halting elongation during translation~\cite{harvey_how_1980}. Consequently, the average efficiency of ribosomes is reduced, a condition referred to as R-limited growth (R-lim)~\cite{hui_quantitative_2015}.
Under these conditions, the R sector increases as translation inhibition becomes more severe, while growth rate decreases. This relationship is captured by the following equation:
\begin{equation}
\phi_\text{R} = \phi_\text{R}^\text{max} - \frac{\lambda}{\nu}, \label{eq:GL_second}
\end{equation}
where~$\phi_\text{R}^\text{max}$ represents the maximum size of the ribosomal sector, and~$\nu$ is the inverse slope of the relationship, Eq.~\ref{eq:GL_second}, which is linked to the specific substrate used and its nutrient quality~\cite{scott_interdependence_2010}. This relationship, referred to as the ``second growth law'', is illustrated in Fig.~\ref{fig1}(b).

As one would expect, the increased demand for ribosomes comes at the expense of other proteome sectors, in particular~$\phi_\text{P}$, highlighting the resource allocation trade-offs necessary to sustain translation under these conditions~\cite{scott_interdependence_2010}.

Motivated by experimental observations of Scott et {\it al.}~\cite{scott_interdependence_2010} and several follow-up studies, it has become a common assumption  that under translation-limiting conditions, the maximum ribosome allocation~$\phi_\text{R}^\text{max}$ is independent of nutrient quality.  This implies a fixed upper limit on ribosome allocation (often quantified as approximately 0.55 in \textit{E.~coli}), and suggests that the remaining proteome fraction consist of an incompressible sector of ``housekeeping'' proteins, called Q-sector. The size of the Q sector is fixed and equal to~$1- \phi_\text{R}^\text{max}$.
\\

\vspace{0.75ex}
\noindent\textit{Growth under unnecessary protein burden.}
Another well-studied growth \edit{condition} involves the burden introduced by the expression of unnecessary proteins~\cite{scott_interdependence_2010}. These non-essential proteins, though not explicitly toxic, can slow down growth by competing for transcriptional and translational resources.
The expression of an unnecessary protein (U) effectively reduces the fraction of cellular resources available for allocation to other sectors, particularly the R sector. This reduction can be modeled as a decrease in the maximal ribosomal allocation:
~${\phi'}_\text{R}^\text{max} = \phi_\text{R}^\text{max} - \phi_\text{U}$, where~$\phi_\text{U}$ represents the sector associated to the unnecessary proteins, whose size depends on their expression level.
This contraction of the ribosomal sector explains the decrease in growth rate, which can be predicted as:
\begin{equation}
    \lambda(\phi_\text{U}) = \lambda(\phi_\text{U}=0)\left(  1-\frac{\phi_\text{U}}{\phi_\text{U}^\text{max}} \right), \label{eq:GL_third (unnecessary)}
\end{equation}
where~$\phi_\text{U}^\text{max} = \phi_\text{R}^\text{max} - \phi_\text{R}^\text{min}$, and~$\lambda(\phi_\text{U}=0)$ is the growth rate under unperturbed conditions. Figure~\ref{fig1}(c) shows this relation between the growth rate and unnecessary proteins expression level.\\

\vspace{0.75ex}
 \noindent\textit{Other means of perturbing growth.}
Other conditions where proteins, enzymes or other key factors in the cellular growth machinery are limiting are possible. For example, Ref.~\cite{you_coordination_2013} realized anabolic limitations (A-lim) by titrating a key enzyme in the ammonia assimilation pathway, identifying the anabolic A sector. Under this limitation, they observed a group of proteins that is overexpressed at the expense of the rest of the proteins, similarly to what happens to the ribosomal sector when chloramphenicol is applied. This overexpressed group, denoted as~$\phi_\text{A}$, is mainly composed of proteins involved in amino acid biosynthesis~\cite{you_coordination_2013, hui_quantitative_2015}.

\begin{mdframed}[backgroundcolor=light-gray]
\refstepcounter{boxcounter} 
{\bf \large \theboxcounter. Computing~$\phi_\text{R}$}
\label{box2:computing_phi_R}

\noindent The ribosomal sector~$\phi_\text{R}$ is a key element in resource allocation models. It is defined as~$\phi_\text{R} = \frac{M_\text{R}}{M_p}$
where~$M_p$ represents the total protein mass, and $M_\text{R}$ is the mass of proteins classified within the ``ribosomal sector''. The exact definition of this sector can vary across different studies (see below). 
Experimentally the sector is measured as the RNA-to-protein mass ratio, denoted by~$r = M_{\text{RNA}} / M_{\text{p}}$~\cite{neidhardt_studies_1960}, where~$M_\text{RNA}$ is the total RNA mass, which includes ribosomal RNA (rRNA), transfer RNA (tRNA), messenger RNA (mRNA), etc. 

From the definitions of~$r$ and~$\phi_R$, the conversion factor is~$C_{r\rightarrow\phi} = M_R / M_\text{RNA}$. By multiplying and dividing by~$M_{\text{rRNA}}M_{\text{rp}}$, which are the total mass of ribosomal RNA and ribosomal proteins of the considered cell culture, respectively, one obtains 
\begin{equation*}\label{eq:conversion_rho1}
     C_{r\rightarrow\phi} = \frac{M_{\text{rRNA}}}{M_{\text{RNA}}}  \frac{M_{\text{rp}}}{M_{\text{rRNA}}}  \frac{M_\text{R}}{M_{\text{rp}}} \,, 
\end{equation*}
and the sector~$\phi_\text{R}$ can be obtained as~$\phi_\text{R} = C_{r\rightarrow\phi} \, r$.
The first factor represents the mass fraction of RNA that is ribosomal (rRNA)~\cite{bremer2008modulation, maaloe1979biological}.
This factor is usually considered to be invariant with the growth rate~\cite{scott_interdependence_2010}. However, a classic study concluded that this may not be the case at slow growth~\cite{maaloe1979biological}.
The second term is the ratio between the mass of proteins and RNA in a ribosome (assuming perfect stoichiometry). 
Values for \textit{E.~coli} are given in Appendix~\ref{sec11:parameters_numbers}, Table~\ref{tab:params_sec_2_GL}.

The R sector sometimes appears as a pure ribosomal sector; 
in that case,~$M_\text{R} = M_\text{rp}$ represents the total mass of {\it ribosomal proteins only}~\cite{Dai2016, Erickson2017, balakrishnan_principles_2022, chure_optimal_2023}. 
However, the R sector can also be defined to include {\it ribosome-affiliated} proteins~\cite{scott_interdependence_2010, bremer2008modulation, Bosdriesz2015, Serbanescu2020new} in addition to the proteins constituting a ribosome (which are likely to be regulated as ribosomal proteins), see Appendix~\ref{sec11:parameters_numbers} Table~\ref{tab:params_sec_2_GL} for values.

\vspace{1ex} \noindent \textbf{Ribosome Number Fraction}

\noindent When working with number fractions instead of mass fractions~\cite{Roy2021, balakrishnan_principles_2022, calabrese_how_2024}, one can use the ratio between the total number of ribosomal proteins (which is different from the \textit{total number of ribosomes}~\cite{calabrese_how_2024}) and total number of proteins:~$\psi_\text{R} = n_\text{rp}^\text{ribo} R/P$, where~$R$ is the total number of ribosomes and~$n_\text{rp}^\text{ribo}$ the number of ribosomal proteins in a single ribosome.
The conversion factor~$C_{r\rightarrow\psi}$ can be found assuming perfect stoichiometry of the ribosomes, i.e. that all rRNA and ribosomal proteins are found in complete ribosomal subunits (note that this assumption is fulfilled by regulatory controls~\cite{Li2014,Taggart2018} but may be broken, e.g., in some perturbed conditions). Thus, the total number of ribosomes~$R$ is the ratio between the total rRNA mass and the mass~$\mu_\text{rRNA}^\text{ribo}$ of rRNA in a single ribosome:~$R = M_{\text{rRNA}}/\mu_\text{rRNA}^\text{ribo}$.
Eventually, we can write 
\begin{equation*}\label{eq:conversion_rho2}
     C_{r\rightarrow\psi} = \frac{\mu_{\text{p}}}{\mu_\text{rRNA}^\text{ribo}}  \frac{M_{\text{rRNA}}}{M_{\text{RNA}}}  n_\text{rp}^\text{ribo} \,, 
\end{equation*}
\edit{where~$\mu_{\text{p}}$ is the average mass of a typical protein, defined as~$\mu_{\text{p}} = M_{\text{p}}/P$. This factor emerges specifically because we are considering number fractions. }
It is also possible to pass from $\phi_\text{R}$ to $\psi_\text{R}$ via the equivalence $\psi_\text{R} = \phi_\text{R} L_\text{p} / L_\text{rp}$, where $L_\text{p}$ and $L_\text{rp}$ are the amino acid lengths of the typical protein and ribosomal proteins, respectively.

\vspace{1ex} \noindent \textbf{Single cell measurements}
Note that the previous paragraphs focus on bulk measurements of the ribosomal sector. However, several studies~\cite{zhang_decrease_2020, Panlilio2021, Iuliani2024} measure~$\phi_\text{R}$ and~$\phi_\text{P}$ employing fluorescent reporters that are subject to the same regulation~\cite{Iuliani2024}. This technique in principle makes possible to estimate the size of the sectors even at the single cell level.

\end{mdframed}

\vspace{0.75ex}
\noindent\textbf{Growth laws across species and beyond bacteria. }
The question on whether and to what extent growth laws hold across bacterial species and also across species of different kingdoms is open. However, sparse results show that they may be quite general~\cite{scott_shaping_2022}. For example, even classic studies showed these behaviors for different fast- and slow-growing bacteria. More interestingly, a growing body of evidence shows that growth laws hold for budding yeast, a central model system for eukaryotes.
A seminal work by Warner and Gorestein~\cite{warner_yeast_1978} experimentally shows that yeast too exhibits the stringent response when cells face amino acids deprivation, like bacteria cells. Their data point out that ribosomal proteins are regulated as a class under this condition. This last piece of evidence allows us to define the ribosomal sector for yeast, i.e. a group of co-regulated proteins corresponding to ribosomes and affiliated.
A more recent work~\cite{metzl-raz_principles_2017} confirms with proteomic data that the first growth law holds in yeast as well, and another work~\cite{xia_proteome_2022} shows growth laws for other sectors using proteomics.
These are remarkable results, considering that the mechanisms regulating resource allocation of budding yeast and {\it E.~coli} are extremely different, and that the same regulatory systems are highly conserved from yeast to many other eukaryotes.

\vspace{0.75ex}
\noindent\textbf{Rationalizing the empirical growth laws}.
Growth laws can be understood as emergent properties arising from (i) autocatalysis—ribosomes making ribosomes, (ii) flux balance, and (iii) resource allocation, all of which are reflected in the proteome composition. 
To better understand the phenomenological laws , Eqs.~(\ref{eq:GL_first})-(\ref{eq:GL_second}), here we use an approach based on three coarse sectors: the ribosome sector $\phi_\text{R}$, the carbon uptake sector included in $\phi_\text{P}$, and the housekeeping sector~$\phi_\text{Q}$. 
Note that the choice of defining three sectors is not arbitrary.
As pointed out before, \edit{when the growth rate is modulated by changing a single limiting factor,}
cellular responses 
can generally be categorized into three behaviors: up-regulation, down-regulation, and maintaining a constant level. Therefore, to construct a minimal model of proteome partitioning that takes into account all the proteins, it is necessary to define at least three distinct sectors.\\

\vspace{0.75ex}
\noindent\textit{First growth law.}
Following the procedure outlined in Refs.~\cite{ecker_ribosome_1963, schleif_control_1967, dennis_macromolecular_1974}, Maaløe~\cite{maaloe1979biological} and then Churchward~\cite{Churchward1982} formalized a mathematical derivation of the carbon-limited growth law:
The rate at which new amino acids are incorporated into proteins equals the average translation elongation rate (in amino acids per unit time),~$\varepsilon$, multiplied by~$R^\text{act}$, the number of ribosomes actively engaged in translation (i.e., elongating a new peptide). Denoting the total protein mass by~$M_p$, this relationship can be expressed as:
\begin{equation}
	\frac{\text{d}M_p}{\text{dt}}= \varepsilon  \mu_\text{aa}  R^\text{act}= \varepsilon \mu_\text{aa} R f_\text{a}.
	\label{eq:maaloe}
\end{equation}
In this equation,~$f_\text{a}$ represents the fraction of ribosomes that are translationally active, $R$ is the total number of ribosomes, and~$\mu_\text{aa}$ is the average mass of an amino acid. Ribosomes not engaged in translation (\textit{inactive} ribosomes) constitute the fraction~$f_\text{i} = 1 - f_\text{a}$.
Under exponential balanced growth,~$\dot{M}_\text{p} = \lambda M_p$ and substituting and rearranging terms yields:
\begin{equation}
	\phi_\text{R} f_\text{a} = \phi_\text{R}(1 - f_\text{i}) = \frac{\lambda}{\varepsilon_\text{R}}, \label{eq:GL_first_fa}
\end{equation}
where~$\phi_\text{R}$ is the ribosomal mass fraction, and~$\lambda$ is the growth rate. More in detail,~$\phi_R$ is the ratio between the ribosomal protein mass $M_\text{R} = R \mu_\text{aa} L_\text{R}$ and the total protein mass~$M_p$, and~$\varepsilon_\text{R} = \varepsilon/L_\text{R}$, where~$L_\text{R}$ is the number of amino acids present in all the proteins making a ribosome (plus, eventually, ribosome affiliated proteins). The quantity~$\varepsilon_\text{R}$ represents the inverse of the time necessary to assemble into proteins all the amino acids constituting a ribosomes, and it is connected to the translation efficiency.
This derivation leads to the phenomenological Eq.~(\ref{eq:GL_first}) if at zero growth all ribosomes are inactive, i.e.,~$\phi_\text{R} (\lambda = 0) = \phi_\text{R}^\text{min}$~\cite{Dai2016}.
The fraction of active (and thus inactive) ribosomes was inferred in~\cite{Dai2016} for \textit{E.~coli} and~\cite{metzl-raz_principles_2017} for \textit{S.~cerevisiae}, and we will provide more details in the next section. 
In both cases,~$f_\text{a}$ has been found to be a function of~$\lambda$ (see Sec.s~\ref{sec3:translation} and~\ref{sec5:degradation}). 

This law is inherently tied to the autocatalytic nature of ribosome production: ribosomes make proteins, including the ribosomal proteins required for assembling new ribosomes. This feedback loop establishes a direct connection between ribosome allocation and growth rate. As the growth rate increases, the demand for protein synthesis rises, necessitating a proportional increase in ribosome production to sustain exponential growth. \\

\vspace{0.75ex}
\noindent\textit{Second growth law.}
 
To study the impact of translation inhibition on the R sector (second growth law), it is essential to consider both the role of catabolism and the feedback from reduced biosynthesis currents.

In a steady state, where all cellular components accumulate at the same rate, protein synthesis and catabolic currents must remain balanced. If this balance is disrupted, precursors such as amino acids would either accumulate or be depleted (see~\ref{box5:aa_deg_recycling} for more details). Mathematically, this balance can be expressed as:
\begin{equation}
J_\text{cat}  = J_\text{syn} = \lambda M_p,
\label{eq:cat_syn_fluxes}
\end{equation}
where $J_\text{cat}$ represents the catabolic flux of precursors required for protein synthesis,
and $J_\text{syn}$ denotes the total biosynthetic flux, which utilizes these precursors. From Eq.~(\ref{eq:maaloe}), we can write~ $J_\text{syn}=\varepsilon_\text{R} M_\text{R}^\text{act}$.
For writing~$J_\text{cat}$, we assume the steady-state catabolic current is proportional to the amount of catabolic/transport proteins, and we can write~$J_\text{cat} \propto M_\text{cat}$. Catabolic proteins belong to the P sector, whose allocation typically shows the opposite trend to that of the R sector~\cite{hui_quantitative_2015}. \edit{Thus, $J_\text{cat} \propto M_P$, which, together with Eq.~(\ref{eq:cat_syn_fluxes}) leads to}
\begin{equation}
\lambda \propto  \frac{M_p}{M_p} \implies \lambda = \nu \phi_\text{P}\,,
\label{eq:nu_phiC}
\end{equation}
where the parameter $\nu$ is the ``nutrient quality'', related to the average flux per enzyme, which depends on the nutrient condition~\cite{Erickson2017}\footnote{If the main nutrient in the medium is a carbon source, nutrient uptake is handled entirely by the catabolic sector, $\phi_\text{C}$. This sector, regulated by cAMP, consists of various substrate-specific transporters (see Sec.~\ref{sec7:shifts} and Refs.~\cite{you_coordination_2013, Hermsen2015, Erickson2017} for more details). However, in richer media —e.g., containing amino acids— the uptake process is less well characterized. In such cases, the precursor supply rate is often assumed to be proportional to the $\phi_\text{P}$ sector, which shows the inverse correlation with the R sector described by Eq.~\eqref{eq:phiQ}. This modeling choice is adopted in Ref.~\cite{Droghetti2025}. Naturally, the proportionality constant $\nu$ must be adjusted depending on whether $\phi_\text{C}$ or $\phi_\text{P}$ is used, in order to match the observed exponential growth rate.}.

Working at fixed nutrient condition, i.e. at fixed~$\nu$, when adding chloramphenicol, the trade-off between the R and P sectors provides an explanation for the second growth law, which describes how~$\phi_\text{R}$ changes under translation inhibition. Using the constraint in Eq.~(\ref{eq:sector_constraint}), this trade-off can be expressed as:
\begin{equation}
\phi_\text{P} = (1 - \phi_\text{Q}) - \phi_\text{R} = \phi_\text{R}^\text{max} - \phi_\text{R}, \label{eq:phiQ}
\end{equation} 
where~$\phi_\text{Q}$ represents the housekeeping sector, assumed to remain constant~\cite{scott_interdependence_2010, scott_emergence_2014}. Substituting this expression into Eq.~(\ref{eq:nu_phiC}) yields the second growth law, Eq.~(\ref{eq:GL_second}).

\subsection*{OUTLOOK} 

The growth laws and proteome sectors described in this section provide a foundational phenomenological framework for understanding the relationship between cellular growth and resource allocation. While this initial framework performs remarkably well in capturing general trends, it opens critical  questions about the underlying mechanisms and motivates further extensions of the theory. 

For instance, beyond the empirical linear relationships between growth rate and proteome allocation the coordinated re-allocation of proteome sectors requires specific regulatory pathways in order to integrate environmental changes into reprogrammed gene expression patterns, but the current framework does not explain how molecular processes, such as ribosome competition and regulation or resource allocation dynamics, give rise to these trends. 

By showing how to build simple prototype models able to deal with these processes, we aim to present in the remainder a toolbox able to better capture the complexity of cellular growth under diverse conditions.

It is important to emphasize that the choice of relevant sectors to include in a model depends critically on the specific perturbation used to modify the growth rate. This section employed a simple description based on the P and Q sectors, defined based on their scaling with nutrient quality. However, in other contexts it becomes necessary to explicitly include different sectors  to accurately capture the system’s behavior. 
For example, not all the P sector performs metabolism, and sec.~\ref{sec7:shifts} will explicitly include a ``catabolic'' sector (with fraction $\phi_\text{C}$)  to accurately capture the dynamics of nutrient shifts. More broadly, the literature currently lacks a strict consensus on the exact functional roles or definitions of various proteome sectors. 
For instance, the Q sector is typically defined empirically as the proteome fraction that remains approximately constant in response to a given perturbation, and is sometimes referred to as the housekeeping sector. However, this definition --and the resulting composition of the Q sector-- can vary substantially depending on the nature of the \edit{growth limitation}~\cite{hui_quantitative_2015, Mori2017}. Moreover, the regulatory mechanisms responsible for maintaining a stable $\phi_\text{Q}$ remain largely uncharacterized~\cite{Ripamonti2025}. To account for this variability, we recommend adopting flexible definitions that are specified clearly, allowing modeling approaches to be tailored to specific experimental conditions and research questions.

\section{Main Course: On active ribosomes and a more mechanistic view of translation} \label{sec3:translation}
 \vspace{-.3cm}
This section introduces a mechanistic layer on the standard theory that focuses on ribosome usage and its role in protein synthesis, looking at the dynamics of ribosome recruitment.

\subsection*{MAIN QUESTIONS AND MOTIVATIONS}
The biomass production flux is expressed as~$J_\text{syn} = \varepsilon  \mu_\text{aa}  R f_\text{a}$, Eq.~(\ref{eq:maaloe}). In this purely ribosome-centered perspective, only  actively translating ribosomes described by the fraction~$f_\text{a}$ play a key role in determining protein synthesis, yet experimentally their direct quantification remains challenging. 
Ribosome-sequestering factors have been identified both in bacteria and in yeast, but their regulation, physiological role across conditions and effective contribution to the fraction of actively translating ribosomes remains unclear~\cite{Dai2020, calabrese_protein_2022, Cherkasov2015, Gemin2024}. 

Experimental evidence in \textit{E.~coli}~\cite{Dai2016} shows that the ribosome elongation rate~$\varepsilon$ depends on the growth condition, varying along both the first and second growth laws, i.e., under changes in nutrient quality or translation inhibition, Fig~\ref{fig2}(a). The fraction~$f_\text{a}$ of active ribosomes across nutrient conditions can be estimated as a function of the growth rate~$\lambda$~\cite{Dai2016}, by integrating experimental data with the phenomenological model described by Eq.~(\ref{eq:maaloe}). This fraction is generally considered an increasing function of~$\lambda$, rapidly saturating to high values at moderate to fast growth rates (e.g., reaching~$\simeq 90\%$ for~$\lambda \gtrsim 0.5$/h in \textit{E. coli}). An illustration of~$f_\text{a}$ is given in Fig.~\ref{fig2}(b). A similar trend has been observed in yeast, where measurements of polysome fractions (complexes of an mRNA with two or more ribosomes) were argued to provide an analogous estimate of ribosome activity~\cite{metzl-raz_principles_2017}. However, that estimate relies on the assumption that the translation elongation rate is constant across conditions. 

Despite these insights, several fundamental questions remain open. The first is how  the total ribosome pool, accessible through the RNA-to-protein ratio~$r$ or by proteomics, relates to the fraction of actively elongating ribosomes. Additionally, the functional role of inactive ribosomes in cellular physiology is not yet fully understood. Finally, the model introduced in the previous section does not explicitly describe ribosome recruitment, i.e., the formation of the transcript-ribosome complex that determines the number of active ribosomes~$R^\text{act}$ in the equation for~$J_\text{syn}$.
Essentially, the basic theory does not account for the binding-unbinding kinetics, where ribosomes process, detach, and rebind to mRNAs (transcripts), forming a dynamic pool of free and bound-active ribosomes.
Moreover, a more mechanistic view of this phenomenon helps identifying what are the limiting (or co-limiting) factors for growth.

Here, we will present two prototype models. First, a purely phenomenological description of~$f_\text{a}$ that neglects ribosome turnover~\cite{Dai2016,Wu2022}. Second, a theory that assumes that the only contribution to active ribosomes is given by the kinetics of binding and unbinding (recruitment and usage) of ribosomes on mRNAs~\cite{calabrese_how_2024}.
Interestingly, this second theory immediately leads to an expression for the growth rate that involves the total pool of messenger RNAs, complicating the picture with respect to the idea that ribosome autocatalysis is the sole driver of cellular growth. The two models are not mutually exclusive, and future contributions will likely have to integrate these and other aspects to fully capture ribosome activity.  

\subsection*{INGREDIENTS AND RECIPES\footnote{The recipe introduced here builds upon ideas and models presented in Refs.~\cite{Dai2016, calabrese_how_2024}.}}

\noindent\textbf{The elongation rate is a function of ribosome allocation}.
In order to proceed, we need to introduce a crucial ingredient that will be further revisited in the following sections: the dependence of the ribosome elongation rate~$\varepsilon$ on growth rate and ribosome allocation. Establishing this relationship is key to understanding how the fraction of active ribosomes is intrinsically linked to the elongation rate before transitioning to a more mechanistic perspective on ribosome recruitment and usage.

\begin{figure}[h!]
    \centering 
    \includegraphics[width = 1\columnwidth]{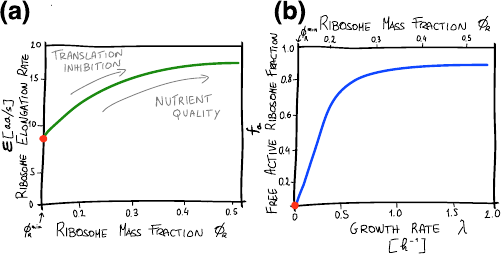}
    \caption{
    \textbf{Translation elongation rate and ribosome activity as function of growth rate.}
    (\textbf{a}) The ribosome elongation rate~$\varepsilon$ as a function of~$\phi_\text{R}$. The values of~$\varepsilon$, obtained by varying nutrient conditions and under sub-lethal antibiotic dosage (chloramphenicol), collapse when plotted as a function of~$\phi_\text{R}$, see~\cite{Dai2016}.  
    (\textbf{b}) The active ribosome fraction $f_\text{a}$ as a function of the growth rate and~$\phi_\text{R}$, predicted by Eq.~(\ref{eq:fa2}).\edit{This figure is inspired by and adapted from Ref.~\cite{Dai2016}.}}
    \label{fig2}
\end{figure}

We first introduce this dependence as an empirical observation.
The ribosome elongation rate,~$\varepsilon$, can be experimentally measured (see, for instance, the SI of Ref.~\cite{Dai2016} for a detailed explanation of the experimental design). In \textit{E.~coli}, across nutrient conditions (i.e. along the first growth law), it has been shown to be solely a function of growth rate, or equivalently of ribosome allocation, following a saturating relation with both quantities, which can be expressed as
\begin{equation}
	\varepsilon = \varepsilon^\text{max} \cfrac{\phi_\text{R}}{\phi_\text{R} + \phi_\text{R}^\text{min}} \,,
	\label{eq:epsilon}
\end{equation} 
where~$\varepsilon^\text{max}$ represents the maximum possible elongation rate, reached at fast growth when $\phi_\text{R} \gg \phi_\text{R}^\text{min}$ \cite{Wu2022}. 
In practice,~$\varepsilon \simeq \varepsilon^\text{max}$ for $\lambda \gtrsim 0.5$/h in \textit{E.~coli}. A sketch of~$\varepsilon(\phi_\text{R})$ is shown in Fig.~\ref{fig2}(a). \edit{Equation~(\ref{eq:epsilon}) can be rationalised mechanistically, taking into account that the speed of the ribosomes is affected by the availability of ternary complexes~\cite{Dai2016}, and considering the empirical relation between ppGpp levels and the ribosome elongation rate~\cite{Wu2022}. We provide a detailed derivation of Eq.~(\ref{eq:epsilon}) in Appendix~\ref{sec11:parameters_numbers}.}\\

\noindent\textbf{Fraction of active ribosomes in linear growth laws}. 
Given this knowledge, we now consider a phenomenological theory for the fraction of active ribosomes~\cite{Dai2016, Wu2022}.
To ensure consistency with the linearity of the phenomenological first growth law~(\ref{eq:GL_first}), the inclusion of the fraction of active ribosomes in~(\ref{eq:GL_first_fa}) imposes the following constraint: 
\begin{equation}
	f_\text{a} = \cfrac{\gamma_\text{0}}{\varepsilon_\text{R}} \left( 1 - \cfrac{\phi_\text{R}^\text{min}}{\phi_\text{R}} \right) \,.
	\label{eq:fa1}
\end{equation}
We recall that~$\gamma_\text{0}$ is a phenomenological parameter quantifying the inverse of the slope of the  ribosomal growth law,  while~$\varepsilon_\text{R}$ is a physical parameter derived in the previous section, Sec.~\ref{sec2:classical}, which is linked to the translation elongation rate by the relation~$\varepsilon_\text{R} = \varepsilon / L_\text{R}$, meaning that it is the inverse of the time needed to sequentially translate the total number $L_R$ of amino acids in a ribosome.

Substituting~$\varepsilon(\phi_\text{R})$ from Eq.~(\ref{eq:epsilon}) into Eq.~(\ref{eq:fa1}), 
we obtain the dependence of~$f_\text{a}$ on the ribosome fraction:
\begin{equation}
	f_\text{a} = \cfrac{\gamma_\text{0} L_\text{R}}{\varepsilon^\text{max}} 
    \left[ 1 - \left( \cfrac{\phi_\text{R}^\text{min}}{\phi_\text{R}} \right)^2 \right] \,.
	\label{eq:fa2}
\end{equation}

Assuming that all ribosomes are active at fast growth, the maximal elongation rate can be related to the slope of the phenomenological growth law as~$\gamma_\text{0} = \varepsilon^\text{max}/ L_\text{R}$.
Thus, in order to be in accordance with a linear relationship~$\phi_\text{R}(\lambda)$, the amount of active ribosomes and the elongation rate must satisfy the relation~$\varepsilon f_\text{a} \phi_\text{R} = \varepsilon^\text{max} (\phi_\text{R} - \phi_\text{R}^\text{min}) $. 
Section~\ref{sec6:mecha_regulation} will discuss how the hypothesis that ppGpp regulates ribosome-sequestering factors~\cite{Wu2022} can lead to Eq.~(\ref{eq:fa2}). 
\\

\noindent\textbf{A mechanistic model for ribosome recruitment leads to a growth rate explicitly depending on the messenger RNA pool}. 
We now take a different direction, and describe an approach that aims to provide a mechanistic explanation for ribosome activity due to their binding and release dynamics from the transcripts, 
assuming for simplicity that there are no sequestration factors~\cite{calabrese_how_2024}. More specifically, the framework  considers (i) ribosome recruitment (initiation) and (ii) ribosome usage (translation elongation) in \textit{E. coli}. 
A related analysis in \textit{B. subtilis} can be found in Ref.~\cite{Borkowski2016}.

To define this model, we reformulate the expression for the protein production rate~$\dot{M}_\text{p}$ by expressing it in terms of the \edit{intensive} ribosomal current~$J_\text{i}^\text{TL}$, due to the ribosomes that are actively translating. The quantity~$J_\text{i}^\text{TL}$ serves as a proxy for the protein synthesis rate of a transcript encoded by the gene~$\text{i}$. It depends on the initiation rate~$\alpha^\text{TL}$, which represents the rate at which a ribosome attempts to bind an mRNA, and the translation elongation rate~$\varepsilon$.
Thus, we can express the variation in total protein mass as
\begin{equation}
    \frac{\text{d}M_p}{\text{dt}}= \lambda M_p = \sum_\text{i}m_\text{i} \, \mu_{\text{aa}} L_\text{i} \, J_\text{i}^\text{TL} (\alpha^\text{TL}, \varepsilon)\,,
    \label{eq:mech_mass_current}
\end{equation}
where the sum runs over all protein-coding genes~$\text{i}$,~$m_\text{i}$ is the number of transcripts 
of gene~$\text{i}$,~$L_\text{i}$ is the number of codons in that mRNA, and~$J_\text{i}^\text{TL}$ is the 
translation rate per mRNA (which we can term ``ribosomal current''). \edit{ The right hand side of Eq.~(\ref{eq:mech_mass_current}) links the translational current to the extensive biosynthetic flux $J_\text{syn}$ introduced in the previous section.}

For the sake of simplicity, we assume that all genes share the same translation initiation rate~$\alpha$ and elongation dynamics~$\varepsilon$. However, this framework can be easily generalized by allowing gene-specific variations in these parameters~\edit{\cite{Wang2021}}. For instance, ribosome elongation speed can in principle vary across genes due to factors such as codon usage bias. Initiation rates can be controlled by properties of the mRNAs sequence~\cite{salis_automated_2009} and, in general, they are determined by the amount of freely available ribosomal subunits via a mass action law. All transcripts thus compete for ribosome recruitment, and are coupled to a common finite pool of unbound and free ribosomes denoted by $R^\text{f}$~\cite{greulich_mixed_2012}. 

The total number of ribosomes per cell can be expressed as
\begin{equation}
    R=R^\text{f}+\sum_\text{i}m_\text{i}\rho_\text{i} L_\text{i},
    \label{eq:mecha_ribosomes}
\end{equation}
in which~$\rho_\text{i}$ is the ribosome density (number of ribosomes on a transcript per length of the transcript). It is possible to make different choices and approximations to model the ribosome density and current; here, we will use concepts from the exclusion process, a prototypical model of one-dimensional traffic studied in statistical mechanics and originally proposed in the context of mRNA translation~\cite{macdonald_kinetics_1968, macdonald_concerning_1969,chou_non-equilibrium_2011}.

In the exclusion process, the current~$J^\text{TL} (\alpha^\text{TL}, \varepsilon)$ depends on both the initiation rate~$\alpha^\text{TL}$ and the elongation rate~$\varepsilon$. However, in many biological conditions the translation process is often initiation-limited~\cite{szavits-nossan_deciphering_2018, balakrishnan_principles_2022}. That is, once a ribosome successfully initiates translation, it tends to travel along the mRNA without significant interference from other ribosomes. In this regime, ribosome traffic (in biological terms ``ribosome collisions'') is negligible and one may approximate~$ J^\text{TL} \sim \alpha^\text{TL}$.

In this situation, we can describe the recruitment of ribosomes to the mRNA as a simple mass-action process and assume the effective initiation rate to be proportional to the concentration of free ribosomes, denoted by~$[R^\text{f}]$. The proportionality constant~$\alpha_\text{0}^\text{TL}$ captures the intrinsic efficiency with which a ribosome binds to the start site. This reasoning leads directly to the relation
\begin{equation}
    \alpha^\text{TL} = \alpha_\text{0}^\text{TL} [R^\text{f}] \,,
    \label{eq:initiation}
\end{equation}
which couples protein synthesis to the pool of free ribosomes.

On a single mRNA, the number of bound ribosomes on a transcript per transcript length (the density~$\rho$, which is set by the relative weight of translation initiation and elongation~\cite{shaw_totally_2003}) is  
\begin{equation}
    \rho = \frac{\alpha_\text{0}^\text{TL}}{\varepsilon} [R^\text{f}] \,,
    \label{eq:ribo_density}
\end{equation}
(see Refs.~\cite{calabrese_protein_2022, calabrese_how_2024} for a simple derivation).

Following the approximation~$J^\text{TL} \sim \alpha^\text{TL}$ 
and using 
Eqs.~(\ref{eq:mecha_ribosomes})-(\ref{eq:ribo_density}), we obtain the following expression for the protein synthesis current:
\begin{equation}
	J^\text{TL} = 
    \alpha_\text{0}^\text{TL} [R] \cfrac{1}{1 + \cfrac{[m]}{K_m}} \,,
    \label{eq:J_TL}
\end{equation}
where we defined~$K_m = \frac{\varepsilon}{L_\text{p} \alpha_\text{0}^\text{TL}}$ and~$L_\text{p}$ is the length of the typical protein. 
This equation retains the dependence on ribosome availability, as in Eq.~(\ref{eq:maaloe}), but it also explicitly captures the competition between mRNAs for ribosomes. The factor~$1/\left( 1 + \frac{[m]}{K_m} \right)$ represents the fraction of ribosomes that remain free and available to initiate translation, while the term~$[m]/K_m$ characterizes the balance between supply and usage of ribosomes, i.e. the time a ribosome spends searching for a transcript,~$1/( \alpha_\text{0}^\text{TL} [m])$, and the time required to complete the  elongation of a typical protein and detach,~$L_\text{p}/\varepsilon$, see also Fig.~\ref{fig3}.

Importantly, the initiation rate varies with mRNA concentration, reaching its maximum~$\alpha_\text{0}^\text{TL} [R]$ for vanishing mRNA concentrations, since all ribosomes are free and available for initiation. However, this is also the regime where elongation is not possible since the fraction of ribosomes on the messenger also vanishes. 
For this reason, from Eq.~(\ref{eq:mech_mass_current}) one obtains a modified version of the usual relation between growth rate and ribosome mass fraction $\phi_\text{R}$, including the concentration of total transcripts $[m]$:
\begin{equation}
	\lambda = 
    \varepsilon_\text{R}  \phi_\text{R} \frac{[m]}{K_m + [m]} \,.
    \label{eq:CFlim_GL}
\end{equation}
The details of the derivation can be found in Ref.~\cite{calabrese_how_2024}. Equation~(\ref{eq:CFlim_GL}) clarifies how, even in absence of ribosome sequestration, the mRNA pool can contribute to the fraction of actively translating ribosomes. The interplay between transcript availability and ribosome binding, and consequently the dependence of the growth rate on the mRNA pool becomes significant when~$[m] \simeq K_m$ (see~\ref{box3:Km}). The following section will expand this model by explicitly incorporating transcription.

\begin{mdframed}[backgroundcolor=light-gray]
\refstepcounter{boxcounter}
{\bf \large \theboxcounter. On the relevance of a ``complex-formation limited'' regime}\\
\label{box3:Km}
\noindent A back-of-the-envelope estimate suggests that the parameter 
$K_m$ appearing in Eq.~(\ref{eq:CFlim_GL}) falls in the range $0.05$ to $0.15 \, \mu$M 
for \textit{E.~coli} and $0.25$ to $0.5 \, \mu$M for \textit{S.~cerevisiae}~\cite{calabrese_how_2024}.

These values are roughly an order of magnitude lower than the typical mRNA concentration in \textit{E.~coli}~\cite{balakrishnan_principles_2022} and yeast~\cite{rob_philips_ron_milo_cell_2015} at fast growth, 
placing the system in 
a condition that is in between a regime in which ribosomes are the main limiting component of growth laws, and another one with strong ribosome-mRNA complex-limited competition.
Following Ref.~\cite{calabrese_how_2024} we name the first regime ``translation limited'' (TL-LIM), and the co-limited regime as ``complex-formation limited'' (CF-LIM).

The estimates for $K_m$ suggest that, at fast growth, small changes from the physiological mRNA levels do not significantly impact growth, whereas a substantial reduction in mRNA 
availability could lead to pronounced effects on the growth rate. Notably, mRNA levels in \textit{E.~coli} can vary over an order of magnitude across growth conditions~\cite{balakrishnan_principles_2022}, and under slow growth conditions one can have~$[m] \approx K_m$, reinforcing the relevance of the CF-LIM regime in determining translational constraints.
\end{mdframed}

\subsection*{OUTLOOK}

This section provides the toolbox for two orthogonal looks on ribosome activity: the overall requirements on ribosome activity due to the growth laws and the prediction of free non-translating ribosomes due to initiation-termination kinetics. 
Sec.~\ref{sec6:mecha_regulation} will address in further detail the molecular mechanisms that govern the interplay between these phenomena. 
It is  often hypothesized that  regulation of ribosome activity ensures that bacterial cells maintain a balance between rapid recovery from slow-growth conditions and optimal resource allocation for steady-state growth~\cite{Mori2017, Dai2020}.
Section~\ref{sec5:degradation} will also show that due to protein degradation, at vanishing growth a significant fraction of ribosomes is expected to remain active but engage in maintenance tasks, such as replacing degraded proteins, thereby not directly contributing to biomass production, and showing up as ``inactive'' in the frameworks presented here.

Mechanistic models of translation offer a different interpretation and classification of active and inactive ribosomes compared to standard theories. Ribosomes that are recruited and bound to mRNAs are the ones contributing to the biosynthesis current, effectively serving as the ``active'' ribosomes in this framework. The model presented here focuses on the competition between different mRNAs and a pool of ``free'' ribosomes, emphasizing the limiting role of translation initiation, particularly in forming the ribosome-transcript complex.
This approach could be further extended to include ``inactive'' ribosomes, aligning it with current knowledge on ribosome hibernating factors (see, e.g. Ref.~\cite{Wu2022}).

The theory proposed here also enables a more detailed accounting of ribosome usage via ribosome density~$\rho$, which depends on both initiation and elongation rates. Notably, a recent study by Balakrishnan et al.~\cite{balakrishnan_principles_2022} suggested that ribosome density remains constant across growth conditions in \textit{E. coli}, indicating a coordinated regulation of elongation and initiation rates. 
Extensions of this model, which decouples the role of initiation and elongation, may be able to provide a description for this regulatory mechanism.

The theory presented in this section neglects ribosome traffic, a factor generally considered negligible~\cite{balakrishnan_principles_2022}. However, sub-lethal doses of antibiotics can transiently bind to 
ribosomes and inhibit elongation, preventing them from contributing to protein synthesis.
Incorporating ribosome traffic into this model could provide valuable insights into the physiology of bacteria growing in sublethal antibiotic levels. In such scenarios, antibiotic-bound ribosomes block the passage of trailing ribosomes, effectively halting the ribosomal current along the transcript and creating bottlenecks in translation~\cite{Kavcic2020, keisers_biologically_2024, keisers_paused_2025}.
Quantifying the role of these stalled ribosomes could help refine our understanding of translation dynamics under antibiotic stress and its influence on growth laws. 

Distinguishing between free and bound ribosomes, as well as between active ribosomes contributing to biomass production and inactive ones that are either sequestered or targeted by antibiotics, remains an experimental challenge. Current estimates often rely on models rather than direct measurement. However, recent advances in single-molecule techniques, such as the ribosome-labeling method described in Metelev et al.~\cite{metelev_direct_2022}, offer promising insights. This technique enables the tracking of ribosomal subunits in live cells, providing real-time information about their dynamics and functional states (free vs. bound). Metelev et al. demonstrated that more than~$90\%$ of ribosomes are engaged in translation under fast growth conditions, in coherence with the frameworks presented in this Tutorial. 

This section described a mechanistic model of mRNA translation that takes into account ribosome usage, moving beyond a purely phenomenological description. This reflects a modeling approach that incorporates mechanistic details to interpret experimental data more directly, while avoiding unnecessary complexity.
Starting from the framework presented in this section, one can develop models that remain tractable and offer a bottom-up description of ribosome allocation. In the next section, we apply the same strategy to transcription, showing parallel trade-offs in resource allocation between ribosomes and RNA polymerases.

\section{Side Dish: Linking mRNA Abundance to Growth} \label{sec4:transcription}
The previous section showed that mRNA abundance influences ribosome recruitment, see Eq.~(\ref{eq:CFlim_GL}), thereby establishing a framework in which both mRNA and ribosomes co-limit protein synthesis and cell growth. Building on this insight, this section extends the model by explicitly incorporating transcription dynamics. This addition provides specific tools to explore how RNA polymerase (RNAP) abundance governs transcript levels, ultimately linking transcriptional capacity to growth.

\subsection*{MAIN QUESTIONS AND MOTIVATIONS}
Ribosomes are widely considered the primary limiting factor for protein synthesis, which is generally accurate under most conditions~\cite{scott_interdependence_2010, scott_emergence_2014, Erickson2017}. However, recent evidence indicates that mRNAs can become the sole limiting component under specific extreme perturbations. For example, inhibiting DNA replication was shown experimentally to make chromosome concentration the limiting factor of transcription in {\it E.~coli}~\cite{makela_genome_2024}, and in arrested budding yeast cells~\cite{neurohr_excessive_2019}. 
Inducing excessive exogenous gene expression in budding yeast with different levels of mRNA degradation was also shown to generate growth rates that depend on mRNA levels~\cite{kafri_cost_2016}.
Furthermore, recent findings suggest that RNA polymerase II (RNA Pol II) acts as a major limiting factor for transcription in budding yeast~\cite{swaffer_RNA_2023}. RNA Pol II is responsible for transcribing protein-coding genes (mRNAs), as well as many non-coding RNAs involved in gene regulation. Specifically, its transcriptional capacity at a given cell size appears to be governed by the mass-action recruitment of free nucleoplasmic RNA Pol II to the genome. This would imply that the transcriptional output is primarily constrained by the availability of free RNA Pol II.

Several theoretical studies have explored scenarios in which transcripts rather than ribosomes limit growth~\cite{Lin2018, Roy2021}. These studies typically focus on conditions where RNA polymerases saturate genes~\cite{Lin2018}  or where RNAP availability constrains biosynthetic fluxes making transcripts the primary bottleneck and leading to RNAP-specific growth laws~\cite{Roy2021}. 

In the previous section, the mRNA concentration~$[m]$ was treated as a fixed parameter derived from empirical data (e.g. using Ref.~\cite{balakrishnan_principles_2022} for \textit{E.~coli}). However, it is straightforward to construct models where~$[m]$ emerges dynamically from the transcription process. Incorporating transcription into the introduced models helps address important questions --such as how transcription-targeting drugs affect growth and how transcription and translation are coordinated-- thus offering deeper insights into the role of transcriptional capacity in cellular physiology. 

\subsection*{INGREDIENTS  AND RECIPIES\footnote{The recipe introduced here builds upon ideas and models presented in Refs.~\cite{Lin2018, calabrese_how_2024, balakrishnan_principles_2022}.}}

\begin{figure}[h!b]
    \centering 
    \includegraphics[width = 1\columnwidth]{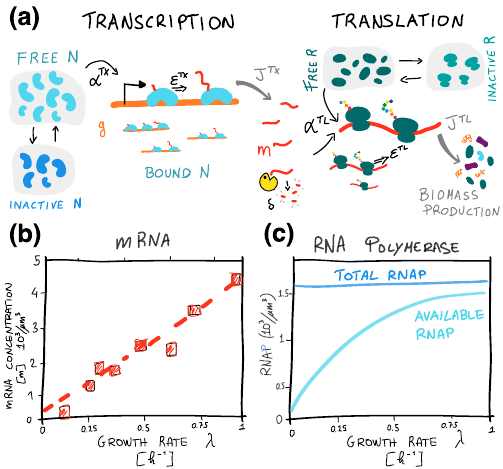}
    \caption{
    \textbf{Scheme and prediction of the transcription-translation model.}
    (\textbf{a}) Scheme of the model recipe considering transcription and translation initiation rates~$\alpha^\text{TX}$ and~$\alpha^\text{TL}$ as dependent on the availability of free RNAPs and ribosomes, respectively. mRNAs are synthesized with a flux~$J^\text{TX}$, then either degraded at rate~$\delta$ or used as templates for translation. Both unbound RNAPs and ribosomes can be sequestered by inactivation factors, reducing their availability for gene expression.
    (\textbf{b}) Illustration of the mRNA growth law~\cite{balakrishnan_principles_2022}, showing the roughly linear relationship between total mRNA concentration and growth rate. (\textbf{c}) While the total RNAP \edit{concentration} remains approximately constant across conditions, sequestration factors modulate the fraction of RNAPs actively engaged in transcription~\cite{balakrishnan_principles_2022}. \edit{Panel (\textbf{b}) is inspired by and adapted from Ref.~\cite{balakrishnan_principles_2022}.}
    }
    \label{fig3}
\end{figure}

\noindent {\bf A basic Transcription-Translation model}. We now formulate a gene expression model considering the time evolution of the abundances of transcripts~$m_\text{i}$ and proteins~$P_\text{i}$ for the gene or sector $i$.
Following for instance~\cite{Lin2018}, this takes the form 
\begin{equation}
  \frac{\text{d}m_\text{i}}{\text{dt}} = g_\text{i} \; J_\text{i}^\text{TX}([N]) - \delta_\text{i} \; m_\text{i} \,,
  \label{eq:transcript_dynamics}
\end{equation}
\begin{equation}
  \frac{\text{d}P_i}{\text{dt}} = m_\text{i} \; J_\text{i}^\text{TL}([R])\,,
  \label{eq:protein_dynamics}
\end{equation}
where the \edit{intensive} fluxes~$J_\text{i}^\text{TX}$ and~$J_\text{i}^\text{TL}$ represent the rate of mRNA synthesis per gene (RNA polymerase flux) and the rate of protein synthesis per mRNA (ribosome flux per mRNA), respectively. In Eq.~(\ref{eq:transcript_dynamics}), the RNA polymerase flux per gene,~$J_\text{i}^\text{TX}$, is multiplied by the copy number of gene i, denoted by~$g_\text{i}$. The term~$\delta_\text{i}$ represents the transcript degradation rate, which cannot  be neglected due to the relatively fast turnover of mRNA. 
Irrespective of their exact functional form, the transcriptional and translational 
fluxes~$J_\text{i}^\text{TX}$ and~$J_\text{i}^\text{TL}$ are determined by the total concentrations of RNA polymerases and ribosomes,~$[N]$ and~$[R]$, respectively, through their corresponding initiation rates. This dependency is explicitly stated in Eqs.~(\ref{eq:transcript_dynamics})-(\ref{eq:protein_dynamics}). A sketch of the model is given in Fig.~\ref{fig3}(a).

Since transcript degradation occurs at a rate orders of magnitude faster than dilution (growth rate)~\cite{Lin2018, balakrishnan_principles_2022,chen_genome_wide_2015}, we can assume a quasi-steady-state condition for mRNA. Under this assumption, we can write~$m_\text{i} = g_\text{i} \; J_\text{i}^\text{TX} / \delta_\text{i}$, which, when substituted into 
Eq.~(\ref{eq:protein_dynamics}), yields:
\begin{equation}
  \frac{\text{d}P_\text{i}}{\text{dt}} = \cfrac{g_\text{i} \; J_\text{i}^\text{TX} \; J_\text{i}^\text{TL}}{\delta_\text{i}}\,.
  \label{eq:protein_JTX}
\end{equation}
Eq.~(\ref{eq:protein_JTX}) couples the transcriptional and translational fluxes, and under balanced growth defines the usual growth laws (see Sec.~\ref{sec3:translation}). 

\begin{mdframed}[backgroundcolor=light-gray]
\refstepcounter{boxcounter} 
{\bf \large \theboxcounter. Saturation regimes for genes and mRNAs}\\
\label{box4:TX-LIM}
\edit{
\noindent Lin and Amir~\cite{Lin2018} developed a transcription/translation theoretical model to describe how growth constraints emerge from the saturation of genes by RNAP and mRNAs by ribosomes. 

In their approach, the transcription rate of a gene $\text{i}$ is determined by the total number of RNAPs and by the gene allocation fraction $g_\text{i}/\left( \sum g_\text{i} \right)$, i.e. the fraction of RNAPs transcribing that particular gene. Similarly, translation depends on the number of active ribosomes and the fraction $m_\text{i}/\left( \sum m_\text{i} \right)$. 

In a ``gene-limited'' regime, genes are saturated with RNAPs, meaning that increasing RNAP availability does not increase mRNA production. Practically, each gene is considered to have an upper bound for the maximal number of concurrently transcribing RNAPs.  Above this threshold, DNA becomes limiting and the transcription rate is determined by the gene number $g_\text{i}$. 
Likewise, in a fully transcription-limited regime, mRNAs are saturated with ribosomes, meaning that additional ribosomes do not enhance protein synthesis. In this regime, growth is constrained by the number of available transcripts $m_\text{i}$ rather than by ribosome concentration, in contrast with the scenario suggested by the first growth law. 

Similarly, it is possible to develop approaches considering limiting steps in different autocatalytic cycles involved in biosynthesis, including the ribosome, RNA polymerase, and transfer RNA (tRNA) cycles~\cite{Roy2021}.
}

\end{mdframed}

\vspace{0.75ex}

\noindent {\bf RNA polymerase turnover and allocation set the transcriptional flux.}
Similarly to the previous section, we assume here that the transcriptional flux is initiation limited, and~$J_\text{i}^\text{TX} \sim \alpha_\text{i}^\text{TX}$, where~$\alpha_\text{i}^\text{TX}$ represents the transcription initiation rate (i.e., RNA polymerase recruitment). 
Following Ref.~\cite{calabrese_how_2024}, we divide the RNA polymerase pool into free and bound enzymes:~$N = N^{\rm f}+\sum_\text{i}  \ell_\text{i} \rho^\text{TX}_\text{i} g_\text{i}$, where the bound RNA polymerases are counted by summing the occupation on the different genes, i.e. the linear density~$\rho^\text{TX}_\text{i}$ multiplied by the gene length~$\ell_\text{i}$, times the copy number~$g_\text{i}$ of that gene. Note that we use the symbol~$\ell$ to denote lengths in base pairs (bp), while we use~$L$ for amino acid-based lengths.
By mirroring the approach used to model translation in the previous section, we first express the transcription initiation rate as a function of the concentration of free and available RNA polymerases, such that~$\alpha_\text{i}^\text{TX} = \alpha_\text{0,i}^\text{TX} [N^\text{f}]$. Next, considering that the RNA polymerase density on gene i is given by the ratio of the transcription initiation and elongation rates~($\rho^\text{TX}_\text{i} = \alpha_\text{i}^\text{TX} / \varepsilon^\text{TX}$), we express the transcriptional flux~$J_\text{i}^\text{TX}([N])$ as
\begin{eqnarray}
	J_\text{i}^\text{TX}([N]) & = & \alpha_\text{0,i}^\text{TX} \cfrac{1}{1 + \sum_\text{j} \ell_\text{j}  \frac{\alpha_\text{0,j}^\text{TX}}{\varepsilon^\text{TX}}[g_\text{j}]} [N] \nonumber\\ 
	& = & \cfrac{\varepsilon^\text{TX}}{\ell_\text{i}} \cfrac{\omega_\text{i}}{[g_\text{i}]} f_\text{bn} [N] \,, \label{eq:J_i_TX}
\end{eqnarray}
where we have defined
\begin{equation}
	\omega_\text{i} = \cfrac{\ell_\text{i} \alpha_\text{0,i}^\text{TX} [g_\text{i}]}{\sum_\text{j} \ell_\text{j} \alpha_\text{0,j}^\text{TX} [g_\text{j}]} \,,
	\label{eq:RNAP_alloc}
\end{equation}
and 
\begin{equation}
	f_\text{bn} = 1- \frac{N^\text{f}}{N} = \cfrac{\sum_\text{i} \ell_\text{i} \frac{\alpha_\text{0,i}^\text{TX}}{\varepsilon^\text{TX}} [g_\text{i}] }{1+\sum_\text{i} \ell_\text{i} \frac{\alpha_\text{0,i}^\text{TX}}{\varepsilon^\text{TX}} [g_\text{i}]} \,.
	\label{eq:fbn}
\end{equation}
The parameter~$\omega_\text{i}$ defined in Eq.~(\ref{eq:RNAP_alloc}) represents the fraction of active RNA polymerases engaged in transcribing genes of type~$\text{i}$. In other words, it serves as the allocation parameter for RNA polymerase activity on gene~$\text{i}$, directly influenced by the promoter strength~$\alpha^\text{TX}_\text{0,i}$. Note that the sum at the denominator is related to the global activity of transcription, similarly to the parameter~$\mathcal{K}$ defined in Ref.~\cite{balakrishnan_principles_2022}.
Equation~(\ref{eq:fbn}) defines~$f_\text{bn}$ as the fraction of RNA polymerase engaged in transcription (bound to the genes).

Note that in the expression for~$J_\text{i}^\text{TX}$ in Eq.~(\ref{eq:J_i_TX}) we explicitly retained its dependence on the properties of gene~$\text{i}$, while a common approximation sets~$J_\text{i}^\text{TL} \approx J^\text{TL}$ for translation~\cite{balakrishnan_principles_2022}. However, transcriptional regulation, which is largely governed by promoter on-rates, is predominant~\cite{balakrishnan_principles_2022}. As a result, transcriptional variability across genes can be significant, and is reflected by~$\omega_\text{i}$ and~$f_\text{bn}$. \\

\noindent {\bf Growth rate depends on ribosome and RNA polymerase allocation. } 
The previous section derived an expression of growth rate as a function of the ribosome fraction~$\phi_\text{R}$, given by Eq.~(\ref{eq:CFlim_GL}), which we recall here for convenience:~$\lambda = \varepsilon_\text{R} \phi_\text{R} [m] / \left(K_m + [m] \right)$. 
Having included gene transcription allows us to state that the total mRNA abundance,~$m$, is proportional to the total amount of RNA polymerases,~$N$. This relationship follows from  Eq.~(\ref{eq:transcript_dynamics}), and, summing over all mRNA classes~$\text{i}$ and assuming~$\ell_\text{i} = \ell$ $\forall \text{i}$, we obtain the expression
\begin{equation}
    m = \cfrac{\varepsilon^\text{TX}}{\delta \ell} f_\text{bn} N. \label{eq:m(d,l)}
\end{equation}
After converting the number of RNAPs into~$\phi_\text{N} = M_\text{N} / M_p$, the protein mass fraction of RNAP, we can express the growth rate as a function of both~$\phi_\text{R}$ and $\phi_\text{N}$:
\begin{equation}
    \lambda = \varepsilon_\text{R} \phi_\text{R} \frac{\phi_\text{N}}{\phi_\text{N} + K_\text{N}} \,,
    \label{eq:lambda_phiN}
\end{equation}
where~$K_\text{N} = K_m \delta / (\varepsilon^\text{TX}_\text{N}  f_\text{bn} [P])$ emerges as a compound parameter incorporating both translational and transcriptional components. The first term,~$K_m$, represents the translational demand as introduced in the previous section. The second part accounts for transcriptional efficiency, describing the mRNA concentration per unit of RNA polymerase fraction, which is given by~$[m] = \frac{\varepsilon^\text{TX}_\text{N}}{\delta}  f_\text{bn} \phi_\text{N} [P]$. 
Similarly to~$\varepsilon_\text{R}$, which represents the inverse of the time required to translate all amino acids in a ribosome, above we defined~$\varepsilon^\text{TX}_\text{N} = \varepsilon^\text{TX} / \ell_N$, where~$\ell_N$ is the total length of genes (in base pairs) encoding RNAP subunits. Thus,~$\varepsilon^\text{TX}_\text{N}$ represents the inverse of the time required to transcribe all RNAP genes.

Balakrishnan and coworkers~\cite{balakrishnan_principles_2022} have shown that in \textit{E. coli} the total mRNA concentration scales linearly with the growth rate~$\lambda$, when modulated by nutrient quality, see Fig.~\ref{fig3}(b). This effect is due to sequestration of RNA polymerases, more precisely of the sigma factor~$\sigma^{70}$, under poor nutrient conditions by the Rsd gene~\cite{balakrishnan_principles_2022}. The sequestration of RNAPs  effectively reduces the pool of available polymerases for transcription, while maintaining an approximately constant RNAP allocation~$\phi_\text{N}$ across conditions, Fig.~\ref{fig3}(c).
%
\edit{Under the assumption that cells adjust their proteome composition~$\bm{\phi}$ to maximize their growth rate, Calabrese et {\it al.}~\cite{calabrese_how_2024} show that Eq.~(\ref{eq:lambda_phiN}), constrained by proteome allocation and flux balance, predicts that mRNA increases linearly with the growth rate, while the fraction of RNA-polymerase proteins stays constant (see Fig.~3 of Ref.~\cite{calabrese_how_2024}). This prediction is in agreement with the experimental observation~\cite{balakrishnan_principles_2022}.
}
\edit{The hypothesis that cells maximize or optimize growth has been widely explored in the field. However, more recent studies suggest that cellular growth is shaped by diverse trade-offs rather than a single optimality principle (see~\ref{box_new:FBA}).}\\

\begin{mdframed}[backgroundcolor=light-gray]
\refstepcounter{boxcounter} 
{\bf \large \theboxcounter. From optimal growth to wider strategies}\\
\label{box_new:FBA}
\noindent \edit{A long-standing view in microbial physiology posits that microorganisms evolve to maximize their growth rate under given metabolic constraints. This idea, formalized e.g. through flux balance analysis (FBA), has been fruitful in finding quantitative relationships between a carbon source uptake rate and maximal biomass production~\cite{Edwards2001}. Follow-up studies linked this principle to cellular regulation and proteome allocation, suggesting that bacteria tune their metabolism to achieve near-optimal growth~\cite{scott_emergence_2014, Chubukov2014, Goyal2010}. These insights have been framed in the broader context of resource allocation and cellular ``economics''~\cite{EPCP2025}.
Recent work, however, highlights clear deviations from strict optimization. Under changing environments, cells often follow  response programs that compromise instantaneous growth in favor of future investment and/or robustness. For example, Balakrishnan et al.~\cite{Balakrishnan2021} showed that {\it E.~coli} transiently expresses nonessential genes during nutrient shifts, leading to suboptimal precursor usage. Likewise, other studies~\cite{Wu2022, Dai2016, Mori2017} found that cells maintain a costly reserve of inactive ribosomes to enable rapid adaptation, contradicting the idea of steady-state growth optimization. Together, these findings suggest that while growth optimization remains a useful guiding concept, microbial strategies are fundamentally shaped by wider trade-offs between efficiency, flexibility, and evolutionary robustness.}

\end{mdframed}

\noindent {\bf Connecting RNAP Allocation to Ribosome Usage and Proteome Composition.}
As we discussed previously, the term ``allocation'' in the context of proteome fractions reflects the idea that cells distribute their biosynthetic resources among different functional sectors to achieve a specific physiological state. This concept is particularly relevant for ribosomes and RNA polymerases. 

To clarify the relationship between ribosome allocation (the fraction of ribosomes engaged in translating a specific class of proteins~$\text{i}$), RNAP allocation (the fraction of RNA polymerases transcribing genes of class~$\text{i}$), and proteome partitions we now formalize these connections and discuss their implications.
Equation~(\ref{eq:protein_dynamics}) and the expression for~$J^\text{TL}$ in Eq.~(\ref{eq:J_TL}) lead to the conclusion that the production rate of proteins belonging to class~$\text{i}$ is proportional to~$R m_\text{i} /m$. Under the simplifying assumption that all mRNAs follow identical dynamics (i.e. have the same initiation and elongation rates, as well as identical lengths), the ratio~$\chi_\text{i} = m_\text{i} /m$ corresponds to the fraction of ribosomes translating the transcripts of class~$\text{i}$.
With this definition, one can specify Eq.~(\ref{eq:maaloe}) for the single sector,
\begin{equation}
    \dot{M}_\text{i} = J_\text{syn} \chi_\text{i} \,, \label{eq:chi_operative_def}
\end{equation}
where~$\chi_\text{i}$ represents the fraction of the total flux belonging to the i-th sector.
For this reason, we can refer to it as the ``ribosome allocation'' parameter. In steady-state exponential growth, this fraction coincides with the proteome partition, i.e.~$\chi_\text{i} = \phi_\text{i}$, meaning that the proteome composition directly reflects ribosome allocation. Sec.~\ref{sec7:shifts} shows that in out-of-steady-state scenarios,~$\chi_\text{i}$ represents a target value for the dynamics of~$\phi_\text{i}$. 

The RNAP allocation~$\omega_\text{i}$ defined in Eq.~(\ref{eq:RNAP_alloc}) depends on both the gene copy number~$g_\text{i}$ and the promoter strength 
of gene type~$\text{i}$, which determine the recruitment efficiency of RNA polymerases. Notably, Eqs.~(\ref{eq:protein_dynamics}), (\ref{eq:J_i_TX}) and (\ref{eq:m(d,l)}) imply that, under the approximation that all genes have identical lengths, the ribosome allocation parameter~$\chi_\text{i}$ is equal to the RNAP allocation~$\omega_\text{i}$ during steady-state growth. 

This leads to the steady-state equivalence
\begin{equation}
    \omega_\text{i} = \chi_\text{i} = \phi_\text{i}\,.
    \label{eq:allocations}
\end{equation}
This results implies that gene-weighted relative promoter activities directly correspond to mRNA fractions, which in turn match protein fractions at steady-state growth. In other words, at steady-state growth, the fraction of RNA polymerases transcribing genes of type~$\text{i}$ directly determines the fraction of ribosomes translating the corresponding mRNAs, which in turn dictates the final proteome fraction. This relationship reinforces the idea that transcriptional regulation and proteome composition are inherently linked through the allocation of biosynthetic resources. 
Note however that the equivalence  assumes uniform translation and transcription rates; relaxing these assumptions yields distorted relationships.

\subsection*{OUTLOOK}

Whether mRNA availability plays a major limiting role in cellular physiology or under certain perturbations remains an open question, with many potential modeling directions.

The simple modeling framework for growth accounting for transcription presented in this section does not explicitly consider the transcription of ribosomal RNA. This is a major simplification for eukaryotic cells, where distinct RNA polymerases specialize in transcribing different RNA classes. Even in {\it E. coli}, where a single RNA polymerase transcribes both rRNA and mRNA, experimental evidence suggests that the RNAP allocation fraction~($\phi_\text{N}$) remains approximately constant across growth conditions~\cite{balakrishnan_principles_2022}. This observation indicates that RNAP allocation does not act as a strong limiting factor in bacterial growth, likely because RNAPs constitute only a small fraction of the proteome (typically a few percent), and competition for different gene categories is likely negligible.
\edit{Note that this statement refers to their limited proteome allocation cost, rather than to optimal expression or robustness of expressed proteins~\cite{Lo2024}.}

Additionally, RNA polymerase and ribosomes are mutually dependent, as each participates in the biosynthesis of components required by the other (ribosomal RNAs and proteins, and the protein subunits of RNA polymerase), leading to a  deeper coupling between transcription and translation~\cite{bremer_parameters_1975, Kostinski2020, Kostinski2021,Roy2021,calabrese_how_2024}. Accordingly, there is evidence that, at least in {\it E. coli}, available RNA polymerase and ribosome abundances are coordinated~\cite{balakrishnan_principles_2022,zhang_decrease_2020}.

\section{Side Dish: Protein degradation} \label{sec5:degradation}
\vspace{-.3cm}
The previous sections discussed the primary biosynthetic mechanisms behind protein production. This section focuses on protein degradation, the process that breaks down existing proteins into amino acids, peptides, or waste products.

\subsection*{MAIN QUESTIONS AND MOTIVATIONS}

One might wonder why a cell would invest resources in degrading proteins it has already synthesized, a process that can also consume energy by consuming ATP~\cite{Gottesman1992}. Yet, protein degradation is widely recognized as a quality control mechanism that can be advantageous by removing unfolded or misfolded proteins, clearing mistranslated proteins that could be harmful to the cell~\cite{Goldberg1972, Goldberg2003, Dukan2000, Manohar2019}, or as a response to severe nutrient restriction~\cite{Vabulas2005}.

In eukaryotes, degradation is widely recognized as a key feature of protein homeostasis~\cite{Stein2019}. In yeast, protein degradation relies on multiple systems conserved across eukaryotes, including mammals, such as the proteasome-ubiquitin system~\cite{Hochstrasser1995} and regulated autophagy~\cite{Nakatogawa2009}.
As in most eukaryotes, autophagy is tightly regulated by the TOR pathway, which also controls ribosome biogenesis (see Sec.~\ref{sec6:mecha_regulation} for more details). 
The precise regulation of degradation and autophagy~\cite{Nakatogawa2009} underscores their importance as critical physiological mechanisms in eukaryotes, linked to both growth and survival during starvation~\cite{Onodera2005, Adachi2017}.

Most bacteria are slow growers, and even fast-growing bacteria have to deal with slow growth conditions~\cite{Koch1971, Li2018}. In such conditions, maintenance and turnover become particularly central in biosynthesis. Indeed, if typical time scale for protein half-life is of order tens of hours, any cell that doubles around that time scale (or slower) has to deal with protein turnover~\cite{calabrese_protein_2022}.
Even in a fast-growing bacterium like \textit{E.~coli}, numerous proteolytic enzymes are involved in protein degradation~\cite{Gottesman1996}. A small fraction of proteins are specifically targeted for degradation to regulate their levels (regulatory degradation), but there is also a basal, non-specific degradation (housekeeping degradation) that is essential for eliminating damaged or abnormal proteins~\cite{Gottesman1996}. 
\\

\noindent\textbf{Neglecting degradation leads to inconsistencies in slow-growth regimes.} 
Protein degradation and turnover are often overlooked in models that describe growth laws (for example in Sec.s~\ref{sec2:classical} and \ref{sec3:translation}).  
This is because the timescale for protein degradation typically falls within the range of 10-100 hours~\cite{Goldberg1974, Maurizi1992}, making it negligible in comparison to the rate at which new proteins are produced under nutrient-rich conditions. However, when the doubling time of the population becomes comparable to the timescale of protein degradation, the interplay between the two starts to influence growth~\cite{Maitra2014, Santra2017}.
Notably, microorganisms often experience prolonged periods of slow or no growth in natural environments~\cite{Koch1971, Li2018, Mori2017}.

The growth laws models discussed so far
contain a crucial inconsistency when describing slow growth conditions, especially as the growth rate approaches zero ($\lambda \to 0$)~\cite{calabrese_protein_2022}. If we consider the first growth law, Fig.~\ref{fig1}(a), as the growth rate decreases a non-zero offset $\phi_\mathrm{R}^\mathrm{min}$ in the ribosome fraction remains, which is a robust feature across species. This offset is typically understood as inactive (non-translating) ribosomes~\cite{scott_interdependence_2010, Mori2017}, implying that, at zero growth, all ribosomes should be inactive, and protein synthesis should cease. Despite that, direct measurements show that while the translation elongation rate decreases with vanishing growth rate (when decreasing nutrient quality), it never reaches zero~\cite{Dai2016, Wu2022}, indicating that not all ribosomes are inactive, contradicting the assumption of complete inactivity at vanishing growth rates (see also Fig.~\ref{fig2}(a)-(b)). 

To solve this inconsistency,
Calabrese et al.~\cite{calabrese_protein_2022} have proposed that at vanishing growth a fraction of ribosomes remains active, and its contribution to biomass production is balanced by protein degradation.
\\

\begin{figure}[h!]
    \centering
    \includegraphics[width=1\columnwidth]{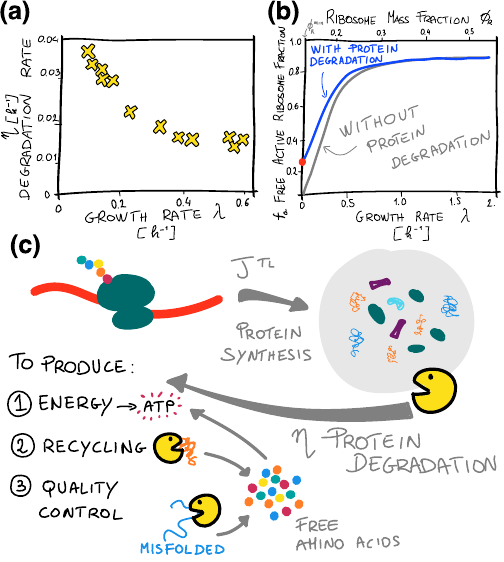} 
    \caption{\textbf{Protein degradation and ribosome activity as function of growth rate.} (\textbf{a}) The growth rate dependence of the degradation rate~$\eta$ for the bacterium \textit{E. coli}. The x-axis represents the steady state growth rate~$\lambda$, while the y-axis the degradation rate~$\eta$. Data points are sketched  from~\cite{calabrese_protein_2022}, the original data can be found in~\cite{Pine1973}.
    (\textbf{b}) The active ribosome franction~$f_\text{a}$ as a function of the growth rate and~$\phi_\text{R}$ for a model including degradation (blue upper line) and for one not including it (grey lower line).
    (\textbf{c}) Scheme of the degradation process. The process degrades proteins possibly to (I) produce energy, (II) obtain amino acids under starvation conditions or (III) function as a quality control, to degrade misfolded or damaged proteins. \edit{This figure is inspired by and adapted from Ref.~\cite{calabrese_protein_2022}.}
    }
    \label{fig:deg}
\end{figure}

\noindent\textbf{Experimental measurements of global protein degradation rates.} 
Incorporating the role of protein degradation into theoretical models requires quantitative data on protein degradation across various growth conditions. Here, we focus on the available experimental information on global protein degradation in the model organisms \textit{E.~coli} and \textit{S.~cerevisiae}.
Despite growing interest in quantitative physiology, there have been no recent systematic studies on protein degradation in bacteria. Early studies from the 1970s and 1980s, such as Refs.~\cite{Nath1970, Pine1973, Mosteller1980}, provided preliminary insights into protein degradation timescales in \textit{E.~coli}, establishing the general understanding that proteins are relatively long-lived, with an average degradation rate~$\eta$ of~$ \simeq 0.02 \, \text{h}^{-1}$. A reanalysis of older data suggests a growth rate-dependent degradation rate, with degradation increasing as the growth rate $\lambda$ decreases~\cite{calabrese_protein_2022} (see Fig.~\ref{fig:deg}(a)).

By contrast, information on the half-lives of individual proteins in yeast has been documented by detailed recent studies~\cite{Christiano2014, MartinPerez2017}. This more recent data on yeast is not always quantitatively consistent across studies, but overall confirms a similar trend, with degradation rates decreasing as the growth rate increases~\cite{Calabrese_dataset_2021}.

Other studies have focused on how degradation is regulated during starvation~\cite{Kraft2008, Sabatier2025} and how it can improve the survival of yeast cells~\cite{Onodera2005, Adachi2017} (also reviewed in~\cite{Nakatogawa2009}). For mammalian cells, a few studies~\cite{Schwanhaeusser2011, An2017} address the role of global degradation in determining and remodeling gene expression and protein levels. Few quantitative models explicitly account for degradation, but there are some exceptions~\cite{Maitra2014, Kempes2016, Kempes2017, calabrese_protein_2022}. A toy model for targeted degradation of ribosomes (ribophagy) has been developed in Ref.~\cite{soubrier_optimal_2024}. In the next paragraphs, we will follow the framework developed in Ref.~\cite{calabrese_protein_2022}.

\subsection*{INGREDIENTS AND RECIPES\footnote{The recipe introduced here builds upon ideas and models presented in Refs.~\cite{calabrese_protein_2022,Droghetti2025}.}}

We provide here a minimal set of ingredients and recipes to build steady-state growth laws models in presence of protein degradation.\\

\noindent\textbf{Degradation current.} Following Ref.~\cite{calabrese_protein_2022}, we write the total mass production $\dot{M_p}$ as the balance between  the biosynthesis (protein translation) flux $J_\text{syn}$ given by~$\varepsilon_\text{R}   f_\text{a} M_\text{R}$,~Eq.~(\ref{eq:maaloe}) 
and the protein degradation flux $J_\text{deg} = \eta M_p$: 
\begin{equation}
    \frac{\text{d}M_p}{\text{dt}} = J_\text{syn} - \eta M_p \,. \label{eq:deg:protein mass}
\end{equation}
For simplicity, we only consider a mean-field description for the average degradation rate~$\eta$ (averaged across different cells and proteins). This assumption can be relaxed to consider that different proteins can be degraded at diverse rates, and one could write~$J_\text{deg} = \sum_\text{i}\eta_\text{i} M_\text{i}$, where the index $\text{i}$ runs over all the distinct expressed proteins. From Eq.~(\ref{eq:deg:protein mass}), we can directly compute the growth rate as
\begin{equation}
    \lambda = \frac{1}{M_p} \frac{\text{d}M_p}{\text{dt}} = \varepsilon_\text{R} f_\text{a} \phi_\text{R} - \eta \ . 
    \label{eq:deg:lambda}
\end{equation}
This relationship shows that degradation has a direct and measurable impact on growth rate when $\varepsilon_\text{R} f_\text{a} \phi_\text{R} \simeq \eta$, and implies that at vanishing growth $J_\text{syn}=J_\text{deg}$.
This means that, when $\lambda=0$, the model admits a non-zero ribosome elongation rate (as measured in Ref.~\cite{Dai2016, Wu2022}) and a ``maintenance'' biomass production that does not contribute to cellular growth but compensates for protein degradation, which implies~$f_\text{a}(\lambda=0) > 0$, Fig.~\ref{fig:deg}(b).

If we consider a variant of the model with $f_\text{a} = 1$ (no inactive ribosomes), all ribosomes contributing to $\phi_\text{R}^\text{min}$ would be actively translating and perform maintenance, even approaching zero growth. By knowing how the ribosome elongation rate varies with $\lambda$ (see Sec.~\ref{sec2:classical}), in this case one could estimate the degradation rate at zero growth $\eta_0 := \eta(\lambda=0) = \varepsilon_\text{R}(\lambda = 0) \,  \phi_\text{R}^\text{min}$.
However, this only qualitatively captures the trend of the degradation rate across growth conditions, and it significantly overestimates~$\eta$~\cite{calabrese_protein_2022}. 
This suggests that~$f_\text{a}(\lambda=0) < 1$, Fig.~\ref{fig:deg}(b), which corresponds to the hypothesis that a part of the ribosomes are sequestered and therefore not available for translation, in line with the current understanding~\cite{Wu2022}.\\

\begin{mdframed}[backgroundcolor=light-gray]
\refstepcounter{boxcounter} 
{\bf \large \theboxcounter. Mass conservation and intracellular free amino acids.}
\label{box5:aa_deg_recycling}
It is important to note that in the presented model the total mass is not conserved, as the mass of the degraded protein is lost. A way to reestablish total (in this case amino acids plus protein) mass conservation is to introduce an equation for amino acids/precursors that assumes the perfect recycling of all the degraded protein mass, which consequently becomes an additional source of amino acids.
This box introduces and expands the modeling strategy presented in Ref.~\cite{Droghetti2025}.

The mass of intracellular amino acids $M_\text{A}$ is determined by the equilibrium between nutrient influx~$J_\text{cat}$, outflux due to biosynthesis~$J_\text{syn}$~\cite{scott_emergence_2014, Droghetti2025}, defined as introduced in Sec.~\ref{sec2:classical}, and the influx given by degraded proteins~$\eta M_p$, corrected with a term defining the recycling efficiency~$\xi_\text{r}$
\begin{equation}
    \frac{\mathrm{d} M_\text{A}}{\mathrm{dt}} = \nu M_\mathrm{P} - \varepsilon_\text{R} f_\text{a} M_\mathrm{R} + \eta \xi_\text{r} M_p\ .
    \label{eq:box-A}
\end{equation}
When~$\xi_\text{r}=1$, the recycling is perfectly efficient and the mass~$M_p +M_\text{A}$ is conserved. Recycling efficiency can be imperfect, for example if proteins are not completely broken down by degradation enzimes.
As done for the proteome mass and the sector, we wish to introduce a intensive quantity representing the amino acids. We therefore define~$\varphi_\mathrm{A} = M_\text{A} / M_\mathrm{p}$,
where $M_\mathrm{p}$ is the total mass of proteins. Assuming a constant dry-mass density~\cite{bremer2008modulation, Erickson2017} leads to the following equation for $\varphi_\mathrm{A}$, 
\begin{equation}
  \frac{\mathrm{d} \varphi_\mathrm{A}}{\mathrm{dt}}
  = \nu \phi_\mathrm{P}  - \varepsilon_\text{R} f_\text{a} \phi_\mathrm{R} + \eta\xi_\text{r} - \lambda \varphi_\mathrm{A} \ ,
    \label{eq:box-aa}
\end{equation}
where the additional term $\lambda \varphi_\mathrm{A}$ accounts for dilution due to cellular growth. \\

\noindent\textbf{Amino acids mass fraction and flux matching.}
A consequence of Eq.~(\ref{eq:box-aa}) is that its steady state condition, given by~$\dot{\varphi_\text{A}} =0$, imposes
\begin{equation}
    J_\text{cat} + J_\text{deg}- J_\text{synt} = \lambda M_\text{A},
\end{equation}
where we multiply~$\dot{\varphi_\text{A}}$ by~$M_p$ in order to get the fluxes. This implies that, even if the degradation flux is neglected, the flux matching between~$J_\text{cat}$ and~$J_\text{syn}$ given by Eq.~(\ref{eq:cat_syn_fluxes}) is no longer valid, as long as the size of the amino acids pool is different from zero. The flux matching condition $J_\text{cat}=J_\text{syn}$ introduced in Sec.~\ref{sec2:classical} holds when protein degradation is neglected and the intracellular amino acid pool is small (i.e., $\varphi_\text{A} \approx 0$).  
Several models take explicitly into account the amino acids/precursor pool~\cite{Shachrai2010, Bosdriesz2015, KoremKohanim2018, chure_optimal_2023,Droghetti2025}.
\end{mdframed}

\noindent\textbf{Accounting for growth-rate–dependent processes.} 
In order to consider these observations in a mechanistic model
and to use them to interpret the data, one has to acknowledge that all the quantities that appear in Eq.~(\ref{eq:deg:lambda}) have a (direct or indirect) dependency on the growth rate~$\lambda$. Specifically, as introduced in Sec.~\ref{sec3:translation} and shown in Fig.~\ref{fig2}, there are direct measurements of the dependence of the ribosome elongation rate~$\varepsilon$ (and thus on $\varepsilon_\text{R} = \varepsilon/L_\text{R}$) on the growth rate~$\lambda$.
The functional form of $\eta(\lambda)$ can be derived from fits of the experimental data --see Refs.~\cite{Pine1973, calabrese_protein_2022} and  Fig.~\ref{fig:deg}(a)-- to identify an empirical dependency as introduced in Ref.~\cite{Droghetti2025}: 
\begin{equation}
  \eta(\lambda)= \frac{\eta_\text{0} + \eta_\infty s_\eta \lambda}{1 + s_\eta \lambda} \,,  \label{eq:deg:eta(lambda)}
\end{equation}
where~$\eta_\text{0}$ and~$\eta_\infty$ represent the protein degradation rate values for~$\lambda = 0$ and~$\lambda \rightarrow \infty$, respectively, and~$s_\eta$ controls the steepness of the hyperbolic function.
The empirical value of the parameters from the fit of data in Ref.~\cite{Pine1973} are given in Tab.~\ref{tab:params_sec_5_deg}.

As explained in Sec.~\ref{sec3:translation}, the fraction~$f_\text{a}$ can be inferred indirectly: one can use the growth rate dependency of~$\eta$ Eq.~(\ref{eq:deg:eta(lambda)}) and Eq.~(\ref{eq:deg:lambda}), this time accounting for the role of varying protein degradation rates across growth conditions:
\begin{equation}
  f_\text{a}(\lambda)= \frac{\lambda + \eta(\lambda)}{\varepsilon_\text{R}(\lambda) \phi_\text{R}(\lambda)}\,. \label{eq:deg:fa(lambda)}
\end{equation}
Using the parameters quoted above, we get a fraction of active (maintenance) ribosomes of roughly 0.25 for \textit{E. coli} at vanishing growth. In other words, when considering the role of degradation, this model predicts that one fourth of the ribosomes expressed at zero growth are active and translationally engaged. Figure~\ref{fig:deg}(b)
shows this fraction as a function of the growth rate, comparing the two cases with and without degradation. Note that the case without degradation is equivalent to the plot shown in Fig.~\ref{fig2}(b).\\

Let us now conclude this section with a few recommendation and observations.
Naturally, inferring $f_\text{a}$ from Eq.~(\ref{eq:deg:fa(lambda)}) strongly depends on the model assumed. As mentioned in Sec.~\ref{sec3:translation}, a similar approach to estimate the fraction of active ribosomes was used in Dai et {\it al.}~\cite{Dai2016}, but it neglected protein degradation.
Furthermore, this procedure does not provide any interpretation of the possible mechanisms for sequestering ribosomes, and the same is true for the growth-dependence of the degradation rate, which is assumed here as a fit of the data, without a mechanistic explanation.

Secondly, one should be careful when introducing degradation and inactive ribosomes into a dynamical model, since these ingredients also affect other equations beyond Eq.~(\ref{eq:deg:lambda}), such as those describing protein production or the amino acid pool introduced in~\ref{box5:aa_deg_recycling}. A good practical approach is to rewrite the equations from the start, making sure to include coherently all the mechanisms one wants to capture in the model.

\subsection*{OUTLOOK}

Several fundamental questions remain open to modeling. For example, why do cells degrade proteins? Is this process primarily a mechanism for conserving energy, or does it play a critical role in maintaining cellular homeostasis through quality control~\cite{Goldberg1972, Goldberg2003, Dukan2000, Manohar2019, Stein2019}? Another pertinent question concerns the impact of degradation on growth: does protein degradation ever become a limiting factor, and under what conditions might it influence cellular fitness~\cite{Lapinska2019}? Addressing these questions would offer new perspectives on the evolutionary and physiological roles of protein degradation.

In the framework of sector models, the ``degradation sector'',  comprising proteases and associated machinery, deserves closer examination. It is unclear whether this sector imposes a significant burden on cellular resources and growth. Investigating how resource allocation is balanced between degradation and other cellular processes, particularly under conditions of high degradation activity, could reveal potential trade-offs and shifts in optimal allocation strategies~\cite{Vabulas2005, RusilowiczJones2022}.

Another direction involves incorporating amino acids recycling mechanisms, as in~\ref{box5:aa_deg_recycling}. We stress that, in the model presented above, the mass is not conserved as the degraded protein mass is lost. Indeed, amino acids resulting from protein degradation may re-enter the intracellular amino acid pool, allowing ribosomes to reuse them in protein synthesis.
The theory presented in this section, e.g. Eq.~(\ref{eq:deg:lambda}), neglects the presence of an intracellular amino acids pool. Instead, when considering the dynamics of free amino acids, as detailed in~\ref{box5:aa_deg_recycling}, the growth rate describes the accumulation of total (protein + amino acids) mass.

Depending on cellular conditions, amino acids could either be fully recycled into new proteins or further processed for energy production. However, the role of energy consumption and production in cell physiology~\cite{Weisse2015} is often neglected in current models, leaving this aspect under-explored.

Finally, new insights from theory and experiments may be gained by examining how cells respond to nutrient shifts that require rapid remodeling of the proteome to synthesize enzymes adapted to the new environment.
In such transitions, existing proteins may be actively degraded to free up resources for newly needed components, thereby accelerating the reallocation of the proteome.
Moreover, in environments where growth is severely impaired (or where starvation is induced), degradation can potentially be the only way to gather the central precursors necessary for new protein synthesis, for example amino acids and ATP. However, the role of degradation during starvation is not fully understood yet, even though this topic is gaining interest in the field.
This dynamic, where degradation accelerate the adaptation to a new environment, is especially relevant in ``catch-22'' scenarios in which cells face metabolic constraints, making it challenging to produce essential enzymes without first alleviating the resource bottleneck through degradation \cite{may2020autophagy, micali2023minorities}.

Answering these questions will shed light on the complex interplay between protein degradation, growth, and resource allocation in bacteria, potentially unveiling new insights into cellular strategies in varying environmental conditions.

\section{À la carte: Mechanistic regulation of ribosome allocation} \label{sec6:mecha_regulation}
In this section we examine simplified models of the regulatory circuits controlling ribosome biogenesis, focusing in particular on the ``stringent response'' --the process by which cells reduce ribosomal protein production when they detect worsening conditions~\cite{warner_yeast_1978, Potrykus2008}. 
Overall, our aim is to illustrate how cells sense changing conditions (through external signals or internal cues) and then adjust protein production and resource usage accordingly.

\subsection*{MAIN QUESTIONS AND MOTIVATIONS}

Resource allocation at different growth rates implies regulation based on environmental and physiological sensing, yet some biosynthesis theories overlook this aspect, relying instead on flux pairing and optimization~\cite{Erickson2017, chure_optimal_2023}. However, cells may optimize other factors rather than growth (e.g. survival, persistence, recycling, energy consumption, etc.)~\cite{Mukherjee2024_plasticity, Zhu2024shaping}. Additionally, mechanistic circuits regulating resource allocation adopt different architectures, each of which may have different operational boundaries and time scales. 

To better describe growth and cellular behavior, and to formulate specific predictions, these theories must be integrated with a description of regulatory circuits. Understanding such regulation is key not only for accurate models but also for targeting cell proliferation and survival in broader biological contexts.
In bacteria and beyond, achieving a quantitative understanding of biosynthetic flux regulation can help develop molecular perturbations and identify drug targets to limit cellular proliferation. \\

\begin{figure}[h!]
    \centering
    \includegraphics[width=1\columnwidth]{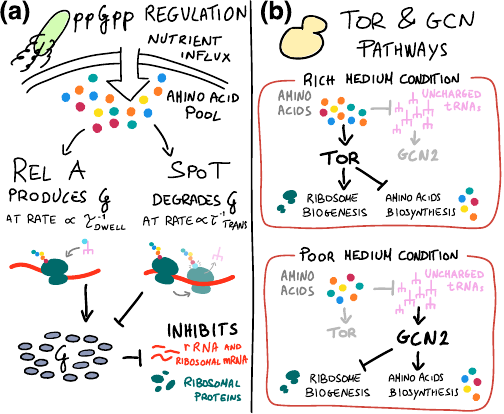} 
    \caption{\textbf{The stringent response in \textit{E.~coli} and \textit{S.~cerevisiae}.}
    (\textbf{a}) scheme for the ppGpp-dependent ribosomal regulation in \textit{E.~coli}. The circuit is believed to sense the translation elongation rate~$\varepsilon= 1/ (\tau_\text{dwell} + \tau_\text{trans} )$, with~$\tau_\text{trans}$ and~$\tau_\text{dwell}$ activating Spot and RelA respectively. The combination of their ppGpp inhibition and production sets the ppGpp level, which in turn governs ribosomal syntesis. 
    (\textbf{b}) scheme of the TOR-GCN mediated response to environmental changes of \textit{S. cerevisiae}. When growing in a rich environment, the TOR pathway is activated to enhance ribosome production (upper panel). On the other hand, when growing conditions worsen, uncharged tRNAs activates the GCN pathway, which enhance the synthesis of proteins related to amino acids and precursors production. \edit{This figure is inspired by and adapted from Refs.~\cite{Wu2022,Joo2010}.}
    }
    \label{fig:mecha_ppGpp_tor}
\end{figure}

\noindent\textbf{The bacterial stringent response.} 
Mechanistic models integrating environmental sensing with resource allocation exist primarily for \emph{Escherichia coli}, where the alarmone (p)ppGpp (guanosine tetra- and pentaphosphate, from now on abbreviated as ppGpp) plays a central role in regulating transcription in response to nutrient  status~\cite{CASHEL1969,Zhu2019, Potrykus2008, marr_growth_1991} --see Fig.~\ref{fig:mecha_ppGpp_tor}(a). ppGpp modulates RNA polymerase activity directly and via $\sigma$-factors, acting preferentially on promoters with specific discriminator sequences~\cite{Zhang2002}.

ppGpp levels are controlled by the RelA-SpoT Homolog (RSH) protein family~\cite{Atkinson2011}. While many bacteria possess a single bifunctional RSH protein (Rel), \emph{E.~coli} uses two: RelA and SpoT. RelA synthesizes ppGpp in response to amino acid starvation, activated by uncharged tRNAs binding to the ribosomal A-site~\cite{Wendrich2002, Magnusson2005}, while SpoT primarily degrades ppGpp, preventing toxic accumulation~\cite{Ferullo2008, Kraemer2019}.

Recent work by Wu and coworkers~\cite{Wu2022} shows that ppGpp levels correlate quantitatively with translation elongation rate across nutrient conditions and for some nutrient shifts. These results suggests that the RelA-SpoT system effectively senses translational speed rather than tRNA charging~\cite{marr_growth_1991, Shachrai2010, Bosdriesz2015}. Models based on this assumption align well with experiments, while those relying on tRNA charging appear to be inconsistent with data~\cite{Wu2022, Droghetti2025}.

Note that ppGpp regulation varies across bacterial species~\cite{Ronneau2019}, and signaling systems contribute to the re-allocation of proteome in response of internal and external perturbations~\cite{Ripamonti2025}.\\

\noindent\textbf{The eukaryotic stringent response.} 
While the regulation of ribosome biogenesis in eukaryotes is more complex than in bacteria, similar principles apply, such as the ribosomal first growth law~\cite{karpinets_rna_2006,metzl-raz_principles_2017} and a stringent response triggered by nutrient stress~\cite{warner_yeast_1978}. In eukaryotes, such as {\it Saccharomyces cerevisiae}, the coordinated regulation of ribosome biogenesis requires coordination among three RNA polymerases: Pol I (transcribing rRNA except 5S)~\cite{Paule1998}, Pol III (5S rRNA, tRNAs), and Pol II (ribosomal proteins and other mRNAs)~\cite{White1998, Kostinski2021, Paule2000}.

The stringent response in yeast consists of two interconnected modules~\cite{warner_yeast_1978, Moehle1991, Joo2010}:
(i) the downregulation of ribosomal gene expression, and
(ii) the upregulation of amino acid biosynthesis genes --see Fig.~\ref{fig:mecha_ppGpp_tor}(b).
These modules are mediated by distinct signaling pathways. 

The first is driven by TOR Complex 1 (TORC1), a central regulator that links environmental nutrient status to ribosome biogenesis~\cite{Gonzalez2017}. Under nutrient-limiting conditions, TORC1 activity is suppressed, leading to reduced transcription of ribosomal protein genes and decreased rRNA synthesis. TORC1 exerts this control through multiple mechanisms: it modulates transcription factors and chromatin remodelers~\cite{Martin2004, Hu2007}, and it promotes rRNA production by regulating RNA polymerase I recruitment~\cite{Mayer2004}. The conservation of TORC1 signaling across eukaryotes --via orthologous pathways in diverse organisms~\cite{Mestre2022, Inoki2006, DeVirgilio2006}-- highlights its fundamental role in coordinating nutrient sensing with cell growth. For further details on how TORC1 balances growth and autophagy, see Refs.~\cite{Wullschleger2006, Nakatogawa2009}.

The second module involves the GCN pathway. Amino acid starvation leads to accumulation of uncharged tRNAs, which activate the GCN2 kinase. GCN2 then promotes translation of the transcription factor GCN4, which induces amino acid biosynthetic genes~\cite{Hinnebusch2005}. Intriguingly, GCN4 may also participate in repressing ribosomal protein gene expression, though the precise mechanisms remain to be elucidated~\cite{Joo2010}.

By translating environmental signals into specific alterations in gene expression, TORC1 ensures that the cell efficiently allocates resources, thereby maintaining a balance between ribosome production and overall growth under varying nutrient conditions.

\subsection*{INGREDIENTS AND RECIPES\footnote{The recipe introduced here builds upon ideas and models presented in Refs.~\cite{Wu2022,Droghetti2025}.}}

Here, we provide a basic recipe for modeling the regulation of ribosomal-sector transcription by ppGpp in \textit{E.~coli}, where current knowledge allows for the design of a quantitative model~\cite{Wu2022,Droghetti2025}. This model consists of three main ingredients.

The first ingredient is the dynamics of ribosome allocation. Following Sec.~\ref{sec4:transcription}, the fraction of ribosomes allocated to translate ribosomal genes, $\chi_\text{R}$, is determined by~$\omega_\text{R}$, the fraction of RNAP transcribing ribosomal mRNAs.
Therefore, we assume that~$\chi_\text{R}$ follows the RNAP partition~$\omega_\text{R}$ 
\begin{equation}
 \chi_\text{R} = \omega_\text{R}  \ .
\label{eq:6chiR_omegaR}
\end{equation}

The second ingredient links $\omega_\text{R}$ to the intracellular ppGpp concentration~$[G] = G/V$, as it has been shown that ppGpp affect ribosome production at the transcriptional level~\cite{Zhang2002,Magnusson2005}.
Specifically, we assume a monotonic relationship
\begin{equation}
\omega_\text{R} = f([G]) \ ,
\label{eq:6_bar_omegaR}
\end{equation}
where higher ppGpp levels repress ribosomal gene transcription. This relationship is commonly described phenomenologically as a Hill function~\cite{marr_growth_1991, Shachrai2010, Droghetti2025}.

The third ingredient is ppGpp production in response to translation. 
The ppGpp concentration~$[G]$ is linked to the translation elongation rate~$\varepsilon$ via the action of RelA and SpoT. Based on the work by Wu et al.~\cite{Wu2022}, we use the empirical relationship
\begin{equation}
[G] \propto \frac{\varepsilon^\text{max}}{\varepsilon(t)} -1 \ ,
\label{eq:6_ppGpp(epsilon)}
\end{equation}
which was shown to hold across steady-state conditions, but also under a nutrient downshift and under translation inhibition~\cite{Wu2022}. 
These three equations connect $\chi_\text{R}$ to the ppGpp concentration.
For steady state, and under the simplifying assumptions of Sec.~\ref{sec4:transcription}, one can assume that~$\omega_\text{R} = \chi_\text{R} = \phi_\text{R}$. The dynamical version of these equation is introduced in Sec.~\ref{sec7:shifts}.

To rationalize the connection between ppGpp levels~$[G]$, the elongation rate~$\varepsilon$ and the ribosome sector size~$\phi_\text{R}$, Wu and coworkers~\cite{Wu2022} have proposed the following simple model, which also factors in the ribosome activity~$f_\text{a}$ --see Eq.~(\ref{eq:fa1}) and~(\ref{eq:fa2}) in Sec.~\ref{sec3:translation}. Equation~(\ref{eq:6_ppGpp(epsilon)}), which is supported by direct measurements, can be combined with the steady state Eq.~(\ref{eq:epsilon}) in Sec.~\ref{sec3:translation}, which we report again for convenience,
\begin{equation}
	\varepsilon = \varepsilon^\text{max} \cfrac{\phi_\text{R}}{\phi_\text{R} + \phi_\text{R}^\text{min}} \,,
	\label{eq:6_epsilonbis}
\end{equation} 
leading to the conclusion that during steady state growth ppGpp represses ribosomes following
\begin{equation}
\omega_\text{R} = \chi_\text{R} = \phi_\text{R} = \phi_\text{R}^\text{min} \frac{[G_\text{0}]}{[G]}, \label{eq:6_G_0}
\end{equation}
where~$[G_\text{0}]$ is the reference value of ppGpp expression in the condition of no growth (full ribosomal repression). 
By further assuming (without direct support from the data) that ppGpp regulates (directly or not) inactive ribosomes as~$\phi_R^{\text{inactive}} = \phi_\text{R}^\text{min} G/G_0$, one immediately gets Eq.~(\ref{eq:fa2}), which we rewrite here for convenience 
\begin{equation}
	f_\text{a} = 
     1 - \left( \cfrac{\phi_\text{R}^\text{min}}{\phi_\text{R}} \right)^2  \,,
	\label{eq:fa2bis}
\end{equation}
where the prefactor was set to one, since the measured value of these parameters coincides in \textit{E.~coli}. These considerations suggest the possibility of a coherent model for \textit{E.~coli} steady-state physiology that includes ppGpp sensing and regulation of the ribosomal sector. However, note that Eq.~(\ref{eq:fa2bis}) implies that all ribosomes will be inactive at zero growth. As discussed in Sec.~\ref{sec5:degradation}, it should be adapted to be consistent with observed translation events at vanishing growth.

Equation~(\ref{eq:6_ppGpp(epsilon)}) is rather simple, but on a careful look it is the result of several coupled processes: translation elongation, amino-acid production and consumption, tRNA charging, and the action of the RelA and SpoT proteins on the build-up and depletion of the ppGpp pool. In order to achieve a mechanistic understanding of this relation one would need to take into account one or more of these processes. Wu and coworkers provide the following mechanistic justification (see ref.~\cite{Wu2022} for details): ppGpp dynamics results from a balance between synthesis and degradation,
\begin{equation}
\frac{\text{d}[G]}{\text{dt}} = f_\text{syn} - [G] f_\text{deg} \,.
\label{eq:6_dG_dt}
\end{equation}
RelA activity (governing~$f_\text{syn}$) is tied to ribosome dwell times on transcripts waiting for charged tRNAs~($f_\text{syn} \propto \tau_\text{dwell}$), while SpoT-mediated degradation~($f_\text{deg}$) relates to the translocation step~($f_\text{deg} \propto \tau_\text{trans}$). Since the elongation rate is defined as
\begin{equation}
\varepsilon = \frac{1}{\tau_\text{dwell} + \tau_\text{trans}} \ ,
\label{eq:6_epsilon(taus)}
\end{equation}
and~$\varepsilon^\text{max}$ is defined as~$\tau_\text{trans}^{-1}$ (in~$h^{-1}$)~\cite{Wu2022}, combining these elements yields Eq.~(\ref{eq:6_ppGpp(epsilon)}),\edit{ see Appendix~\ref{sec11:parameters_numbers} for more details on the derivation.}

\subsection*{OUTLOOK}
While much remains to be uncovered about the molecular circuits governing cellular growth, recent work in \textit{E.~coli} has laid a foundation for quantitative growth models integrating environmental sensing with resource allocation. However, significant gaps persist in understanding how these paradigms may extend to other bacteria and eukaryotes. For instance, while ppGpp is a well-characterized regulator of the stringent response in \textit{E.~coli}, its role varies across bacterial species~\cite{Zhu2019}, and analogous systems in eukaryotes --such as the TOR and GCN pathways-- operate through functionally convergent yet distinct mechanisms~\cite{Gonzalez2017}. Future work must resolve whether these differences reflect evolutionary adaptations, fundamental constraints in regulatory architecture, or a combination of both. Additionally, the precise triggers and sensors of nutrient stress in eukaryotes remain elusive, particularly in linking translational cues (e.g., uncharged tRNAs) to transcriptional reprogramming. Bridging these gaps will require comparative studies across phylogenetically diverse organisms to uncover universal design principles of growth regulation.  

On the modeling front, the recent advances connecting ppGpp dynamics to translation elongation in \textit{E.~coli} demonstrate the power of hybrid  approaches to bridge the power of coarse-grained models with the specific predictions of mechanistic descriptions. However, several open questions remain, such as the regulatory logic of SpoT and its interplay with RelA, and a full understanding of the physiology under perturbed ppGpp conditions, challenging long-standing assumptions about ppGpp homeostasis~\cite{Zhu2019NAR,Mu2023,Zhu2024}. Extending such models to other systems will require integrating additional signaling pathways (e.g., cAMP-CAP in carbon metabolism~\cite{you_coordination_2013,Kochanowski2021}) and accounting for species-specific circuit topologies. Beyond bacteria, eukaryotic growth regulation involves multilayered coordination between TOR, GCN, and chromatin-based mechanisms, posing both conceptual and experimental challenges. A unified/comparative framework for sensing in biosynthetic regulation --one that spans bacteria and eukaryotes-- could not only refine predictive models of cellular behavior but also inspire new strategies for targeting proliferation in biomedical and biotechnological contexts.

\section{À la carte: Non-steady states and shifts}\label{sec7:shifts}

The theory presented in previous sections was largely defined only during periods of steady-state growth. This section instead focuses on the cell's dynamic reaction to external perturbations, in particular on how to model the reallocation of protein synthesis after a nutrient shift.

\subsection*{MAIN QUESTION AND MOTIVATIONS}

Thus far, our discussion has centered on steady-state exponential growth, a condition that, however, is an idealized abstraction. In natural environments, fluctuating conditions force cells to constantly adapt, making balanced exponential growth difficult to achieve or maintain, with significant implications for fitness~\cite{Kussell2005, Mori2017, Biselli2020}.
Understanding the kinetics of cellular responses to external perturbations is also relevant in the context of antimicrobial drugs and antibiotics (or in general growth-inhibitory drug treatments)~\cite{Deris2013, greulich_growth-dependent_2015, Dai2016}. These responses are typically mediated by one or more intracellular mechanisms, as discussed in the previous section, Sec.~\ref{sec6:mecha_regulation}.
Studying these mechanisms in steady-state conditions is a daunting process: even if the behavior observed during steady-state growth is the result of the action of these circuits, under this condition all the relevant molecular players are balanced and the causal relations between them are not evident, complicating the modeling process. 
Instead, under non-steady conditions is easier to gain insight into the relations that give rise to the growth laws, which components are limiting and under which conditions they are, and the importance of the external nutrients on cellular growth~\cite{Bosdriesz2015, Erickson2017, Shachrai2010, Bren2013, Wu2022}.

In recent years, numerous studies have focused on non-steady conditions, aiming to collect quantitative data and/or design models that can explain such data. There are various ways of perturbing steady-state growth, from changing the external nutrient source~\cite{Dennis1974, Bosdriesz2015, Erickson2017, KoremKohanim2018, Panlilio2021, chure_optimal_2023, Droghetti2025} or allowing nutrients to be consumed until depletion~\cite{Bren2013, metzl-raz_principles_2017, Droghetti2024_stv}, to designing experiments where cells need to rearrange their metabolic pathways between conditions~\cite{Basan2020}.

Literature works on this topic employ a variety of modeling strategies, which can be broadly classified as ``top-down'' and ``bottom-up''.
A top-down approach starts from empirical observations and phenomenological relationships, bypassing the need to explicitly model molecular interactions and regulatory circuits, potentially uncharacterized. It does not require detailed kinetic parameters but, as drawback, it cannot provide mechanistic insights into the underlying regulation. A bottom-up approach, by contrast, starts at the molecular level, constructing biochemical network models that explicitly describe regulatory components and their interactions. This strategy is advantageous when testing specific hypotheses about the architecture of regulatory mechanisms or exploring how molecular parameters shape cellular responses. However, it is inherently more challenging, as it often requires a large number of parameters that may not be experimentally accessible.

The top-down approach is exemplified by the study by Erickson and coauthors~\cite{Erickson2017}, which predicts kinetic responses to nutrient shifts using phenomenological regulatory knowledge and flux balance conditions, avoiding the need to explicitly model intracellular reaction rates. A related strategy is seen in optimization-based approaches, which assume that cells adjust their physiology to achieve a defined goal. For example, in Ref.~\cite{chure_optimal_2023}, a phenomenological model is developed assuming that cells match metabolic and protein synthesis fluxes during adaptation, while Ref.~\cite{Giordano2016} formulates a model based on growth rate optimization principles.

Conversely, the bottom-up approach aims to explicitly describe molecular circuits. A key example is Ref.\cite{Bosdriesz2015}, which models ppGpp sensing and RelA-mediated regulation using biochemical reaction networks~\cite{Wendrich2002}. A similar molecular approach is employed in Ref.~\cite{Shachrai2010}, where regulatory interactions governing ribosome synthesis are modeled explicitly.

A hybrid between the top-down approach and a model including mechanistic aspects is proposed in Ref.~\cite{Droghetti2025}, which, building upon the model by Erickson and coworkers~\cite{Erickson2017}, additionally describes the ppGpp sensing of the elongation translation rate using the relation discovered in Ref.~\cite{Wu2022} (described in Sec.~\ref{sec6:mecha_regulation}), and the kinetics of the amino acid pool (discussed later in Sec.~\ref{sec10:communities}). 

In this section, we have chosen to illustrate how a top-down approach can be used to model nutrient shifts in \textit{E. coli}, focusing on the kinetics of proteome sectors and resource allocation during transitions out of steady state.

\subsection*{INGREDIENTS AND RECIPES\footnote{The recipe introduced here builds upon ideas and models presented in Ref.~\cite{Erickson2017}.}}

We start by providing some broad guidelines to help build a model under non-steady conditions, taking a ``macroscopic'' ODE approach describing global averages of key cellular quantities (ribosomes, metabolic enzymes and key regulators). Then, we will introduce an example of a top-down dynamical model.\\

\noindent\textbf{General considerations on out-of-steady-state modeling.}
If the aim of the model is to study the adaptation of the protein synthesis during out-of-steady-state perturbations, specifying the kinetics of the biosynthesis flux~$J_\text{syn}(t)$ is essential. Previous sections show different way of modeling~$J_\text{syn}(t)$, which always depends on the ribosomal sector~$\phi_\text{R}(t)$. Therefore,~$\phi_\text{R}(t)$ must be included with its allocation parameter~$\chi_\text{R}(t)$, which in this case it  will be a function of time --with direct dependence or via the dependence on other physiological parameters. In this context, the allocation functions are also called ``sector regulatory functions'', as they regulate the size of the sector in out-of-steady-state scenarios by setting the time-dependent target of~$\phi_\text{i}$ (see~\ref{box6:target}).
Different types of perturbations elicit distinct cellular responses (Sec.~\ref{sec2:classical} and Refs.~\cite{scott_interdependence_2010, hui_quantitative_2015, xia_proteome_2022}), determining which additional specific components should be incorporated into the model. It is important to assess how the perturbation affects the system and how the system detects and responds to it, as this aids in selecting these additional components (fluxes, sectors/molecules, rates etc.) and identifying growth-limiting factors to focus on. For instance, in a carbon downshift, where bacteria transition from fast to slow growth, growth-limiting sectors such as carbon uptake must be explicitly modeled to capture experimental observations~\cite{Hermsen2015, Erickson2017}. Conversely, in other perturbations, such as nutrient upshifts or drug treatments, the emphasis might shift to other sectors or regulatory molecules.
The (experimental) knowledge of the regulatory mechanisms that govern the proteome regulatory functions~$\chi_\text{i}$ (and the other quantities incorporated in the model) will determine the level of detail of our description. Depending on this knowledge, ``top-down'' or  ``bottom-up'' modeling approaches can be employed. A bottom-up approach is generally preferred for testing specific architectures or generating predictions, while a top-down approach may be safer to characterize general behavior and experimental data. A hybrid approach is sometimes useful to study the interplay between specific circuits and global physiology.

\begin{figure}[!bt]
    \centering
    \includegraphics[width=0.8\columnwidth]{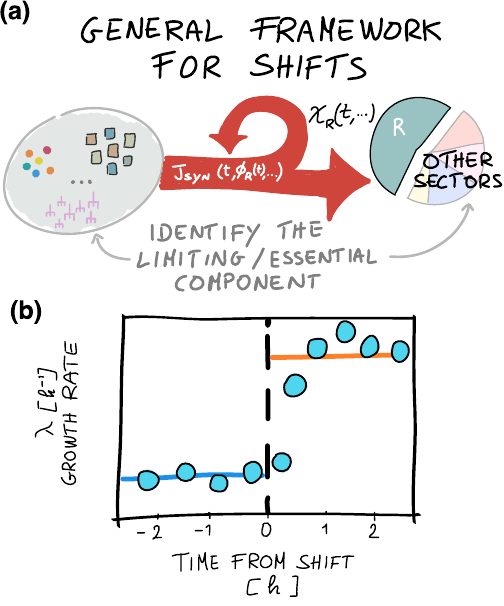} 
    \caption{\textbf{General flux framework for non-steady conditions.} 
    (\textbf{a}) The fundamental biosynthesis flux~$J_\text{syn}(\phi_\text{R}(t), t,...)$, operated by the ribosomal sector~$\phi_\text{R}(t)$ is represented by the red arrow. A part of $J_\text{syn}$ is redirected towards the ribosomal sectors according to the regulatory functions~$\chi_\text{R}(t,...)$, while the remaining protein mass produced belongs to the other sectors. 
    In addition to these fundamental quantities, the panel shows a pool of precursors that determines the biosynthesis flux (via resource availability) and the regulatory functions (via sensing mechanisms). This pool can represents amino acids, tRNAs, the external nutrients, regulatory molecules and so on, depending on the model demands. Depending on the specific perturbation applied, it could be necessary to add other ingredients to the base one-flux framework. In this case, one has to identify which is the limiting process for growth and add its description.
    (\textbf{b}) Example of the dynamic of the growth rate in a dynamical environment (in particular, a nutrient upshift from a glucose base medium to a rich medium with casaminoacids), from Ref.~\cite{Droghetti2025}. The plot shows the time course of the exponential growth rate, which goes from the preshift value (highlighted with the blue line) to the post shift one (orange line). \edit{Panel (\textbf{a}) is inspired by Ref.~\cite{Erickson2017}.}
    }
    \label{fig:shift}
\end{figure}

Figure~\ref{fig:shift}(a) shows a scheme of the essential model ingredient: (I) the biosyntheis flux~$J_\text{syn}(t)$, (II) the ribosomal sector~$\phi_\text{R}(t)$, and (III) the ribosomal regulatory function~$\chi_\text{R}(t)$.
Additionally, the scheme displays the other protein sectors, synthesized by the ribosomes, and a pool of precursor (which can be amino acids, elongation factors, tRNA etc) from which the synthesis flux and the regulatory function(s) depend on.
Figure~\ref{fig:shift}(b) shows an example from Ref.~\cite{Droghetti2025} of the dynamics displayed by the istantaneous growth rate of a batch culture when a rich component (in this case casaminoacids) is added to the previous carbon medium.

In the following, we present a simple top-down approach for studying nutrient shifts, derived from Ref.~\cite{Erickson2017}.\\

\noindent\textbf{A top-down model for nutrient shifts.}
The model in Ref.~\cite{Erickson2017} describes the rearrangement of protein synthesis during nutrient up- and down-shifts, specifically when the carbon substrates present in the external medium change because either (I) one of the sugars runs out before the others or (II) a new sugar is added to the solution.
Hence, next to the  biosynthesis flux mediated by the ribosomes, the carbon uptake sector that mediates the nutrient uptake flux~\cite{you_coordination_2013, Hermsen2015}, must be included.
Indeed, in a downshift, the carbon uptake becomes the limiting factor for growth. When a nutrient is depleted or replaced with a lower-quality alternative, the cell must not only expand its carbon uptake sector to manage the inferior nutrient but also completely reconfigure it, as different carbon sources require distinct uptake proteins.

The carbon-intake and catabolism flux is defined by
\begin{equation}
J_\text{cat}(t) = \Sigma_\text{i} \nu_\text{i}(t) M_{\text{C}_\text{i}}(t)
\label{eq:FCR_J_C} ,
\end{equation}
where the sum covers all available carbon sources to the cell, each characterized by an uptake rate/nutrient capacity $\nu_\text{i}$ and a substrate-specific catabolic proteins of mass~$M_{\text{C}_\text{i}}$, whose corresponding sectors~$\phi_{\text{C}_\text{i}}$ are part of the much broader sector~$\phi_\text{P}$, and are specifically regulated by cAMP~\cite{you_coordination_2013} (see also Sec.~\ref{sec10:communities}). 
The biosynthesis flux, in this context, is defined as
\begin{equation}
J_\text{syn}(t) = \sigma(t) M_{R}(t) \label{eq:FCR_Jsyn}
\end{equation}
where~$M_\text{R}(t)$ denotes the mass of the ribosomal proteins (sector~$\phi_\text{R}$), and, following Ref.~\cite{Erickson2017}, $\sigma(t)$ represents the translational activity. Linking to the notations used in this Tutorial, $\sigma = \lambda / \phi_\text{R} =\varepsilon_\text{R} f_\text{a}$.
At the level of the studied timescales (which correspond to the synthesis timescales, order of tens of minutes) we can assume that the synthesis flux instantaneously matches the catabolic one~\cite{Erickson2017}:~$J_\text{cat}(t) = J_\text{syn} (t)$. As explained in Sec.~\ref{sec5:degradation} and in~\ref{box5:aa_deg_recycling}, this assumption holds as amino-acids are not explicitely considered.
During a nutrient shift, the uptake flux undergoes a sudden change determined by the set of external nutrient qualities~$\{ \nu_\text{1}... \ \nu_\text{i}... \ \nu_\text{N} \} \rightarrow \{ \nu_\text{1}... \ \nu_\text{j}... \ \nu_\text{N} \}$, which in turn causes a rapid change in the translational activity~$\sigma$ (the proteome composition~$\Phi = \{ \phi_\text{R}, \phi_{\text{C}_\text{i}}... \ \}$ will change with slower time scales). This rapid change serves as the signal informing the cell of external environmental variation~\cite{Bosdriesz2015, Erickson2017, Mori2017, Wu2022, Droghetti2025}.

The final element to consider is the synthesis and regulation of the proteome sectors involved, described by the equation
\begin{equation}
\frac{\text{d}M_\text{j}(t)}{\text{dt}}=\chi_\text{j}(t) J_{\text{syn}}(t) , \ \text{j} \in [\text{R}, \text{C}_\text{i}]   \label{eq:FCR_synthesis} ,
\end{equation}
where~$\chi_\text{j}$ represents sector-specific regulatory functions. These functions determine the fraction of the total biosynthesis flux that is directed towards different sectors, as presented in Sec.~\ref{sec3:translation} and derived in~\ref{box6:target}. \edit{In principle the regulatory functions depend on details of the nutrient sensing, however, adopting a top-down approach, one can describe them using information from the steady state growth laws, as Eq.~\eqref{eq:GL_first}, bypassing the regulatory details illustrated in~\ref{box6:target}.} Specifically, \edit{this approach 
leads to the following definitions,}
\begin{equation}
\chi_\text{R}(t) = \frac{\phi_\text{R}^\text{min}}{1 - \sigma(t) / \gamma_\text{0}} , \label{eq:FCR_xR}
\end{equation}
\begin{equation}
\chi_{\text{C}_\text{i}}(t) = h_\text{i} \left( 1 - \frac{\sigma(t)}{\lambda_\text{C}} \chi_\text{R}(\sigma(t))  \right) \ , \label{eq:FCR_xCat}
\end{equation}
\edit{
where Eq.~(\ref{eq:FCR_xCat}) derives directly from the first growth law, Eq.~\eqref{eq:GL_first}, and Eq.~(\ref{eq:FCR_xCat}) from the one of the catabolic sector~$\phi_\text{C}$ (see Appendix~\ref{sec11:parameters_numbers}). In there, $\lambda_\text{C}$ is the maximum growth rate achievable on carbon limited media~\cite{you_coordination_2013, Hermsen2015} while~$h_\text{i}$ is an index function that equals to~$\phi^\text{max}_{\text{C}_\text{i}}$ when the corresponding substrate is present and is 0 otherwise~\cite{Erickson2017}.
Appendix~\ref{sec11:parameters_numbers} reports a detailed derivation of Eqs.~(\ref{eq:FCR_xR})-(\ref{eq:FCR_xCat}). This approach entails a quasi-steady state approximation for all the regulatory steps (compare to~\ref{box6:target}) linking the effective translation rate~$\sigma(t)$ to the regulatory functions~$\chi_\mathrm{R}$ and~$\chi_{\text{C}_\text{i}}$~\cite{Erickson2017}.
}

In summary, this top-down approach effectively regulates proteome sector synthesis through sector-specific regulatory functions derived from steady-state relations, offering a coarse-grained representation of cellular biosynthesis dynamics. A simple way to include mechanistic aspects in this model is to specify the biochemical networks relative to the different steps linking allocation of resources to nutrients (see Refs.~\cite{Bosdriesz2015, Droghetti2025} and Sec.~\ref{sec6:mecha_regulation}).

\begin{mdframed}[backgroundcolor=light-gray]
\refstepcounter{boxcounter}
{\bf \large \theboxcounter. Dynamical equations for~$\chi$,~$\phi$,~$\omega$.}
\label{box6:target}

\noindent\textbf{Equation for~$\chi(t)$.} In Sec.~\ref{sec4:transcription} we have defined the regulatory function~$\chi_\text{i}$ as the ratio of the number of transcripts coding for proteins belonging to the i-th sector to the total number of transcripts:~$\chi_\mathrm{i}(t) = m_\mathrm{i}(t) / m(t)$.
Therefore, to study the dynamics of~$\chi_\mathrm{i}$, we need to consider the dynamics of transcripts, which can be described by the following equation, 
\begin{equation}
  \frac{\mathrm{d}m_\mathrm{i}(t)}{\mathrm{dt}}
  = \frac{\varepsilon^\text{TX}}{\ell} \omega_\mathrm{i}(t)[N]f_\text{bn} - \delta m_\mathrm{i}(t) \ ,
  \label{eq:7_dm_i_dt}
\end{equation}
where $\varepsilon^\text{TX}$ and $\delta$ are constants representing RNAP elongation rate and mRNA degradation rate, respectively,~$[N]f_\text{bn}$ is the total number of bound RNAP (see Sec.~\ref{sec4:transcription}), and~$\ell$ is the average gene length. This equation is obtained by combining Eq.~(\ref{eq:transcript_dynamics}) with Eqs.~(\ref{eq:J_i_TX} - \ref{eq:RNAP_alloc}), all introduced in Sec.~\ref{sec4:transcription} and by making the additional assumption that all transcripts have the same degradation rate ($\delta_\text{i} = \delta \ \forall \text{i}$) and the same gene lenght ($\ell_\text{i} = \ell \ \forall \text{i}$). Reference~\cite{calabrese_how_2024} shows the complete derivation. 
Eq.~(\ref{eq:7_dm_i_dt}) represents the dynamics of the i-th class of transcripts. \edit{Note that Ref.~\cite{Lin2021} shows that this mean-field equation can change in presence of extrinsic noise.} To obtain the total transcript dynamics we sum over i to obtain:
\begin{equation}
  \frac{\mathrm{d} m(t)}{\mathrm{dt}} = \frac{ \varepsilon^\text{TX}}{\ell} [N]f_\text{bn} - \delta m(t) \ .
  \label{eq:7_dm_dt}
\end{equation}

By combining Eq.~(\ref{eq:7_dm_i_dt}),~(\ref{eq:7_dm_dt}) and the definition of~$\chi_\text{i}$, we obtain (via the chain-rule of the derivative):
\begin{equation}
  \frac{\mathrm{d} \chi_\mathrm{i}(t)}{\mathrm{dt}} =
   \frac{1}{m} \frac{\varepsilon^\text{TX}}{\ell} [N]f_\text{bn} [ \omega_\mathrm{i}(t) - \chi_\mathrm{i}(t)] \ .
  \label{eq:7_dchi_i_dt}
\end{equation}


\noindent \textbf{Equation for~$\phi(t)$.} The following describes the argument introduced in Ref.~\cite{Erickson2017}. The dynamics of the sector sizes~$\phi_\text{i}$ is given by
\begin{equation}
  \frac{\mathrm{d} \phi_\mathrm{i}(t)}{\mathrm{dt}} =
  \lambda(t) [ \chi_\mathrm{i}(t) - \phi_\mathrm{i}(t)] \ .
  \label{eq:7_dphi_i_dt}
\end{equation}
\edit{
The above equation is derived by applying the chain-rule of the derivative to the definition of a  proteome fraction~$\phi_\text{i} = M_\text{i}/ M_p$, inserting the already introduced equation for the sector- and total-mass accumulation~$\dot{M}_\text{i} = J_\text{syn} \chi_\text{i} = \lambda M_p \chi_\text{i}$, introduced in Sec.~\ref{sec2:classical} and~\ref{sec3:translation}.\\
}
\noindent\textbf{Equation for~$\omega(t)$.}
Observe that also~$\omega_\text{i}$ can display a characteristic time to adjust to the changing regulations\cite{Droghetti2025}. In this case, one can define a ``target'' for~$\omega_\text{i}$, denoted as~$\bar{\omega}_\text{i}$, which in the case of ribosomal proteins is a function of the ppGpp concentration~$[G]$, Eq.~(\ref{eq:6_bar_omegaR}), and will reach the steady state value~$\bar{\omega}_\mathrm{i}^*$ when also~$[G]$ arrives at steady state. The dynamics of the RNAP allocation is described by
\begin{equation}
  \frac{\mathrm{d} \omega_\mathrm{i}(t)}{\mathrm{dt}} =
  \frac{1}{\tau_{\omega}} [ \bar{\omega}_\mathrm{i}(t) - \omega_\mathrm{i}(t)],
  \label{eq:7_domega_i_dt}
\end{equation}
where~$\tau_\omega$ is the typical RNAP adjusting time.\\

\noindent\textbf{Steady state growth.} During steady state growth, the usual equality holds~$\bar{\omega}_\text{i}^* = \omega_\text{i}^* = \chi_\text{i}^* = \phi_\text{i}^*$, where the apex * denotes the steady state value of the various quantities.
\end{mdframed}

\subsection*{OUTLOOK} 
Studying non-stationary regimes, such as onset of exponential growth, of stationary phase, starvation, and the resumption of growth~\cite{Monod1949}, is crucial for understanding bacterial fitness and evolutionary adaptation maximizing survival and reproduction along complex life-history paths.
By integrating diverse datasets and theoretical frameworks, we can gain a deeper insight into the complex dynamics of adaptation across different organisms. 
In recent years, significant progress has been made in understanding nutrient shifts in bacteria, through data and models~\cite{Dennis1974, Bosdriesz2015, Erickson2017, KoremKohanim2018, chure_optimal_2023, Wu2022, Droghetti2025}. Different models offers advantages and limitations, and many can be further refined by incorporating additional mechanisms such as degradation, regulation of translation, mRNA dynamics, and resource recycling.
However, other out-of-steady-state growth conditions still deserve attention. 
In particular, less attention has been given to scenarios where growth is temporarily arrested. Examples include shifts toward starvation conditions~\cite{Bren2013, metzl-raz_principles_2017, Droghetti2024_stv, Gervais2024} and transitions between growth conditions that involve substantial lag times~\cite{Basan2020}.  Other scenarios involes study where the environment is constantly fluctuating between different states~\cite{Julou2022, Murugan2021} that can include the presence of antibiotics and the emergence of tollerance and resistance~\cite{Fridman2014}.

For eukaryotes, particularly in the context of out-of-steady-state conditions, the development of quantitative models remains underexplored.  
There is a wealth of out-of-steady-state data available for budding yeast, starting with seminal studies from the Botstein lab~\cite{Gasch2000, Ronen2005, Brauer2008}. A recent paper from the Barkai lab~\cite{metzl-raz_principles_2017} collected data on the transition into stationary phase. These efforts provide a foundation for further research and model development in this area.

\section{Intermezzo: Density homeostasis and osmo-metabolic regulation of volume}
\label{sec8:density}

This section considers volume growth as a complex and dynamic process, driven by osmotic pressure balance on fast time scales (seconds to tens of minutes) and by the synthesis of cellular mass (e.g., proteins, lipids) on larger time scales (from tens of minutes to the cell cycle)~\cite{Cadart2019}.

\subsection*{MAIN QUESTION AND MOTIVATIONS}

Achieving balanced exponential growth --where all cellular components increase proportionally at the same rate,~\ref{box1:exp_balanced}-- requires the regulation of intracellular density, ensuring that the cell's accumulated mass from biosynthesis keeps precise pace with its expanding volume. In turn, intracellular density, by setting the concentration of all cellular components within the cytoplasm, is a critical parameter influencing cellular function by modulating biomolecular activities, macromolecular crowding, diffusion, phase behavior of the cytoplasm and mechanical properties of the cell~\cite{vandenBerg2017, Neurohr2020, odermatt_variations_2021}.
Therefore density homeostasis, or the preservation of the concentration of all cellular components during growth~\cite{Lin2018,Cadart2019, Neurohr2020, fu_cells_2024,Srivastava2025}, is key to maintaining optimal biochemical rates and translation capacity~\cite{Pang2023,dai_slowdown_2018,Delarue2018}. 

While we have already discussed the assumption that cell mass dynamics is approximated by protein production see Sec.~\ref{sec2:classical}, volume dynamics follows different behaviors depending on time scales~\cite{Cadart2019}. At the time scale of a cell cycle, volume has to follow mass in balanced growth conditions (or the cell would be indefinitely diluted or concentrated~\cite{Ginzberg2015}). At shorter time scales, cell volume depends on an intersection of biophysical processes that include dynamic equilibrium of ion pumping, mechanical balance between osmotic and hydrostatic pressure, and maintenance of an overall neutral charge~\cite{rollin_physical_2023}. The main question arises: how do cells mechanistically achieve density homeostasis, especially when mass and volume growth rates can decouple at short time scales or under perturbations, leading to changes in cellular density?

This question is central because understanding density regulation can reveal how cells optimize biochemical processes, respond to stress, and maintain growth across varying conditions. The ``Ingredients and Recipes'' section below describes a theoretical framework to address this question.

\subsection*{INGREDIENTS AND RECIPES\footnote{The recipe introduced here builds upon ideas and models presented in Refs.~\cite{Srivastava2025,rollin_physical_2023}.}}

\noindent
\textbf{Generic phenomenological model for cellular density homeostasis}. We start by considering a general and phenomenological model of density homeostasis, from Ref.~\cite{Srivastava2025}, where the dry mass density~$\rho_\text{DM}$ dynamics is determined by the balance between independent but density-coupled mass growth rate~$\lambda_M$ and volume growth rate~$\lambda_V$,
\begin{equation}
\label{eq.8_drhodt}
  \frac{\text{d}\rho_\text{DM}}{\text{dt}} = \left(\lambda_M - \lambda_V \right) \rho_\text{DM} \ .
\end{equation}
Importantly, this model differs from the standard assumption of autonomous exponential volume growth, which simplifies density dynamics by assuming constant production and dilution rates, and overlooks the possible couplings and regulatory links between mass and volume. 

Instead, we treat the volume and mass growth rates~$\lambda_M$ and~$\lambda_V$ as dynamic variables, potentially dependent on density through functions~$\mu_M(\rho_\text{DM})$ and~$\mu_V(\rho_\text{DM})$, which can in principle be measured experimentally, representing density-dependent target mass and volume growth rates, respectively. The rate of change of the growth rates can be expressed as
\begin{equation}
\label{eq.8_mass_growth}
  \frac{\text{d}\lambda_M}{\text{dt}} =
  \left[ \mu_M(\rho_\text{DM}) - \lambda_M  \right] \theta_M \ ,
\end{equation}
and
\begin{equation}
\label{eq.8_volume_growth}
  \frac{\text{d}\lambda_V}{\text{dt}} =
  \left[ \mu_V(\rho_\text{DM}) - \lambda_V  \right] \theta_V \, ,
\end{equation}
with $\theta_M$ and $\theta_V$ the timescales to relax to the respective steady-state $\mu$ values.

Given this model, we can ask for the conditions for density homeostasis, defined as stable fixed point~$\rho_\text{DM}^*$. 
The model predicts density homeostasis if two conditions are met. First, we must have~$\mu_M(\rho_\text{DM}^*) =  \mu_V(\rho_\text{DM}^*)$ (at~$\rho_\text{DM}=\rho_\text{DM}^*$, mass and volume growth rates are equal). Second, we must have that 
\begin{equation}
\label{eq.8_rho*}
    \left. \left(\frac{\partial}{\partial \rho}\frac{\text{d}\rho_\text{DM}}{\text{dt}}\right)\right|_{\rho = \rho^*}
          = \frac{\partial}{\partial \rho} \mu_M(\rho_\text{DM})
          - \frac{\partial}{\partial \rho} \mu_V(\rho_\text{DM}) |_{\rho =\rho^*} < 0  \, ,
\end{equation}
where the partial derivative is in $\rho^*_\text{DM}$.

The derivative of the density variation with respect to density at the fixed point is negative, ensuring stability~\cite{Srivastava2025}.

If we assume linear relationships~($\mu_M(\rho_\text{DM}) = a_M + b_M \rho_\text{DM}$ and~$\mu_V(\rho_\text{DM}) = a_V + b_V \rho_\text{DM}$), homeostasis requires~$b_M < b_V$. This includes cases where~$b_M < 0$ and~$b_V \geq 0$ , or~$b_M$ and~$b_V$ are both positive but with $b_M < b_V$. In these cases, the homeostatic target density~$\rho_\text{DM}^*$ is precisely at the intersection of the~$\mu_M(\rho_\text{DM})$ and~$\mu_V(\rho_\text{DM})$ curves. This also holds approximately for non-linear relationships with a non-vanishing linear term if linearized around~$\rho_\text{DM}^*$, i.e., if the fluctuations around the homeostatic density can be assumed to be sufficiently small. 
The four parameters~$a_M$, $b_M$, $a_V$ and~$b_V$ were measured directly from joint dynamic measurements of mass and volume of mammalian cells in Ref.~\cite{Srivastava2025}, by taking volume-specific derivatives of the mass and volume time series and comparing them to the mass-to-volume ratio in the corresponding frame. By this approach, the model predicts a value for the homeostatic density~$\rho_\text{DM}^*$, which can be compared to experimental measurements. 

In conclusion, this phenomenological framework from Ref.~\cite{Srivastava2025} provides a basic understanding of the dynamics of cellular density homeostasis, highlighting the importance of density-coupled mass and volume growth rates in achieving stable density. It also suggests simple experimental tests for looking at the presence of density regulation and asking whether the compensatory effects are related to volume or biosynthesis.\\

\noindent\textbf{An osmo-metabolic volume growth model including osmotic pressure balance and the Pump-Leak mechanism.} Beyond the phenomenological level, we can ask whether we can provide a mechanistic framework for cell volume~\cite{Cadart2019}. We start by describing in a simplified way how on fast time scales (minutes to tens of minutes) cellular volume $V$ is determined by the osmolyte balance inside and outside the cell. The so called ``pump-leak model'' first pioneered by Tosteson et Hoffman~\cite{tosteson_regulation_1960} describes the basic aspects of ion transport and osmotic equilibrium.  Regarding the dry mass $M$ of the cell, as the majority of the dry mass is composed of proteins~\cite{Gasch2000}, it can be assumed again (to a first approximation) to be set by protein production. 

The pump-leak model was formulated as a model of ionic transport and proposed as a way for the cell to regulate its volume through active cation transport~\cite{Kay2017}, but in its general formulation it is complicated by the complex ion dynamics in a cell. 
Recently, Rollin and coworkers~\cite{rollin_physical_2023} have simplified and adapted this model to predict the volume of a growing cell and derived an expression the cell density~$\rho_\text{DM} = M/V$ under different conditions. Following this approach, we describe a cell as a compartment with a dilute solution (where van't Hoff's law governing osmotic pressure~\cite{rodenburg_vant_2017} can be applied) containing ionic species that can be actively transported, as well as other impermeant osmolytes. We take three main assumptions: (i) electroneutrality, (ii) osmotic balance (mechanical balance between osmotic and hydrostatic pressure), and (iii) balance of ionic fluxes.

In a bacterial cell, the most abundant cations are generally actively pumped outside the cell, except for potassium which remains as the most abundant positively charged species~\cite{rob_philips_ron_milo_cell_2015}. This is because maintaining a low intracellular concentration of certain cations, such as sodium and calcium (but not potassium), is crucial for various cellular processes~\cite{GALERALAPORTA_ion}. Conversely, the anions are not pumped but they are mostly regulated by passive transport and ion exchange systems~\cite{Terradot2024}. However, the most abundant source of negative charges comes from proteins, which need positive counterions for electroneutrality~\cite{ProteinCharge, SMITH1992124}.
To simplify this complex scenario we will assume, following Rollin and coworkers, that only two species of monovalent positive and negative ions~($\pm$) are present~\cite{rollin_physical_2023}. 

Electroneutrality imposes charge neutrality inside the cell. The non-ionic impermeant osmolytes inside the cell, with concentration~$c$ and average charge~$z$, are balanced with the positive and negative ions of concentrations~$n_+$ and~$n_-$, respectively, such that
\begin{equation}
    \label{eq.8_electroneutrality}
    c z + n_- = n_+ \, .
\end{equation}

The osmotic balance condition imposes that the osmotic pressure difference~$\Delta \Pi$, caused by the difference in impermeant osmolyte concentration inside and outside the cell, balances the hydrostatic pressure difference $\Delta P$. Thus we write
\begin{equation}
    \label{eq.8_osmotic}
    \Delta P = \Delta \Pi = \Pi_{cell} - \Pi_{ext} = k_\text{B} T (c + n_+ + n_- - 2n_0) \, ,
\end{equation}
where~$n_0$ is the outside anionic and cationic concentration (equal by electroneutrality), $k_\text{B}$ the Boltzmann constant, and $T$ the temperature. \edit{This assumes that each ion counts as an independent osmolyte, \textit{i.e.} we neglected Manning condensation~\cite{rollin_physical_2023}, which could play a role for counterions to a single highly negative macromolecule like RNA. Furthermore, for simplicity we will set~$\Delta \Pi = 0$ in the following, and such that tension from the cell surface (due to, e.g., membrane and cell wall) is neglected~\cite{Cadart2019}}.
This is a rough approximation for microbes, as for example in bacteria membrane and cell-wall tension are strong, and turgor pressure is reported to play a relevant role in cell growth~\cite{ROJAS}.
The surface tension contribution can be described by the Young-Laplace equation~\cite{Young_Laplace}.   
Finally, the balance of ionic fluxes shows the equilibrium imposed by the passive motion of ions through channels and their active pumping. 
\begin{figure}[!bt]
    \centering
    \includegraphics[width=1\columnwidth]{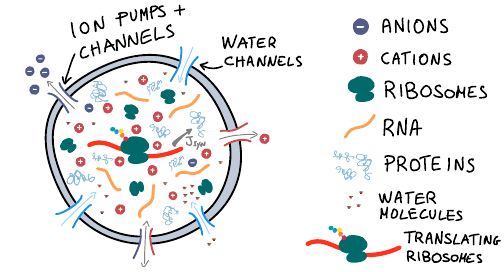} 
    \caption{ \textbf{Pump-leak model of the cell density.} Cell volume is set according to the concentrations of osmolytes in the cell. These osmolytes are mainly composed of cations, necessary as counterions to negatively charged amino acids and other large molecules (\textit{e.g.} proteins, RNA). To maintain electroneutrality, anions are pumped out of the cell \edit{(passive channels and active pumps are schematically represented in the figure)}. The ratio of amino acids to proteins impacts the number of counterions and through this the cell volume. The synthesis flux $J_{syn}$ is thus an integral part of the cell volume in the pump-leak model~\cite{tosteson_regulation_1960}. \edit{This figure is inspired by Ref.~\cite{rollin_physical_2023}}
    }
    \label{fig7}
\end{figure}

The ion concentrations are set  by a balance between chemical potential differences~$\Delta \mu_\pm = k_\text{B} T \log(n_\pm / n_0)$ (for the two ion species labelled by~$\pm$), electric potential difference~$U$ applied on the ions of average charge~$z_\pm$, and the ratio of ion pumping rate~$\sigma_\pm$ to permeability~$\kappa$, $\sigma_\pm/\kappa$.  This balance equation can be expressed as
\begin{equation}
n_\pm = n_0 \exp ( \beta Q_\pm )\ ,    
\end{equation}
where~$\beta = 1/k_\text{B} T$ and ~$Q_\pm = \sigma_\pm/\kappa - z_\pm U $ summarizes the contribution of active pumping and membrane potential difference. Multiplying the two equations for~$n_+$ and~$n_-$, we get
\begin{equation}
    \label{eq.8_ionic}
    n_- n_+ = n_0^2 a \, ,
\end{equation}
where the factor~$a$ represents the pumping efficiency~$\exp[ \beta (Q_+ + Q_-)]$, most of which represents the pumping current of abundant cations outside into
the cell. This can be approximated as the outwards pumping of mostly protons and sodium ions (together with calcium ions) by most pumps, but the model can be adapted to any monovalent ions considered.

From Eqs.~(\ref{eq.8_electroneutrality}) and~(\ref{eq.8_ionic}), we can predict the cytoplasmic ion concentration to be 
\begin{equation}
    n_{\pm} = \frac{\pm c+ \sqrt{\left( c z \right)^2 - 4n_0^2 a}}{2} \, .
\end{equation}
Following again ref.~\cite{rollin_physical_2023}, we approximate the pumping to be infinitely efficient, setting~$a=0$. In this situation~$n_- n_+ = 0$ and there can only be negative or positive ions in the cell.  We further assume that there are only cations in the cell and they act as counterions for proteins 
\begin{equation}
    \label{eq.metabolite_conc}
    c = \frac{2n_0}{1+z} \, .
\end{equation}
As we mentioned previously, this is a highly simplified description, since in reality (especially in bacteria) mostly \textit{positive} ion species are pumped out, and the main counterion for proteins can be considered to be potassium. However, as we will demonstrate below, the current assumptions provide valuable insights and serve as a productive starting point for understanding the underlying mechanisms, while also inviting scrutiny for exploring complementary assumptions.

The concentration~$c$ of non-ionic osmotically active species depends on an osmotically active volume~$V-V_\text{ex}$. In a simple model of a compartment containing osmolytes having a finite volume,~$V_\text{ex}$ can be interpreted as the total volume of all nonsolvent molecules in the compartment~\cite{rollin_physical_2023}. For a cell, this volume is more complex to interpret, but can clearly contain a contribution from entire cellular compartments that contain water. 
Importantly,~$V_\text{ex}$ is measurable from the offset of the line relating volume changes to concentration changes in osmotic shocks (called ``Ponder plot''~\cite{venkova_mechano-osmotic_2022}).

We can write
\begin{equation}
    c = \frac{C}{V - V_\text{ex}} \, ,
    \label{eq:8_Vext}
\end{equation}
with $C$ representing the number of active osmolytes. 

Rollin et al.~\cite{rollin_physical_2023} further hypothesize that the active osmolytes are dominated by a metabolic product, such as free amino acids or precursors in the cell, so that we can replace~$C$ with the number of these species (for simplicity of language referred to as ``amino acids'' in the following) $A$ and their average charge~$z_A$, such that we can write the volume as a function of the protein and free amino acid number abundance. This expression then gives the cell density~$\rho_\text{DM}$, the ratio between the total dry mass, assumed to be set by the protein content and the volume, given by 
\begin{equation}
    \label{eq.8_rho_pump_leak}
    \rho_\text{DM} = \frac{\mu_\text{p} P}{V_\text{ex} + A \left( 1 + z_A \right)/2n_0} \, ,
\end{equation}
with~$\mu_\text{p}$ being the average protein mass.

Section~\ref{sec2:classical} introduced the catabolic flux ~$J_\text{cat}= \nu M_p$ defined in Eq.~(\ref{eq:nu_phiC})
and the protein synthesis flux, as defined in Eq.~(\ref{eq:GL_first_fa}), $J_\text{syn} = \varepsilon_\text{R} f_\text{a} M_\text{R}$ (neglecting protein degradation). Using these expressions, we can determine the number of free amino acids, set by the offset of~$J_\text{cat}$ and~$J_\text{syn}$. Considering the ratio of free to bound amino acids~$\varphi_\text{A}$, we then have~$\dot{\varphi_\text{A}} = \nu \phi_\text{P} - \varepsilon_\text{R} f_\text{a} \phi_\text{R} (1+\varphi_\text{A})$, as in~\ref{box5:aa_deg_recycling}.
In the case of exponential growth, as introduced in~\ref{box1:exp_balanced}, $\varphi_\text{A}$ is constant, and the total number of free amino acids then increases exponentially over time, following the same rate as the number of proteins~$P$,
\begin{equation}
    A = P \frac{ \nu \phi_\text{P} - \lambda_\text{M} }{\lambda_\text{M}} \, .
\end{equation}
The volume growth rate~$\lambda_V$ is then set by the number of free amino acids.

Additionally, in exponential growth where~$\frac{\text{d}P}{\text{dt}} \propto P$, the cell density~$\rho_\text{DM}$ remains constant and
\begin{equation}
    \rho_\text{DM} = \frac{\mu_\text{p}}{V_\text{ex}/P  + \left( J_\text{cat}  - \lambda_\text{M} \right) \left( 1 + z_A \right)/2n_0 \lambda_\text{M}} \, .
\end{equation}
\noindent
Along the first growth law, growth rate~$\lambda $ and the other parameters are nutrient-independent (see Sec.~\ref{sec2:classical}). Consequently, the model predicts that the dry-mass density~$\rho_\text{DM}$ should be constant across conditions, provided that the dry mass density of the osmotically inactive part of the cell is also constant~\cite{Srivastava2025}. Hence, any experimentally observed deviation from a constant dry-mass density must originate from factors outside this minimal description, such as finite varying pumping efficiency, shifts in osmolyte composition, or changes in the osmotically inactive volume~\cite{venkova_mechano-osmotic_2022}.

\subsection*{OUTLOOK}

We now attempt to place these tools in a wider context of experimental results. 
Population-average behavior of steadily-growing \textit{E.~coli} shows quite constant dry mass density across growth conditions~\cite{kubitschek_independence_1984, balakrishnan_principles_2022}. However, density homeostasis breaks down at the single-cell level due to fluctuations and cell-cycle trends~\cite{oldewurtel_robust_2021, odermatt_variations_2021, Srivastava2025}, caused by cell shape changes or phases where protein synthesis decouples from volume growth. Perturbations in mass/protein density can emerge during the cell cycle or be artificially induced, generally associating with detrimental effects on growth processes~\cite{neurohr_excessive_2019, terhorst_environmental_2023,dai_slowdown_2018}.

The variety of scenarios where mass and volume dynamics uncouple is vast, including nutrient limitation, genome mutations, and antibiotics, highlighting the need for mathematical models that account for their interplay~\cite{rollin_physical_2023, oldewurtel_robust_2021, Srivastava2025}. 
Importantly, the basic osmo-metabolic framework defined in the previous section can be modified to explore different scenarios and compare different hypotheses, exploring their consequences. We list below three areas where this appears to be particularly interesting. 

First, the assumptions made in the previous section refer to population averages in balanced growth conditions, and lead to conclude that density remains constant with these approximations. However, in perturbed non-steady conditions, this may not be the case, and the framework for predicting volume expansion is easily adapted to these conditions.
One example is the case of non-exponential growth, or any condition that is not balanced growth~\cite{Lin2018, neurohr_excessive_2019, rollin_physical_2023}. In such conditions, the number of free amino acids no longer follows the protein number. As the intake of amino acids is no longer proportional to their consumption, the cell density can change accordingly. 
\edit{More widely, recent studies found that deviations from exponential growth occur systematically across the cell cycle of different bacteria~\cite{Nordholt2020, Kar2021,Messelink2021}, and have proposed explanations based on different possible limiting factors. These works show that cell volume can grow sub-exponentially, with growth rate  depending on cell-cycle progression and environmental conditions. Such findings call for models capable of describing mass–volume coupling beyond the steady-state assumption. In parallel, recent experimental evidence points to a feedback between intracellular crowding and protein synthesis and degradation rates~\cite{Chen2024}, suggesting that cytoplasmic viscosity itself may act as a regulatory variable. Incorporating these effects within the osmo-metabolic framework presented here provide a route to mechanistically link intracellular physical state, biosynthetic capacity, and growth dynamics under non-balanced conditions.}

A second important point is the assumption that amino acids can be considered the most abundant osmolyte, thus setting cell volume. 
A recent study by Mukherjee and coworkers~\cite{mukherjee_homeostasis_2024} provides evidence that puts this point into question, leading to the conclusion that the main contributors to active osmolytes in \textit{E.~coli} are not only free amino acids and their associated counterions, but also the (mainly potassium) counterions of RNA (which is mainly ribosomal RNA) and proteins. 
Their findings show a correlation between an increase in counterions and an increase in total RNA, alongside shifts in dry mass density when altering the charge of overexpressed proteins, suggesting that protein charge plays a role in protein density. 
By a two time-scale model where the turgor pressure inside the cell driven by the increasing osmolytes pushes on the cell wall, triggering surface growth, they propose a mechanism of biomass density homeostasis via ribosomal counterions, where perturbations in pressure (mediated mainly by ribosomal RNA counterions) coordinate cell wall expansion in a way that is linked to ribosome content scaling with growth rate (the first growth law). Such an approach also adresses how cell wall expansion can be limiting in cell volume growth.

Thirdly, the (dilute-solution) Van't Hoff picture for intracellular osmotic pressure used in our recipe may likely be too naive. Indeed, the cell cytoplasm is a complex medium where crowding, disorder, viscoelasticity and polyelectrolyte-gel properties can coexist, comprising complex organelles with the same properties. On one hand, the translational degrees of freedom of abundant small particles clearly contribute to its entropy~\cite{Buda2016, venkova_mechano-osmotic_2022}, hence to osmotic presure equilibrium. On the other hand, we lack simple effective models summarizing relevant contributions to intracellular entropy from more complex processes. To give a concrete example, the physical interpretation of the osmotically inactive volume $V_\text{ex}$ from Ponder plots made with cells may pose some caveats (see our considerations in the previous section). 
A deeper physical understading of this entropy, hence of osmotic equilibrium, may also be important to produce growth- and size-homeostasis models of water-containing organelles such as the eukaryotic nucleus~\cite{Deviri2022, rollin_physical_2023, Pennacchio2024, Biswas2023}. 

More widely, the importance of intracellular density is underscored by its impact on various biological processes~\cite{vandenBerg2017, Delarue2018, dai_slowdown_2018, Pang2023}. 
Increases of cellular concentration are generally proposed to lead to a reduction to the tRNA diffusion times as diffusion slows down according to multiple papers measuring the translation elongation~\cite{klumpp_molecular_2013, dai_slowdown_2018}. However, direct measurements of the diffusion coefficients of particles of the size of tRNAs do not show significant differences~\cite{alric_macromolecular_2022}.
In any case, decreased translation leads to a reduction in protein synthesis and often to a regulated cellular response, for example by an increased ribosome pool as measured in bacteria~\cite{dai_slowdown_2018}, or an activation of the Environmental Stress Response pathway as observed in yeast~\cite{terhorst_environmental_2023}.
Hence, gene-regulatory response by the cell likely plays a major role in preserving cell density.  

Finally, Pang and Lercher~\cite{Pang2023} recently investigated the optimal cytosolic volume occupancy of differently-sized players in a biosynthetic model, including the effects of diffusion and perturbed chemical equilibria, revealing a tradeoff between the allocation to large ribosomal complexes, which enhances metabolic enzyme saturation, and smaller metabolic macromolecules, which facilitate unhindered tRNA diffusion, and showing that this optimality principle predicts the slight variation in cytosolic density across different growth conditions in bacterial cells observed by a recent study~\cite{oldewurtel_robust_2021}.
This idea is in general agreement with the earlier hypothesis of Dill and coworkers~\cite{dill_physical_2011}, according to which the generally preserved density is a result of the trade-off between collision rates (that increase with density) and diffusion rates (that decrease with density).

\section{Dessert: Crosstalk between growth and cell-cycle progression}\label{sec9:cell_cycle}
This section introduces the dimension of cell-cycle progression to our discussion of cell growth physiology. Our goal is to present a bare-bone quantitative framework for describing the complex interplay between cell growth and cell-cycle progression - a relationship that remains largely unexplored.

\subsection*{MAIN QUESTIONS AND MOTIVATIONS}
Cells across all domains of life have evolved specific mechanisms to tightly coordinate cell-cycle progression with growth and biosynthesis~\cite{Cadart2019, Kaldis2016, bruggeman_searching_2020, Amir2019, Bertaux2018, Skotheim2013, Fishov1995}. 
This coordination is crucial for maintaining cellular homeostasis, and disruptions in this coordination are often associated to various diseases and pathological conditions~\cite{Ginzberg2015, Laplante2012, neurohr_excessive_2019, Neurohr2020}. Furthermore, the ability of cells to adapt to environmental stresses, such as fluctuations in nutrient availability or exposure to growth-inhibitory stimuli, depends on their capacity to reprogram growth and division processes in a coordinated manner~\cite{Skotheim2013,Ovrebo2022}. 
Despite the significance of this coordination, our quantitative understanding of how it operates in different biological contexts remains incomplete. Fundamental questions arise, such as ``In what ways do alterations in cell growth and protein synthesis influence cell cycle progression?'' and ``How does interfering with the cell cycle affect cellular growth?''. 
These questions have yet to be fully explored from a quantitative standpoint.

Below we extract the main ingredients from recent quantitative attempts to fill this gap in the context of \textit{E.~coli} cell growth and division~\cite{Serbanescu2020new, Serbanescu2021, Bertaux2020, Pandey2020, ElGamel2024}. 
These studies integrate established models of cellular growth and cell division control — each extensively studied independently~\cite{jun_fundamental_2018} (see~\cite{Xie2022} for a recent review of these themes in eukaryotes) — into a ``Unified-Growth-Division'' (UGD) recipe whose main ingredients we are now going to discuss. 

\subsection*{INGREDIENTS AND RECIPES\edit{\footnote{The recipe introduced here builds upon ideas and models presented in Refs.~\cite{ElGamel2024,Serbanescu2020new,Serbanescu2021}.}}}

We now introduce a UGD recipe inspired by sigle-cell measurements of \textit{E.~coli} growth and division. 
The growth laws discussed in the preceding sections capture population-averaged behaviour, but division is intrinsically a single-cell event. 
Therefore, here we analyze growth, division and proteome partitioning of an individual cell as it progresses through its $j$-th cell cycle. Throughout the following paragraph, all quantities refer to the cell cycle $j$ of a single cell within a lineage, but for simplicity of notation, we omit the cycle index. 

\begin{figure}[!bt]
    \centering
    \includegraphics[width=1\columnwidth]{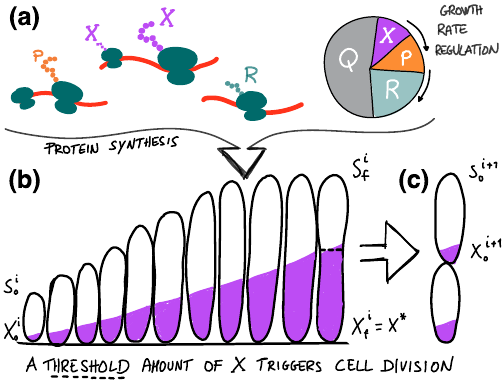} 
    \caption{\textbf{Key ingredients of the unified-growth-division framework.}
    (\textbf{a}) A division sector, $X$, representing proteins central to cell division control, is integrated into the standard proteome allocation framework. The synthesis rate of the divisor protein mass, regulated at the gene expression level, is coupled to the cellular growth rate, consistent with proteome allocation theory. 
    (\textbf{b}) Cell-cycle-level description of the dynamics of cell size, $s$, and divisor protein copy number, $X$. Cell size increases exponentially throughout the cell cycle, while divisor proteins accumulate due to a combination of size-specific and constant synthesis rates, as well as a degradation rate, Eq.~(\ref{eq:macro}). 
    (\textbf{c}) A threshold-accumulation rule completes the model: cell division is triggered once a threshold amount of divisor protein is synthesized. After division, the divisor protein amount can be reset to zero or to a defined fraction (e.g., halved to follow the symmetry of the division process). 
    }
    \label{fig8}
\end{figure}

The first ingredient in constructing a UGD model is the integration of a dedicated cell division sector, see Fig.~\ref{fig8}(a), denoted by $X$, into a single-cell revisitation of the proteome allocation theory discussed in Secs.~\ref{sec2:classical} and \ref{sec5:degradation}. This division sector contains proteins required for executing and completing the cell division process, e.g., those involved in septum formation and constriction for bacteria.
The dynamics of protein expression for this sector, along with those of other proteome sectors ($i \in \lbrace X, Q, P, R \rbrace$) can be described by the following equation:
\begin{equation}
      \frac{dM_i}{dt} =
        J_\text{syn} \chi_i -
        \eta_i M_i \ .
\label{eq:micro}
\end{equation}
Eq.~\eqref{eq:micro} describes how the mass of each proteome sector changes according to its sector-specific protein-synthesis flux, see Eq.~(\ref{eq:chi_operative_def}), and the corresponding degradation rate, cf. Eq.~(\ref{eq:deg:protein mass}).

The second key ingredient of the model is a cell-cycle-level description of cell growth and division, Fig.~\ref{fig8}(b). For simplicity, we assume that cell division is tied to the production of a single protein component, $X$, which is considered the bottleneck for division \cite{Harris2016}. While this assumption simplifies the analysis, more detailed and specific mechanistic models can certainly be developed. Hence, our model is based on two central variables: the size of the single cell $s$ —which can represent either total cell mass $M$ or volume $V$— and the division protein copy number $X$. Following ElGamel and Mugler~\cite{ElGamel2024},  the time evolution of $s$ and $X$ can be described by the following set of equations:
\begin{equation}
\begin{aligned}
  \frac{ds}{dt} &= \lambda s \ , \\
  \frac{dX}{dt} &= k + k_X s - \eta_X X \ .
\end{aligned}
\label{eq:macro}
\end{equation}
Here, the cell size $s$ is assumed to grow exponentially at rate $\lambda$. The copy number of division proteins $X$ evolves through three processes: (i) basal synthesis at a constant rate $k$; (ii) size-dependent synthesis, proportional to cell size, at rate $k_X s$; and (iii) degradation at rate $\eta_X X$. 

Bacteria growing exponentially can maintain size homeostasis by tightly regulating their division timing~\cite{Amir2014, Osella2014}. Therefore, to complete the model, a division rule is required to define the condition under which a cell decides to divide and initiates a new cycle. This leads to the third key component of the UGD framework: the implementation of cell-division control strategies~\cite{Amir2014, Grilli2018, Micali2018, Micali2018b, Zheng2020,TaheriAraghi2015}.  

The molecular mechanisms linking cell size and division in \textit{E.~coli} remain only partially understood. In this recipe, we model this coupling by specifying a division rule based on the accumulation dynamics of the division protein $X$. \edit{A widely used~\cite{Fantes1975, Mahaffy1989, Donachie03, Basan2015, Harris2016, Serbanescu2020new, Bertaux2020, Pandey2020, Si2019, Serbanescu2021, Panlilio2021} 
example is a threshold-accumulation mechanism:} the cell synthesizes the division proteins during the cycle until their copy number reaches a threshold $X^*$, which then triggers division, see Fig.~\ref{fig8}(c). Upon division, the cell size is halved while the division protein is reset to zero: 
 \begin{equation}
 X(\tau) = X^* \implies 
 \begin{cases}
    s(\tau) \to s(\tau)/2 \\  
    X(\tau) \to 0
 \end{cases} \ .
 \label{eq:divisionstrategy}
\end{equation}
Note that while some studies proposed a reset-to-zero model~\cite{Harris2016,Ojkic2019,Panlilio2021}, similar results can be achieved with a division protein that reach a constant threshold (without fluctuations) and perfectly divide in half: $X^* \implies X(\tau) \to X^*/2$~\cite{Si2019,ElGamel2024}. \edit{While the threshold accumulation control is supported in some contexts~\cite{Si2019,Panlilio2021}, other studies report departures from a single-threshold picture, indicating additional layers of regulation \cite{Micali2018b,Mannik24,Witz19}. Finally, threshold formulations have been generalized to cells growing non-exponentially, where adder behavior arises when the threshold accumulator is produced at a rate proportional to the cell’s growth~\cite{Nieto2025}.}

This coarse-grained model links proteome allocation and cell growth to division timing, forming the foundation for more detailed mechanistic modeling. Specifically, the set of equations describing proteome sector dynamics Eq.~(\ref{eq:micro}) and those describing cell-cycle dynamics Eq.~(\ref{eq:macro}) are connected through the proportionality between cell size and total mass: \begin{equation}
  s \propto M = M_P + M_R + M_Q + M_X \ .
  \label{eq:micromacro}
\end{equation}
If size is interpreted as volume, this relationship implies constant cell density. Note that while this assumption holds at the population-averaged level, it may not be valid for individual cells or under certain perturbations~\cite{oldewurtel_robust_2021, bremer2008modulation} (see Sec.~\ref{sec8:density}).

The three sets of equations—Eqs.~\eqref{eq:micro}, \eqref{eq:macro}, and \eqref{eq:divisionstrategy}—together with the constraint in Eq.~\eqref{eq:micromacro}, constitute a minimal yet sufficient framework for addressing cell growth and division control in a unified manner. 
Importantly, we are assuming Eq.~\eqref{eq:micro}, which has been largely discussed for population averages, holds for single cells. However, both the precise forms of the synthesis fluxes and the applicability of the first growth law at the single-cell level remain unresolved. In other words, a detailed understanding of the emergence of population growth laws from single cells is an open problem~\cite{susman2018,Pavlou2025,Kiviet2014,Panlilio2021}. \\

\noindent\textbf{Phenomenological classification of cell-size control strategies}.
Here, we guide the reader into the interpretation of the empirical fluctuations observed in size and growth into a cell size-strategy. In exponentially growing cells, size homeostasis reflects a tight coordination between cell growth and cell-cycle progression~\cite{Osella2014}. At the single-cell level, one can view the final division size $s_f^j$ in cycle $j$ as a function of the size at birth $s_0^j$, $s_f^j = s_f^j(s_0^j, \text{noise terms})$. 
Because the empirical size fluctuations are small, the control function can be approximated by a linear expansion around the mean birth size~\cite{Amir2014, Grilli2017}, i.e. in powers of $\delta s_0^j \equiv s_0^j -\langle s_0 \rangle$
\begin{equation}
    s_f^j \approx  \langle s_f \rangle + \dfrac{d s_f^j}{d s_0^j}|_{\langle s_0 \rangle} \delta s_0^j + \text{noise terms}.
    \label{eq:linearexp}
\end{equation}

Introducing a `size-control parameter' $\zeta_G = 1-s_f^j(\langle s_0\rangle)$ then yields
\begin{equation}
    \delta s_f^j = (1-\zeta_G) \delta s_0^j + \text{noise terms}.
    \label{eq:zetaG}
\end{equation}

Fundamentally, the entire complexity of the control function, recapitulated by the linear-order size-control parameter, $\zeta_G$, becomes this way directly accessible experimentally by computing the correlation (or equivalently the slope of a binned average) between $s_f^j$ and $s_0^j$ observed in single-cell data~\cite{Grilli2017,Grilli2018}, 
\begin{equation}
    1-\zeta_G = \dfrac{\langle \delta s_f^j \delta s_0^j \rangle}{\langle \delta s_0^j \delta s_0^j \rangle} \ .
    \label{eq:zetaG2}
\end{equation}

Historically, three typical strategies are identified by the value of $\zeta_G$: (\emph{i}) \emph{Timer} ($\zeta_G=-1$)  in which division follows a constant duration with no direct size control; (\emph{ii}) \emph{Sizer} ($\zeta_G=1$), in which cells divide upon reaching a critical size; and (\emph{iii}) \emph{Adder} ($\zeta_G=0$), in which cells add a constant (birth-size independent) size each cycle. \\

\noindent\textbf{Connecting control strategies with divisor protein dynamics.} The above classification is phenomenological, based on the observed correlation between birth and division sizes. Following Refs.~\cite{Serbanescu2020new,ElGamel2024}, we now show how these strategies naturally arise from the accumulation dynamics of the divisor protein $X$, Eq.~(\ref{eq:macro}), and the threshold-accumulation division rule, Eq.~(\ref{eq:divisionstrategy}). \\

\noindent
\emph{Timer (no control)}.
If division proteins are only produced at a constant size-independent rate $(k_X=0)$, and degradation is negligible $(\eta_X=0)$, then the solution of Eqs.~\eqref{eq:macro} reads $X(t) = x_0 + k t $, which combined with the threshold accumulation division rule and the reset-to-zero scheme yields a size-independent duration of the cell cycle
\begin{equation}
    \tau  =  \dfrac{X^*}{k},  
\end{equation} 
where $X^*$ is the threshold level of $X$. This corresponds to the “timer” regime, with division occurring after a constant time from birth. \\

\noindent
\emph{Sizer and Adder behaviors.} If, instead, $X$ is synthesized only in a size-dependent manner $(k=0)$, then applying a threshold condition $X(\tau)=X^*$ leads to  
\begin{equation}
  X^* = \dfrac{k_X}{\lambda + \eta_X} \left( s_f^j - s_0^j 2^{-\frac{\eta_X}{\lambda}}\right) \label{eq:threshold}.
\end{equation} 
Depending on the strength of the degradation term, we can interpolate between sizer and adder behaviors~\cite{ElGamel2024}. Indeed, if the degradation term can be neglected, $\lambda \gg \eta_X$ (e.g. fast growth conditions), then 
\begin{equation}
    s_f = s_0 + \dfrac{\lambda}{k_X} X^* = s_0 + \Delta,
    \label{eq:adder}
\end{equation} 
which add a constant size $\Delta$ each cycle (adder behavior). 
If instead degradation is much faster than cell growth, $\lambda \ll \eta_X$,  then 
\begin{equation}
    s_f =  \dfrac{\eta_X}{k_X} X^* = \text{const}.
\end{equation} 
which indicate a constant threshold for the final size (sizer behavior).

Empirically, \emph{E.~coli} cells are known to regulate their size mostly by adding a constant volume between consecutive cell divisions~\cite{TaheriAraghi2015}, thus implementing the adder strategy. We saw how this control strategy can naturally emerge within the framework from the molecular dynamics of the division protein when degradation is negligible. 
Notably, under slow growth conditions where degradation is no longer negligible relative to the growth rate, the system is predicted to deviate from the adder regime. In the extreme case $\lambda \ll \eta_X$, the implemented control strategy transitions to a sizer mechanism, \edit{consistent with experimental observations of deviations from adder at slow growth \cite{Wallden2016, Nieto20}. 
In addition to degradation effects, several other mechanisms have been proposed to explain why \textit{E. coli} adopts a more sizer-like division strategy at slow growth, such as a non-linear dependence of FtsZ production rate on cell size and the requirement for a minimal ‘commitment’ size before initiating division \cite{Nieto24}. 
Moreover, slow growth reduces the coupling between replication initiation and division so that division is no longer tightly constrained by the replication cycle \cite{Colin2021}. 
Altogether, these factors naturally weaken the effective adder behaviour at slow growth, in agreement with recent analyses of single-cell data \cite{Si2019,Nieto24}.}\\

\noindent\textbf{Growth laws and trade-offs between protein sectors.} 

So far, we have shown that a simple UGD model captures the single-cell empirical division strategies that maintain size homeostasis. Moreover, by assuming that the classical growth laws hold at the single-cell level, we establish a direct correspondence between population-averaged growth laws and the intracellular allocation of resources within individual cells. In the sections that follow, we identify the relationships tying sector X to the other biosynthetic sectors and derive the corresponding growth laws for this new division sector. 

\noindent
\emph{Trade-Offs between Ribosomes and Division Protein Synthesis}. We now apply the UGD model to explicitly highlight the trade-off between the synthesis of ribosomes and division proteins. In doing so, we follow the approach developed in Refs.~\cite{Serbanescu2020new, Serbanescu2021}. 
Similarly to the discussion in Sec.~\ref{sec2:classical}, one can balance flux of amino acids production and consumption and write $\nu \phi_P = \epsilon_R f_a \phi_R 
$, see Eqs.~(\ref{eq:maaloe}-\ref{eq:phiQ}). Note that this equation assumes no degradation, and that the total free amino acid mass is a negligible fraction of the total mass of the cell. Therefore, using the constraint that the protein sectors sum to 1, one gets
\begin{equation} 
\phi_X = \phi^{\text{max}}_R - \phi_R -\frac{\epsilon_R}{\nu} f_a \phi_R 
  \label{eq:serbanescu7}
\end{equation}
Eq.~(\ref{eq:serbanescu7}) describes a negative correlation between expressing the ribosomal and division sectors in response to nutrient or translational perturbations, consistent with recent experimental findings~\cite{Mori2021}. 
Since the synthesis rates of growth and division proteins are proportional to the ribosome and division sectors, respectively, this negative correlation highlights a trade-off in allocating ribosomal resources towards growth or division processes (see Fig.~1(f) in Ref.~\cite{Serbanescu2020new}). 

\noindent
\emph{Growth law for the division sector}. As we just discussed, our simple UGD model predicts that the larger the fraction of ribosomes making division proteins the smaller the fraction of ribosomes making ribosomes. In other words, there should be a negative correlation between the growth rate and the division protein sector, something we would call "the division sector growth law". To show that, we note that the ribosome sector is related to the growth rate via the first growth law $\lambda = \epsilon_R f_a \phi_R$. 
Eq.~(\ref{eq:serbanescu7}) allows to related the growth rate $\lambda$ to $\phi_X$,
\begin{equation}
  \lambda = \frac{\nu \varepsilon_R f_a}{\nu+f_a \varepsilon_R} \left( \phi_R^{max} - \phi_X \right). 
  \label{DIVGL}
\end{equation}
This result makes explicit how allocating more proteome fraction $\phi_X$ to division proteins lowers the growth rate $\lambda$, thus establishing the prediction of a ``division sector growth law'' \cite{Serbanescu2020new}. \\ 
\noindent
\emph{``Schaechter-Maaloe-Kjeldgaard'' (SMK) growth law}. 
We end our survey by showing how this UGD model can deal with the foundational cell size growth law discovered by Schaechter, Maaløe, and Kjeldgaard~\cite{Schaechter1958}. \edit{This classic work demonstrated that fast-growing cells had higher levels of key macromolecules like RNA
and proteins compared to slower-growing cells. }As a consequence, larger cells are typically associated with faster growth rates. Specifically, we are interested in the observation that the population-averaged cellular size increases monotonically with growth rate, in an approximately exponential fashion, although deviations from the exponential trend have recently been reported, particularly at slow growth~\cite{Zheng2020}. 

We note that the standard explanation for the SMK growth law is based on the chromosome cell cycle, and in particular with the replication-initiation event~\cite{Donachie1968,cooper_origins_1993,Zheng2020}. However, for the sake of modeling, it is interesting to note that the UGD model (which does not include the chromosome cycle) can reproduce an increase of the average cell size as a function of the growth rate as follows. We first note that in an exponentially expanding population the average cell size can be expressed as $\langle s \rangle = 2 \log 2 \langle s_0 \rangle$~\cite{jun_fundamental_2018}. Combining this observation with Eq.~(\ref{eq:threshold}), recalling that 
$\langle s_f \rangle = 2 \langle s_0 \rangle$ we can write 
\begin{equation} \langle s \rangle = \frac{\lambda + \eta_X}{\tilde{k}_X
    \left(2-2^{-\frac{\eta_X}{\lambda}}\right)} \ . \label{eq:SKM}
\end{equation} 
Following Ref.~\cite{Serbanescu2020new}, we have defined $\tilde{k}_X \equiv k_X/(\log 2 X^*)$, and we have assumed again that we can interpret the single-cell quantities as population averages, but we note once more that this assumption is untested. Note that since $\lambda \propto \phi_R$ and $k_X \propto - \phi_R$ the average cell size increases with ribosome abundance, a trend observed in experiments. Notably, upon determining the model parameters and making explicit the growth rate dependence of Eq.~\ref{eq:SKM} Serbanescu and coworkers~\cite{Serbanescu2020new} derived the expression 
\begin{equation}
    \langle s \rangle \approx \dfrac{\lambda}{1.26 - 0.38 \lambda}\ ,
\end{equation}
which reproduces the monotonic increase of average cell size with the nutrient-imposed growth rate and describes the experimental data with no further fitting, with a trend that deviates from the exponential assumed previously~\cite{cooper_origins_1993,Zheng2020,Serbanescu2020new}. While this approach is interesting, the biological consensus is still that the chromosome plays some role in setting cell size. A similar approach attempting to integrate the chromosome and cell-cycle views was proposed in Ref.~\cite{Zheng2020}. This model is essentially a zero-reset accumulator whose production rate is aware of the average replication-segregation timing of the chromosome.    

\subsection*{OUTLOOK}
The unification of growth and division (UGD) processes within a single mathematical framework is still in its early stages. Building on recent bacterial studies~\cite{Serbanescu2020new, Serbanescu2021, Bertaux2020, Pandey2020}, and in particular the approach of Ref.~\cite{Serbanescu2020new}, we showed how a unified model can account for multiple experimental observations related to cell growth and division control (see also Chapter 14 in Ref.~\cite{EPCP2025} for further details).
Crucially, a comprehensive UGD framework must connect the dynamics of the single-cell proteome, including a dedicated sector for division proteins and mechanisms of cell-cycle progression control, to population-level growth laws. However, the theoretical bridge between these microscopic and macroscopic scales remains to be constructed. Ultimately, such a framework should predict functional relationships for the divisome sector, elucidating how the allocation and dynamics of division proteins determine key division metrics.
The simple recipe presented here only refers to bacteria and simplifies the underlying cell-cycle decision processes by omitting other important cell-cycle regulatory mechanisms, such as initiation of DNA replication. In the following we briefly outline promising directions for future work.

\noindent\emph{Expanding beyond chromosome-agnostic cell-cycle decisions.}
The decision to divide has been modeled using a division sector of the proteome which is expressed constitutively accross the cell cycle and accumulate up to a thereshold before division~\cite{Panlilio2021}.  
Although this has been effective, the underlying decision-making process in bacterial cell division is far more complex. In particular, we already mentioned that the model we discussed as a recipe does not capture the interplay between cell division and chromosome replication~\cite{Zheng2020}\edit{, and perturbation studies show that shifts in replication timing and cell size challenge purely division-sector explanations~\cite{Zheng2016, Si2017, Colin2021}. Indeed multiple} both processes are known to contribute to the decision to divide~\cite{Micali2018, Colin2021, TIRUVADIKRISHNAN2022}. Incorporating these details could offer a more comprehensive picture of the division process. 

\noindent\emph{Going beyond steady-state conditions.}
Investigating non-steady-state conditions is crucial for distinguishing between competing mechanisms of cell-cycle regulation~\cite{Harris2016, bakshi2021tracking, Si2019, Wallden2016}. In a recent study~\cite{Panlilio2021}, long-term microfluidic experiments monitored cell growth and cell-cycle progression during nutrient upshifts. Fluorescent markers were used to track the dynamics of both the $R$ and $P$ sectors throughout these shifts. Notably, highly complex, multi-timescale dynamics were observed in division-related variables—such as inter-division time, division rate, added volume, and the ratio of added-to-initial volume—while the overall division control strategy remained remarkably robust during the nutrient shift.

By comparing theoretical predictions with experimental data—using growth measurements and fluorescent proxies for the relevant sector—the work suggests two key points: (i) only a threshold-accumulation model provides a satisfactory minimal model to fit the data, and (ii) a putative divisor protein must be associated with the $P$-sector to reproduce the observed trends, indicating that it is constitutively expressed.
Interestingly, this finding aligns with the steady-state regulation of the $X$ sector predicted by the framework of Serbanescu and coworkers~\cite{Serbanescu2020new} that we used to build our recipe, further supporting the role of the $P$ sector in division control. 

Promising directions for future studies rely on investigating the dynamic interplay between the divisor protein and the proteome performing chromosome maintenance, replication and segregation, which could yield deeper mechanistic insights into how cells coordinate replication and division at the molecular level. Additionally, considering sublethal antibiotic conditions may reveal how bacterial populations adapt their growth and division strategies to enhance survival and connect with the second growth law. 

\noindent\emph{Scaling up to more complex organisms.} 
A key direction for future research is determining how this framework can be adapted to more complex organisms, such as yeast and mammalian cells. Whether a unified framework can be developed and effectively applied to these systems remains an open question~\cite{Ovrebo2022}. Extending the current framework would likely require incorporating control mechanisms for additional cell-cycle decisions beyond the division decision discussed in this section. For instance, in both budding yeast \textit{S.~cerevisiae} and higher eukaryotes, it is observed that much of the cell-size control takes place at the G1-S transition, where DNA replication is initiated, rather than at cell division~\cite{Skotheim2013,Soifer2016,Zatulovskiy2020,Xie2020}. Moreover, there is increasing evidence that differential proteome scaling plays a crucial role in driving the cell-cycle progression of more complex organism, as well as in the emergence of cell-cycle-related phenomena such as senescence~\cite{swaffer_RNA_2023,Jones2023,Xie2022,Chen2020}. A modeling framework able to deal with differential protein  scaling and with the onset of cell-cycle progression stages could help us clarify different phenotypes of lower and higher eukaryotic cells.

\section{Digestif: Ecological implications of proteome allocation and growth laws} \label{sec10:communities}

The previous sections introduced mathematical models for cellular growth physiology, focusing on how cells allocate their proteome across essential processes and how that allocation shapes overall cellular functions. These models treat cells as isolated entities that acquire resources from an environment presumed to remain unchanged. In reality, microorganisms actively transform their surroundings by consuming nutrients, releasing metabolic by-products, and adjusting their internal proteome accordingly~\cite{Huelsmann2024, Narla2025, dal_bello_resourcediversity_2021}. Far from acting as passive consumers, they engage in complex interactions with other cells, compete for shared resources, form cooperative networks, and evolve context-dependent notions of nutrient quality~\cite{Abreu2023}.

\edit{As a brief conceptual outlook (rather than a comprehensive review),} this section integrates an ecological dimension into our proteome allocation framework by presenting three concise recipes that serve as starting points for modeling microbial communities. Each recipe illustrates a different mode of interaction: competition for shared substrates, cross feeding via secreted metabolites, and the ecological interpretation of nutrient quality, which links proteome investment during current growth to anticipatory adjustments that enhance fitness in future environments. 
\edit{Our goal here is to sketch how the growth-law and proteome-allocation perspective may inform ecological thinking, as an outlook that identifies opportunities and open questions.
This section intentionally departs from the previous sections - both in topic (from physiology to ecology) and in style (from worked to a mixture of worked and exploratory `recipes') - to illustrate how growth-law principles may extend beyond physiology. We leave detailed community modeling to future work and view this area as a promising direction for near-term advances. 
}

\subsection*{MAIN QUESTIONS AND MOTIVATIONS}
Natural bacterial communities consist of many species that interact through both antagonistic and facilitative mechanisms ~\cite{faust2012microbial,smith_classification_2019,Kost2023,vandenBerg2022}.
Antagonistic interactions --such as competition for nutrients or direct killing-- allow one species to increase in relative abundance at the expense of others. Facilitative interactions occur when one population benefits another, for example by releasing utilizable metabolites, detoxifying the environment, or protecting partners from antimicrobial agents or predators. Because microbes continually sense and modify their surroundings, their ecological decisions must also anticipate future environmental conditions. Here, we outline three modelling approaches for interacting microbial systems and provide practical guidelines for embedding proteome-allocation theory into ecological frameworks that aim to predict species coexistence, resource flows, and growth trade-offs in microbial communities.

First, we focus on nutrient competition. To elucidate how resource usage shapes community dynamics, species distributions and extinction risk, ecologists have developed consumer-resource models that track the coupled dynamics of bacteria (consumers) and the nutrients (resources) they exploit~\cite{macarthur_species_1970, chesson_macarthurs_1990}. In these models, typically bacteria compete for externally supplied substrates, and researchers use them to ask how the number and composition of carbon sources constrain or expand community richness~\cite{dal_bello_resourcediversity_2021}.

The majority of these models do not take into account the physiological details of consumers in different growth conditions. In contrast, the cell-physiology models introduced in the previous sections treat isogenic populations and ask how they partition their internal resources under fixed environments. Integrating these approaches could significantly enhance our understanding of community functioning and improve our predictive power regarding ecosystem responses~\cite{pacciani-mori_constrained_2021}.

Second, we focus on a fundamentally different ecological interaction which involves metabolic \textit{cross-feeding} --that is, the exchange of nutrients among species via metabolite secretion and uptake. 
Indeed, bacteria not only deplete nutrients but also modify their environment by leaking metabolites, thereby enriching their surroundings~\cite{dsouza_ecology_2018, dsouza_experimental_2016, ona_obligate_2021,huelsmann2024framework}. Some bacteria rely on these secreted compounds for growth because they lack the biosynthetic capability to produce one or more essential organic molecules. These organisms, known as auxotrophs, are contrasted with prototrophs, which synthesize all the compounds they require. Amino acid auxotrophies are prevalent in bacterial communities and often play a critical role in community stability~\cite{dsouza_less_2014,starke2023amino}.
Given the ubiquity of amino-acid auxotrophies, it is important to understand how leakage both influences and is influenced by proteome regulation. Building on this insight, we argue that incorporating leakage fluxes into physiological models may provide a more robust framework for predicting nutrient sharing in microbial communities.

Third, another key aspect, relevant to both physiology and ecology, is the concept of ``nutrient quality''. It is well established that bacteria exhibit different growth rates in different media --a foundation of the growth laws~\cite{scott_interdependence_2010}. Traditionally, these discrepancies were attributed to the metabolic cost of processing specific nutrients. However, recent studies suggest that nutrient quality is influenced not only by the proteome cost of the biosynthetic pathways required to metabolise a substrate but also by the ecological adaptability, wherein cells strategically invest in functions that contribute to growth and survival in future environments~\cite{Mukherjee2024_plasticity}.
Understanding these adaptive proteome allocation decisions in an ecological context opens a new window into how cells prepare for, respond to, and shape future environmental conditions, ultimately influencing patterns of coexistence, resource partitioning, and community stability.

\subsection*{INGREDIENTS AND RECIPES{\footnote{The recipe introduced here builds upon ideas and models presented in Refs.~\cite{pacciani-mori_constrained_2021, Mukherjee2024_plasticity}}}}
\noindent\textbf{Proteome allocation in consumer-resource frameworks}.
We begin by presenting a model that couples bacterial growth laws and proteome allocation with a consumer-resource model for a bacterial community competing for share nutrients. We do not aim to present all the specific results obtained in the literature --which is still in its early stages--, but instead we outline a set of minimal ingredients that are necessary to assemble a coarse-grained general theory for a consumer-\textit{proteome}-resource model.  In doing so, our starting point and main inspiration is the work of Pacciani-Mori and colleagues~\cite{pacciani-mori_constrained_2021}. 

MacArthur’s consumer-resource model describes how multiple species compete for shared resources by explicitly considering both population sizes and resource concentrations~\cite{chesson_macarthurs_1990}. We consider a system composed of $n_\text{S}$ species and $n_\text{r}$ resources, where the population size of species $\text{i}$ is denoted by $S^{(\text{i})}$ and the concentration of resource $\text{j}$ by $r_\text{j}$. In general, we use the superscript $^{(\text{i})}$ to indicate species-specific variables and the subscript $_\text{j}$ for the resources.  The dynamical equations for the system are given by
\begin{align}
	\frac{\text{d}S^{(\text{i})}}{\text{dt}} &= S^{(\text{i})} \left( \lambda^{(\text{i})}(\{r_\text{j}\})- d^{(\text{i})} \right),  \label{eq:species_growth} \\
	\frac{\text{d}r_\text{j}}{\text{dt}} &= k_\text{j} - \sum_{\text{i}=1}^{n_\text{S}} u^{(\text{i})}_\text{j} S^{(\text{i})}.
    \label{eq:resource_dynamics}
\end{align}
Here, $\lambda^{(\text{i})}(\{r_\text{j}\})$ is the per-capita growth rate of species $\text{i}$, 
which depends on the concentrations of all available resources $\{r_\text{j}\}$ and 
reflects interspecies competition. The parameter $d^{(\text{i})}$ includes the maintenance cost and death rate of species $i$; $u^{(\text{i})}_\text{j}$ is the per-capita uptake rate of 
resource $\text{j}$ by species $\text{i}$ --which may not directly translate into biomass increase due to maintenance; and $k_\text{j}$ is the constant supply rate of resource $\text{j}$~\cite{chesson_macarthurs_1990}. 

Assuming that proteome allocation adjusts on a timescale much faster than that 
of resource and species dynamics, we can consider a quasi-steady state for proteome 
fractions. Under this assumption, the first growth law applies to 
each species:
\begin{equation}
    \label{eq:X_1st_community}
    \phi^{(\text{i})}_\text{R} = \frac{\lambda^{(\text{i})}}{\gamma_\text{0}^{(\text{i})}} + \phi_\text{R}^{(\text{i}), \text{min}} \,,
\end{equation}
where the parameters~$\phi_\text{R}^{(\text{i}), \text{min}}$ and~$\gamma_\text{0}^{(\text{i})}$ can be species-dependent~\cite{karpinets_rna_2006, scott_interdependence_2010}.
If different species consumes a distinct subset of resources, then there is no conceptual difference from the models introduced in previous sections. However, when two or more species compete for the same resources, Eqs.~(\ref{eq:species_growth})-(\ref{eq:resource_dynamics}) introduce 
nontrivial coupling between the growth laws of different species via shared resource dynamics.

We now outline how constraints from proteome allocation and assumptions on multiple resource-utilization pathways could enhance the predictive power of ecological consumer-resource models.
To account for the uptake of multiple resources, the P sector $\phi^{(\text{i})}_\text{P}$, including nutrient uptake proteins, is further subdivided, similarly to what it is done in Sec.~\ref{sec7:shifts}.
The growth rate $\lambda^{(\text{i})}$ can then be written as a linear combination of substrate-specific resource uptake sectors $\phi^{(\text{i})}_{\text{P}_\text{j}}$ as
\begin{equation}
\lambda^{(\text{i})} = \sum_\text{j} \nu^{(\text{i})}_\text{\text{j}}(\{r_\text{j}\}) \phi^{(\text{i})}_{\text{P}_\text{j}}  
\label{eq:lambda_i}
\end{equation}
where $\nu^{(\text{i})}_\text{j}$ is the nutrient quality for resource $\text{j}$ of species $\text{i}$, that in principle depends on the whole set of resources $\{r_\text{j}\}$, and $\sum_\text{j} \phi_\text{P$_\text{j}$} = \phi_\text{P}$.
Equation~(\ref{eq:lambda_i}) has been obtained in the assumption that resources do not significantly inhibit or enhance one another, and the individual contributions to growth of each resource add up in proportion to the fraction of the proteome allocated to each resource. Observe that the co-utilization of substrates, although common, it is not always the case~\cite{Hermsen2015, okano_hierarchical_2021} --a notable substrate that is preferentially consumed is glucose.

This formulation highlights how the growth laws~(\ref{eq:X_1st_community}) for the different species, whose abundances evolve according to Eq.~(\ref{eq:species_growth}),  are intrinsically coupled through the shared dependence on the resource pool, whose dynamics is governed by Eq.~(\ref{eq:resource_dynamics}).

Following classical proteome allocation models (see Sec.~\ref{sec2:classical}), each species $\text{i}$ allocates its proteome among three main sectors --the P sector $\phi^{(\text{i})}_P$, the R sector $\phi^{(\text{i})}_R$, and the Q sector $\phi^{(\text{i})}_Q$-- such that $\phi^{(\text{i})}_P + \phi^{(\text{i})}_R + \phi^{(\text{i})}_Q = 1$. 
Combining this constraint with  Eqs.~(\ref{eq:X_1st_community})-(\ref{eq:lambda_i}) leads to the following relation:
\begin{equation}
        \sum_{\text{j}}\phi^{(\text{i})}_{\text{P}_\text{j}}\left( 1+\frac{\nu^{(\text{i})}_\text{j}}{\gamma^{(\text{i})}_\text{0}}\right) = \phi_P^{(\text{i}), \text{max}}
        \label{eq:crp_proteome}
\end{equation}
where we have defined  $\phi_P^{(\text{i}), \text{max}} = 1-\phi^{(\text{i})}_Q-\phi_R^{(\text{i}), \text{min}}$.

All together, these equations provide the first steps towards a modified consumer-resource model that incorporates key aspects of proteome allocation and species-specific growth physiology. 
Pacciani-Mori and coworkers~\cite{pacciani-mori_constrained_2021} further introduced the proteome allocation dynamics into this framework, so that as competition between consumers depletes a resource, species shift a greater fraction of their proteome toward its uptake. In this way, they show that multispecies coexistence under purely competitive conditions requires a balance between maintenance rate and the total proteome fraction devoted to metabolism, and that both faster reallocation of proteomic resources and higher resource quality widen the range of resource-supply conditions that support large biodiversity. \\

\begin{figure}[!tb]
    \centering
    \includegraphics[width=1\columnwidth]{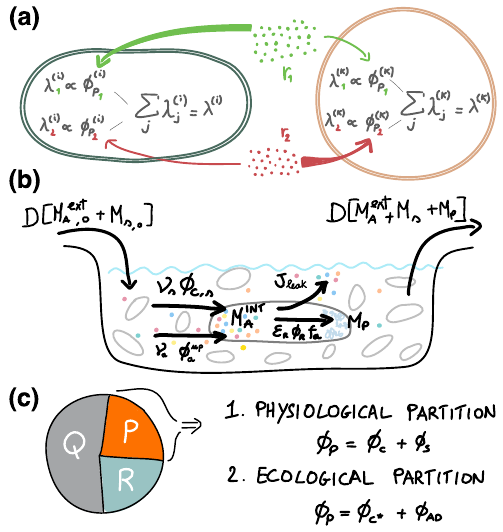} 
    \caption{\textbf{Ecological context for growth laws.} 
    \textbf{(a)} Scheme of the main nutrient fluxes in a consumer-resource framework. Different species can feed on different resources at varying uptake rates, resulting in a fine equilibrium between protein allocation, competition, and resource availability. 
    \textbf{(b)} Fluxes scheme for amino acids cross-feeding in a chemostat setting. The panel illustrates the main fluxes determining cell concentration and protein allocation in a chemostat setting. In this case, in addition to the constant influx of substrate, the external amino acids concentration~$[M^\text{ext}_\text{A}]$ is replenished also from the amino acids leaked from the cells, via the leakage flux~$J_\text{leak}$.
    \textbf{(c)} Scheme of the possible partition of the P sector~$\phi_\text{P}$, following the new insight on nutrient quality presented in Ref.~\cite{Mukherjee2024_plasticity} and in the third recipe of this section. 
    }
    \label{fig9:communities}
\end{figure}

\noindent\textbf{Amino acid cross-feeding}.
We previously included concepts from single-species growth laws into a multispecies consumer-resource framework to capture competition for shared nutrients. We now turn to cooperative interactions and, specifically, we focus on amino-acid cross-feeding. 
As bacteria grow, they leak amino acids into the environment, allowing auxotrophic strains to grow and establishing a network of dependencies that drives community-level nutrient exchange~\cite{morris_black_2012}.

Building on the catabolic and protein-synthesis fluxes identified in the previous sections as key determinants of intracellular amino-acid pools (Sec. \ref{sec2:classical},\ref{sec5:degradation},\ref{sec7:shifts}), we now introduce a leakage flux to couple internal and external amino-acid concentrations. This extra term lets us devise a ``cross-feeding recipe'' that ties proteome-allocation strategies directly to amino-acid exchange among cells, their surroundings, and other species.  
Reyes-González et al. were the first to link microbial growth laws with amino acid cross-feeding mediated by secreted metabolite \cite{reyes-gonzalez_dynamic_2022}. Here, we propose a simplified, novel version inspired by their model. Although introducing an original framework is unusual for a review, we believe that presenting this prototype in a recipe-style format will be useful to the field.

Let us consider an isogenic bacterial population capable of synthesizing all amino acids, growing in a \textit{chemostat}. 
\edit{A chemostat is a continuous-culture device in which fresh medium enters at a constant dilution rate $D$ while an equal volume of culture (cells plus residual nutrients and leaked metabolites) is simultaneously removed, keeping the total volume $V_\text{tot}$ fixed (for a concise overview of modeling in batch, chemostat, and spatial settings, see \cite{Allen2018})}
In this setting, the concentration (calculated over the total volume of the flask) of nutrients and cells can reach a nonzero steady state only if the dilution rate is lower than the maximum growth rate the cells can achieve in that environment~\cite{Novick1Szilard50}.

To model amino acid cross-feeding in a chemostat, we need to describe the dynamics of both internal and external amino acid concentrations (rather than absolute masses). Concentrations directly drive diffusion, dilution, uptake, and leakage kinetics via a Michaelis-Menten or Monod dependence of the uptake rates. Specifically, we need to specify the dynamics of (i) the external amino acids concentration~$[M^\text{ext}_\text{A}] = M^\text{ext}_\text{A} /V_\text{tot}$, (ii) the external nutrient concentration 
$[M_\text{s}] = M_\text{s} / V_\text{tot}$ (here $s$ stands for carbon substrate), (iii) the total protein concentration~$M_p / V_\text{tot}$ --assumed to be proportional to the cell concentration-- and (iv) the internal amino acids concentration~$M^\text{int}_\text{A} / V_\text{tot}$. Note that the volume~$V_\text{tot}$ always refers to the volume of the entire culture (cells and medium).

The dynamics of the external nutrients is described by 
\begin{equation}
\frac{\text{d}[M^\text{ext}_\text{A}]}{\text{dt}} = 
D [M^\text{ext}_\text{A, 0}] - D[M^\text{ext}_\text{A}] - \frac{M_p}{V_\text{tot}} \nu_\text{a} \phi_{a}^\mathrm{up} + \frac{J_\text{leak}}{V_\text{tot}} , 
\label{eq:10.extAAleak}
\end{equation}
\begin{equation}
\frac{\text{d}[M_\text{s}]}{\text{dt}} = 
D [M_\text{s, 0}] - D[M_\text{s}] - \frac{M_p}{V_\text{tot}} \nu_\text{s} \phi_\text{C, s}.
\label{eq:10.extGLU}
\end{equation}
The first two terms in both equations represent the influx and the dilution term, respectively. The parameters~$[M^\text{ext}_\text{A, 0}]$ and~$[M_\text{s, 0}]$ set the concentration of the medium from the incoming flux. The third terms in Eqs.~(\ref{eq:10.extAAleak}) and (\ref{eq:10.extGLU}) represent the cellular uptake of external amino acids, mediated by the amino acid uptake sector $\phi_{a}^\mathrm{up}$, and of the carbon substrate, mediated by the catabolic sector~$\phi_\text{C, s}$. Note that usually these two uptakes are considered together in~$\nu \phi_\text{P}$\footnote{Also in other models catabolism and uptake are considered separately~\cite{hui_quantitative_2015, scott_shaping_2022} and mediated by two different sectors~$\phi_\text{C}$ and~$\phi_{a}^\mathrm{up}$. This distinction should be made carefully depending on the modeling goal. For example, when modeling a species that cannot synthesize a particular amino acid, it is important to distinguish between carbon-source import and amino-acid import and catabolism to identify which flux is growth-limiting, and thus a model with $\phi_\text{C}$ and $\phi_{a}^\mathrm{up}$ is preferable.}.
Lastly, the last term in Eq.~(\ref{eq:10.extAAleak}) represents the overall cellular leakage of amino acids~$J_\text{leak} / V_\text{tot}$, which in general will depend on the total internal and external amino acid concentration.

Importantly, by adding a leakage term for intracellular amino acids into the surrounding medium, we effectively create a second, endogenously produced resource pool, even when no amino acids are supplied in the inflowing feed ($[M^\text{ext}_\text{A, 0}]=0$). Ecologically, amino acids become a ``public good'' and can potentially sustain strains that lack biosynthetic capacity: auxotrophic cells can persist by scavenging the overflow from prototrophs \cite{morris_black_2012,dsouza_less_2014}. 

The kinetics governing leakage remain poorly characterized. To model this process, one can consider two broad scenarios: proteome-independent leakage and proteome-dependent leakage.
In the first case, leakage is modeled as a passive process
that depends solely on the intracellular amino acid concentration, independent of the proteome composition~\cite{dal_co_emergent_2019, dal_co_short-range_2020, vliet_spatially_2018}.
Alternatively, leakage may be actively controlled and regulated by a dedicated sector of the proteome, potentially involving specific transporters or export mechanisms. 
Such regulation has been proposed, for example, in Ref.~\cite{reyes-gonzalez_dynamic_2022}.    
Both approaches are worth mentioning since the precise contributions of different leakage mechanisms to the extracellular amino acid pool remain largely unknown~\cite{mckinlay_are_2023}.

We now turn to the equations for cellular growth. The total protein mass dynamics is given by the equilibrium between synthesis~$\varepsilon \phi_\text{R} f_\text{a}$ and dilution,
\begin{equation}
    \frac{\text{d}M_p / V_\text{tot}}{\text{dt}} =
    \frac{M_p}{V_\text{tot}} [\varepsilon \phi_\text{R} f_\text{a} - D] \, , \label{eq:10.Mp}
\end{equation}
while the concentration of the internal amino acids is described by
\begin{equation}
    \frac{\text{d}M^\text{int}_\text{A} / V_\text{tot}}{\text{dt}} =
    \frac{M_p}{V_\text{tot}} [ \nu_\text{s}\phi_\text{C, s} + \nu_\text{a} \phi_{a}^\mathrm{up} - \varepsilon \phi_\text{R} f_\text{a}]  - \frac{J_\text{leak}}{V_\text{tot}} \, . \label{eq:10.aaINTleak}
\end{equation}
Note that this last equation presents the uptake and protein synthesis fluxes, plus the leakage, but the dilution term $D$ is not explicitly present. This is because it is not directly the internal amino acids that are being diluted, but the overall cell population, as in Eq.~(\ref{eq:10.Mp}).
\edit{In Eq.~(\ref{eq:10.aaINTleak}) 
both \textit{de novo} amino acid synthesis from the carbon substrate ($\nu_s \phi_{C,s}$) and direct amino-acid uptake ($\nu_{a} \phi_{a}^\mathrm{up}$) contribute to the single intracellular amino-acid pool. This simplified additive form is appropriate when carbon is supplied and external amino-acid levels are tunable, and it could be generalised to consider imperfect substitution\footnote{{External amino acids can bypass parts of biosynthesis, whereas in their absence cells synthesize amino acids from carbon, nitrogen, and sulfur sources. Conversely, when the carbon substrate is absent but specific amino acids are present, growth is only sustainable if those amino acids can be catabolized to provide both energy and carbon skeletons.}.}}

Eqs.~\eqref{eq:10.extAAleak}-\eqref{eq:10.aaINTleak} provide a starting point for including proteome allocation models into the dynamics of amino acid cross-feeding. 
Once each term in these equations is specified --recognizing that the uptake and biosynthesis rates depend on internal and external amino-acid and carbon substrate concentrations ($[M_\text{A}^{\mathrm{ext}}]$, $[M_\text{A}^{\mathrm{int}}]$,$[M_\text{s}]$)-- the system can be solved, for a given set of environmental conditions and proteome-sector allocations, to predict the extracellular amino-acid concentration.
\edit{Crucially, in our formulation the exchange flux emerges from growth-law–based allocation via the intracellular amino-acid pool, coupling uptake, biosynthesis, translation demand, and a distinct leakage term.} This provides a quantitative prediction of the amino-acid concentration to which other cells are exposed \edit{across environmental conditions}, thus allowing for exploring and predicting ecological interactions.

In more complex scenarios, one can consider multiple distinct amino acid pools, each governed by its own uptake and leakage kinetics, and allow different strains or species to exhibit unique physiological responses while collectively modulating the shared extracellular environment. These responses, in turn, modulate the shared extracellular environment in a coupled manner. 
While we have framed the dynamics in a chemostat, so that Eqs.~\eqref{eq:10.extAAleak}-\eqref{eq:10.aaINTleak} admit steady-state solutions, the same mathematical framework can be readily adapted to other experimental configurations or transient regimes.
Ultimately, explicitly tracking extracellular amino acid concentrations enables the formulation of cross-feeding models, in which cells import complementary amino acids leaked by others to meet their biosynthetic needs. Such a framework would naturally capture cooperative metabolic interactions and competition for shared metabolic intermediates. \\

\noindent\textbf{A mechanistic understanding of nutrient quality}.
The nutrient quality~$\nu$, first introduced in Sec.~\ref{sec2:classical}, is a key determinant of bacterial growth, typically linked to the energetic yield and protein costs associated with carbon uptake and catabolism~\cite{you_coordination_2013, hui_quantitative_2015, Hermsen2015, Mukherjee2024_plasticity}. However, the precise factors setting this parameter and the resulting growth rate remain unclear. 
Inspired by the work by Mukherjee and coworkers~\cite{Mukherjee2024_plasticity}, we introduce recent results that shed new light on the interpretation of the nutrient quality, opening also for the incorporation of ecological insight into the proteome framework. For this recipe, we will focus on growth on carbon source only, \edit{and we remind the reader the phenomenological definition of nutrient quality, given by the observed growth rate on a given substrate,
\begin{equation}
    \lambda = \nu \phi_\text{P} \rightarrow \nu = \frac{\lambda}{\phi_\text{P}} \ .
\end{equation}
}\\

\noindent\textit{Insights into the~$\nu(\lambda)$ relation.}
The above relationship is derived without perturbing the expression of the \edit{uptake} sector, and this could hide the molecular mechanism that gives rise to it.
Mukherjee and coworkers~\cite{Mukherjee2024_plasticity} investigate in this direction, asking whether the relation between growth rate and nutrient quality is only determined by the characteristic of the substrate, namely the metabolic cost of processing a certain substance, or if it is determined by an equilibrium between the expression of metabolic enzymes and the rest of the proteome. In the second case, bacteria would be genetically programmed to grow more slowly on certain substrates.

To test this, they engineered an \textit{E. coli} strain where the promoters for the metabolic enzymes for glucose --a ``high quality'' substrate~\cite{scott_interdependence_2010, Hermsen2015}-- and mannose --a ``low quality'' one-- were swapped.
Specifically, they replaced the promoters of transporters and metabolic enzymes necessary for mannose metabolism with the glucose import promoter (P-ptsg). Additionally, to prevent repression due to a lack of glucose, the glucose-specific transcriptional regulator, Mlc, was knocked out. Lastly, to ensure that the mannose influx enters the glycolysis pathway, mannose-6-phosphate isomerase was placed under a strong constitutive promoter (P-tet).
As a result, they obtain a strain that performed equally well on both glucose and mannose, displaying the same growth rate as the wild type growing on glucose~\cite{Mukherjee2024_plasticity}.
However, this enhanced growth is accompanied by ecological trade-offs, such as longer lag phases during diauxic shifts and reduced survival after seven days of starvation. 
These two results opened questions on the relation between growth, carbon uptake, and protein sectors related to stress resilience.\\

\noindent\textit{A new partition of~$\phi_\text{P}$.}
\edit{
Both these observations --the improved growth on mannose and the trade-off between growth and survival-- can be rationalised within a proteome allocation framework. As already discussed in Secs.~\ref{sec2:classical} and ~\ref{sec7:shifts}, and previously in this section, the P sector includes different categories of proteins, including those involved in catabolism, nutrient import and related processes.
How one chooses to ``fine-grain'' the P sector, however, directly affects both the definition and interpretation of the nutrient quality~$\nu$ (see the footnote in this section or in Sec.~\ref{sec2:classical}).
Previously in this section, when focusing on amino-acid cross-feeding, we considered only the contributions of carbon and amino-acid uptake, and separated their respective role in $\phi_\textrm{P}$, thus defining $\nu_\textrm{s}$ and $\nu_\textrm{A}$.
Mukherjee \textit{et al.}~\cite{Mukherjee2024_plasticity}, instead, subdivide the P sector into distinct components that capture both the biochemical essence of nutrient quality and adaptability dynamics. This is the key element of this recipe.

Based on physiological functions, the P-sector~($\phi_\text{P}$) can be further divided into two sub-sectors:
\begin{equation}
    \phi_\text{P} = \phi_\text{C} + \phi_\text{S}, \label{eq:10.phiP_partition_physio}
\end{equation}
where~$\phi_\text{C}$ includes proteins responsible for nutrient catabolism (including both transporters and enzymes), while $\phi_\text{S}$ comprises proteins expressed in response to nutrient and antibiotic stresses.

While it is generally accepted that the stress sector S is part of P, since fast-growing bacteria respond less efficiently to stress~\cite{Biselli2020}, the exact extent of~$\phi_\text{S}$ remains under debate. 

The C sector, however, is a collection of different enzymes specialized for different carbon substrates (Refs.~\cite{you_coordination_2013, Hermsen2015} and Sec.~\ref{sec7:shifts}). It can be further divided into a core catabolic sector~$\phi_{\text{C}^\star}$, which includes the enzymes directly responsible for nutrient uptake and metabolism under the current growth condition, and a residual sector,$\phi_{\text{C}^\text{rest}}$, incorporating enzymes that are not actively contributing to growth in that condition.
By definition, these two components satisfy $\phi_\text{C} = \phi_{\text{C}^\star} + \phi_{\text{C}^\text{rest}}$. The allocation of $\text{C}^\text{rest}$ reduces the lag time that bacteria display when switching to alternative carbon sources.
The allocations ~$\phi_{\text{C}^\text{rest}}$ and~$\phi_\text{S}$ both contribute to cellular adaptability: they do not support growth in the current growth condition, but they represent a preparatory investment for future environmental changes. These fractions can be grouped together into a single adaptability sector~$\phi_\text{AD} = \phi_{\text{C}^\text{rest}}+\phi_\text{S}$.

Thus, based on ecological functions, one can partition the P sector into two sub-sectors:
\begin{equation}
    \phi_\text{P} = \phi_{\text{C}^\star} + \phi_\text{AD}. \label{eq:10.phiP_partition_eco}
\end{equation}
Since $\phi_{\text{C}^\star}$ is, by definition, the fraction of the P sector directly contributing to growth under the current carbon source, we can write~$\phi_{\text{C}^\star} = f_{\text{C}^\star} \phi_\text{P}$. Accordingly, the growth rate satisfies
$\lambda = \nu \phi_{\text{P}} = \nu^\star \phi_{\text{C}^\star} = \nu^\star f_{\text{C}^\star} \phi_{\text{P}}$, where $\nu^\star = \nu/f_{\text{C}^\star}$ represents the effective nutrient quality associated with the active catabolic fraction.  
This framework thus provides a new interpretation of the nutrient quality $\nu$ as introduced in Sec.~\ref{sec2:classical}. Rather than representing a fixed biochemical property, the nutrient quality here emerges from from two components: the intrinsic substrate efficiency($\nu^\star$) and the cell’s internal allocation strategy within the P sector~($f_{\text{C}^\star}$).}
%
\edit{
A smaller adaptability sector~$\phi_\text{AD}$ allows a larger investment in the core catabolic sector $\phi_{C^\star}$, thereby supporting faster growth. Instead, a larger~$\phi_\text{AD}$ favors survival and adaptability at the cost of growth~\cite{Omer2025}. 

This trade-off has important consequences for bacteria's fitness~\cite{Kussell2005, Biselli2020} and underscores a strong interplay between ecological conditions and proteome regulation. The results show that, when exposed to different substrates, bacteria often display complex regulation of the stress and adaptive sectors, presumably reflecting the ecological contexts in which these traits evolved.
}

\subsection*{OUTLOOK}
We examined three scenarios demonstrating how growth physiology only exists in the presence of interactions with the environment and within a community. Next, we assess the strengths and limitations of the approaches we discussed and introduce additional relevant perspectives.

We introduced a framework for extending consumer-resource models by incorporating constraints that account for how consumer proteome partitioning influences system behaviour --even in systems composed of just a few species and resources. Although the model presented here uses three broad proteome sectors, future extensions could include a richer set of sectors and more precise descriptions of the first growth law (see sections~\ref{sec1:introduction}-\ref{sec6:mecha_regulation}). 
For example, integrating a stress sector might enhance predictions when resources are fluctuating over time. Integrating the degradation dynamics proposed in Sec.~\ref{sec10:communities} could improve model predictions under slow growth conditions. Finally, the model can include perturbations from the second growth law and be adapted to measure competition for resources under sublethal levels of antibiotics. 

In addition, the approach of embedding proteome allocation constraints can also be adapted to other ecological models, such as predator-prey systems~\cite{wang_modeling_2021}. Predatory interactions in bacterial communities are also prevalent~\cite{johnke_multiple_2014, guo_predation_2023}, and predators and prey may allocate their proteome for different strategies.

Regarding resource sharing among species, this problem remains poorly explored by the scientific community working on theoretical models of microbial communities~\cite{Huelsmann2024, Narla2025, dal_bello_resourcediversity_2021}. Even though we know potential mechanisms by which bacteria can externalise amino acids, we do not know the kinetics of these processes and their relative contributions to amino acid leakage~\cite{mckinlay_are_2023}. This is a significant limitation of models that were built using this framework. Nevertheless, they allow us to explore the dynamics of microbial communities engaged in cross-feeding. 
Starting from a single-species model, one can investigate how more complex bacterial communities that share amino acids emerge through a bottom-up approach. One can also construct a multi-resource system (a model with two or more amino acids, which would further allow the exploration of auxotrophic species that are entirely dependent on extracellular amino acids secreted by other species (see~\cite{reyes-gonzalez_dynamic_2022} as an example). We believe this set of models could yield valuable predictions on how obligatory cross-feeding is maintained across various environments and under which conditions bidirectional versus unidirectional cross-feeding is favoured.
While this recipe focuses on amino acid cross-feeding, similar models can be developed for catabolic cross-feeding in which extracellular molecules are taken up, metabolically transformed intracellularly, and subsequently leaked. 
Finally, we discussed the idea that proteome allocation in a given environment can be shaped by ecological trade-offs, such as investing in preparedness for future conditions. This insight lays the groundwork for building a general framework capable of predicting lag time distributions across different environmental fluctuations~\cite{Mukherjee2024_plasticity}.

Two additional directions, not covered here, are worth mentioning. First, the work of Greulich and coworkers builds on the growth laws to predict differential antibiotic inhibition rates in slow- versus fast-growing conditions~\cite{greulich_growth-dependent_2015}. Incorporating antibiotics into the frameworks described here, while accounting for strain- and species-specific susceptibilities, could open promising new directions. Second, although this section focuses on ecology, the recipes can be readily adapted to include evolutionary dynamics --for instance, by allowing mutations that bias proteome partitioning.

\section*{Acknowledgments} 
\label{sec:Acknowledgments}
Rossana Droghetti (R.D.), Mattia Corigliano (M.C.), Ludovico Calabrese (L.Ca.), Philippe Fuchs (P.F.), Abhishek Vaidyanathan (A.V.), Johannes Keisers (J.K.), Gabriele Micali (G.M.), Marco Cosentino Lagomarsino (M.C.L.) and Luca Ciandrini (L.Ci.) led the drafting of key parts of this review: L.Ci., M.C.L. and G.M. drafted the Introduction; R.D.,  M.C. and L.Ci. Section II; L.Ci. Section III; L.Ca and L.Ci Section IV; R.D. and L.Ca. Section V; R.D. and M.C.L. Section VI; R.D. Section VII; P.F. and L.Ca. Section VIII; M.C. and G.M. Section IX; A.V. R.D.
L.Ci and G.M Section X; R.D. and J.K. compiled appendix and tables. R.D., G.M., M.C.L., and L.Ci. then integrated, revised and polished the full manuscript. All authors provided critical feedback, read and approved the final version. 

G.M. acknowledges support from Human Technopole grant no. 782 (G.M. and A.V.). M.C.L., M.C. and R.D. acknowledge funding from AIRC - Associazione Italiana per la Ricerca sul Cancro AIRC IG grant no. 23258 (M.C.L., M.C. and R.D.).  R.D. was supported by the AIRC fellowship Italy Pre-Doc 2022 (ID 28176). M.C. was supported by the AIRC fellowship Italy Pre-Doc 2022 (ID 28177).
L.Ci. acknowledges support from the French National Research Agency (REF: ANR-21-CE45-0009) (L.Ci., P.F., J.K.).

\clearpage
\newpage

\clearpage
\appendix
\section{Wine pairing: Derivations and Numbers}\label{sec11:parameters_numbers}

This appendix collects all the variables and relevant quantities introduced in the sections of this cookbook\edit{, together with the derivation of a few equations}.

\edit{

\subsection{Derivation of Eq.~(\ref{eq:epsilon})}
The ribosome elongation rate $\varepsilon$ (the average rate at which a ribosome steps from one codon to the next) is determined by two fundamental microscopic timescales. The first is the dwell time $\tau_\text{dwell}$, which corresponds to the waiting time for the arrival of a cognate EF–Tu–GTP–aminoacyl–tRNA ternary complex (TC) to the ribosomal A site. The second is the translocation time $\tau_\text{trans}$, which is the time required for the ribosome to complete peptide-bond formation and move to the next codon once the correct tRNA has been accommodated.
Under the assumption that these two processes occur sequentially and independently at each codon, the total time required to elongate by one codon is given by
\begin{equation}
\tau_\text{step} = \tau_\text{dwell} + \tau_\text{trans},.
\end{equation}
The elongation rate is therefore the inverse of this total stepping time:
\begin{equation}
\varepsilon = \tau_\text{step}^{-1}
= \frac{1}{\tau_\text{dwell} + \tau_\text{trans}}.
\label{eq:eps_basic}
\end{equation}
Assuming that the waiting time $\tau_\text{dwell}$ is limited by the diffusion and capture of ternary complexes, it becomes inversely proportional to their concentration $[TC]$. This concentration must in principle be computed by accounting for both codon usage and the effective availability of each ternary-complex species (see Note 3 in the SI of Ref.~\cite{Dai2016}).
Furthermore, Ref.~\cite{Dai2016} argues that  the total concentration of ternary complexes scales proportionally with the ribosomal proteome fraction $\phi_\text{R}$. Under these assumptions it follows that $\tau_\text{dwell} = A / \phi_\text{R}$, where $A$ is a proportionality constant. Substituting this relation into Eq.~\eqref{eq:eps_basic}, and keeping $\tau_\text{trans}$ as a constant, yields the following Michaelis–Menten–like expression for the elongation rate across steady-state conditions,
\begin{equation}
\varepsilon
= \varepsilon^\text{max}
\frac{\phi_\text{R}}{\phi_\text{R} + C} \ ,
\end{equation}
where $C = A\,\varepsilon^\text{max}$ is an empirical constant that can be estimated as in Ref.~\cite{Dai2016}, and is found to be approximately equal to the minimal ribosomal fraction $\phi_\text{R}^\text{min}$.

A further step becomes possible by connecting the ribosome elongation rate to ppGpp levels, as discussed in Section~\ref{sec6:mecha_regulation}. Wu \textit{et al.} found an empirical relationship between the ppGpp concentration $[G]$ (normalized by its value at zero growth, $[G_0]$) and the elongation rate~$\varepsilon$:
\begin{equation}
\frac{[G]}{[G_0]} = \frac{\varepsilon^\text{max}}{\varepsilon(t)} - 1 .
\label{eq:G_ratios}
\end{equation}
Combined with Eq.~(\ref{eq:eps_basic}), this relation implies that the ratio of the dwell to translocation times satisfies
\begin{equation}
    \frac{\tau_\text{dwell}}{\tau_\text{trans}} = \frac{[G]}{[G_0]}\,,
\end{equation}
and starting from Eq.~(\ref{eq:G_ratios}), we can express the elongation rate directly as a function of the ppGpp concentration,
\begin{equation}
     \varepsilon = \varepsilon^\text{max} \frac{1}{ 1 + [G]/[G_\text{0}] }\,.
     \label{eq:epsilon_derivation_2}
\end{equation}

Wu \textit{et al.}\cite{Wu2022} assume that ribosome expression is inversely proportional to the ppGpp concentration,
\begin{equation}
  \phi_\text{R} \propto \frac{1}{[G]}.
\end{equation}
Since~$\phi^\text{min}_\text{R}$ corresponds to the ribosomal fraction at a concentration $[G_\text{0}]$ it follows that
\begin{equation}
    [G]/[G_\text{0}] = \phi^\text{min}_\text{R} / \phi_\text{R} \ .
\end{equation}
Combining this expression with Eq.~(\ref{eq:epsilon_derivation_2}) gives 
\begin{equation}
\varepsilon = \varepsilon^\text{max} \cfrac{\phi_\text{R}}{\phi_\text{R} + \phi_\text{R}^\text{min}}\ .
\label{eq:epsilon_derivation}
\end{equation}
which corresponds to Eq.~(\ref{eq:epsilon}). Another useful derivation of this relationship can be found in the Supplementary Information of Ref.~\cite{zhu_distantly_2025}.

\subsection{Derivation of Eq.~(\ref{eq:J_TL}) and Eq.~(\ref{eq:CFlim_GL})}
For convenience, we first restate Eq.~(\ref{eq:J_TL}), which gives the total translation flux as a function of ribosome concentration and mRNA availability:
\begin{equation}
    J^{\mathrm{TL}}
    \;=\;
    \alpha_{0}^{\mathrm{TL}} \,[R]\,
    \frac{1}{1 + \dfrac{[m]}{K_m}} \,.
    \label{eq:J_TL_derivation}
\end{equation}
The ribosome initiation rate on an mRNA is given by
Eq.~(\ref{eq:initiation}),  
\begin{equation}
    \alpha^{\mathrm{TL}}
    = \alpha_{0}^{\mathrm{TL}}\, [R^{\mathrm{f}}] \,,
    \label{eq:initiation_deriv}
\end{equation}
that is, initiation occurs at a rate proportional to the pool of free
ribosomes $R^{\mathrm{f}}$.
From ribosome traffic models~\cite{shaw_totally_2003}, the density of
ribosomes on a transcript --i.e.\ the number of elongating ribosomes per
unit transcript length-- is determined by the balance between initiation and
elongation.
This leads to Eq.~(\ref{eq:ribo_density}), rewritten here for convenience:
\[
    \rho = \frac{\alpha_\text{0}^\text{TL}}{\varepsilon} [R^\text{f}] \,,
\]
see Refs.~\cite{calabrese_protein_2022, calabrese_how_2024} for a more detailed derivation.
Assuming that all mRNAs share the same typical length~$L_\mathrm{p}$, the total number
of ribosomes is the sum of free and translating ribosomes:
\begin{equation}
    R
    = R^{\mathrm{f}}
      + L_\mathrm{p}\,m\,\rho
    = R^{\mathrm{f}}
      \left(
        1 + \frac{\alpha_{0}^{\mathrm{TL}}\,L_\mathrm{p}\,m}{\varepsilon}
      \right),
    \label{eq:R_total_relation}
\end{equation}
which provides a direct relationship between free ribosomes
$R^{\mathrm{f}}$ and the total ribosome pool~$R$.
As explained in the main text, the protein synthesis rate is often approximated with the initiation rate, $J^{\mathrm{TL}} \sim \alpha^{\mathrm{TL}}$, and by replacing $R^\mathrm{f}$ from Eq.~(\ref{eq:R_total_relation}) into Eq.~(\ref{eq:initiation_deriv}) one obtains Eq.~(\ref{eq:J_TL_derivation}).

By multiplying and dividing Eq.~(\ref{eq:J_TL_derivation}) by $K_m$,
with
\[
    K_m \;=\; \frac{\varepsilon}{L_{\mathrm{p}}\,
        \alpha_{0}^{\mathrm{TL}}},
\]
we obtain
\begin{equation}
    J^{\mathrm{TL}}
    \;=\;
    \cfrac{\varepsilon}{L_\textrm{p}} \,[R]\,
    \frac{1}{K_m + [m]} \,.
    \label{eq:J_TL_derivation2}
\end{equation}
The biomass production $\dot M_p$ is related to the translational current through Eq.~(\ref{eq:mech_mass_current}), which we can write as
\begin{equation}
    \frac{\text{d}M_p}{\text{dt}}= \lambda M_p = m \mu_\textrm{aa} L_\textrm{p} J^{\mathrm{TL}} \,.
\end{equation}
Substituting Eq.(\ref{eq:J_TL_derivation}) yields 
\begin{equation}
    \lambda M_p =  \mu_\textrm{aa} \varepsilon \,R\,
    \frac{[m]}{K_m + [m]}\,.
    \label{eq:dMpderivation}
\end{equation}
To proceed, we relate the total number of ribosomes $R$ to the total mass
of ribosomal proteins $M_{\mathrm{rp}}$.  If each ribosome contains
$n^{\mathrm{ribo}}_{\mathrm{rp}}$ ribosomal proteins of average length
$L_{\mathrm{rp}}$, then
\[
    M_{\mathrm{rp}}
    \;=\;
    n^{\mathrm{ribo}}_{\mathrm{rp}}\,
    R\,
    L_{\mathrm{rp}}\,
    \mu_{\mathrm{aa}}\,.
\]
Dividing Eq.~(\ref{eq:dMpderivation}) by $M_p$ and using
$\phi_{\mathrm{R}} = M_{\mathrm{rp}}/M_p$ gives
\begin{equation}
	\lambda = 
    \varepsilon_\text{R}  \phi_\text{R} \frac{[m]}{K_m + [m]} \,,
    \label{eq:CFlim_GL_derivation}
\end{equation}
where we defined the effective elongation constant per ribosome,
\[
    \varepsilon_{\mathrm{R}}
    \;=\;
    \frac{\varepsilon}{n^{\mathrm{ribo}}_{\mathrm{rp}}\,L_{\mathrm{rp}}}\,,
\]
i.e.\ the elongation rate normalized by the total number of amino acids in
a ribosome.  Equation~(\ref{eq:CFlim_GL_derivation}) corresponds to
Eq.~(\ref{eq:CFlim_GL}) in the main text.

\subsection{Derivation of Eq.~(\ref{eq:6_ppGpp(epsilon)})}
This subsection derives Eq.~(\ref{eq:6_ppGpp(epsilon)}) following Ref.~\cite{Wu2022}. We report Eq.~(\ref{eq:6_ppGpp(epsilon)}) here for convenience,
\begin{equation}
[G] \propto \frac{\varepsilon^\text{max}}{\varepsilon(t)} -1 \ .
\label{eq:6_ppGpp(epsilon)_derivation}
\end{equation}
The derivation starts from Eq.~(\ref{eq:6_dG_dt}). As explained in Sec.~\ref{sec6:mecha_regulation}, one can assume that ppGpp synthesis is proportional to the waiting time of the ribosomes on the transcript (i.e.~$f_\text{syn} \propto \tau_\text{dwell}$), while degradation is proportional to the translocation time,~$f_\text{deg} \propto \tau_\text{trans}$, see Ref.~\cite{Wu2022}. With these assumptions, Eq.~(\ref{eq:6_dG_dt}) becomes
\begin{equation}
    \frac{\text{d}[G]}{\text{dt}} \propto \tau_\text{dwell} - [G] \tau_\text{trans} \ ,
\end{equation}
implying that the steady state equation for~$[G]$ reads
\begin{equation}
    [G] \propto \frac{\tau_\text{dwell}}{\tau_\text{trans}} \ .
\end{equation}
Then, by using the definition of~$\varepsilon$ in terms of~$\tau_\text{dwell}$ and~$\tau_\text{trans}$ given by~$\varepsilon = (\tau_\text{dwell} + \tau_\text{trans})^{-1}$, Eq.~(\ref{eq:6_epsilon(taus)}), we obtain the above expression for the ppGpp concentration, Eq.~(\ref{eq:6_ppGpp(epsilon)_derivation}).\\

\subsection{Derivation of Eqs.~(\ref{eq:FCR_xR}) and~(\ref{eq:FCR_xCat})}
This subsection presents a step-by-step derivation of the the regulatory functions~$\chi_\text{R}$ and~$\chi_{\text{C}_\text{i}}$ under the steady-state assumption across nutrient conditions, introduced in Ref.~\cite{Erickson2017}, and expressed by Eqs.~(\ref{eq:FCR_xR}) and~(\ref{eq:FCR_xCat}), which we rewrite below for convenience,
\begin{equation}
\chi_\text{R}(t)= \phi^*_\text{R}(\sigma(t)) = \frac{\phi_\text{R}^\text{min}}{1 - \sigma(t) / \gamma_\text{0}} \ , \label{eq:FCR_xR_derivation}
\end{equation}
\begin{equation}
\chi_{\text{C}_\text{i}}(t)= \phi^*_{\text{C}_\text{i}}(\sigma(t)) = h_\text{i} \left( 1 - \frac{\sigma(t)}{\lambda_\text{C}} \chi_\text{R}(\sigma(t))  \right) \ . \label{eq:FCR_xCat_derivation}
\end{equation}
Since at steady state the regulatory functions~$\chi_\text{i}$ can be described as functions of the translational activity~$\sigma$,~$\chi_\text{i} = \chi^*_\text{i}(\sigma)$, and~$\phi^*_\text{i} = \chi^*_\text{i}$. We assume that these relations hold when the translational activity varies with time as~$\sigma(t)$ (see also Sec.~\ref{sec7:shifts}). 
Given these assumptions, Eq.~(\ref{eq:FCR_xR_derivation}) is immediately obtained by solving for~$\phi_\mathrm{R}(\sigma)$ the system
\begin{eqnarray}
  \phi_\mathrm{R} = \phi_\mathrm{R}^\text{min} + \frac{\lambda}{\gamma_\mathrm{0}}\\
  \lambda = \sigma \phi_\mathrm{R} \ ,
\end{eqnarray}
where the first equation corresponds to the first growth law, Eq.~(\ref{eq:GL_first}), and the second defines the translational activity.

In order to derive Eq.~(\ref{eq:FCR_xCat_derivation}) one has to model the behavior of the carbon metabolism sectors $\mathrm{C}_\text{i}$ along the first growth law (i.e. when the growth rate is controlled by the quality of carbon source). Refs.~\cite{you_coordination_2013, Hermsen2015}, provide the following simple emprically grounded model,
\begin{equation}
    \phi_{\mathrm{C}_\text{i}} = h_\text{i} \left(  1 - \frac{\lambda}{\lambda_\text{C}} \right) \ ,
\end{equation}
where~$\lambda_\text{C}$ is a maximum growth rate attainable in carbon-limited conditions~\cite{you_coordination_2013, Hermsen2015}, and~$h_\text{i}$ is an index function equal to~$\phi^\text{max}_{\text{C}_\text{i}}$ when the corresponding single substrate the available one, and~0 otherwise~\cite{Erickson2017}.
Eq.~(\ref{eq:FCR_xCat_derivation}) follows by substituting~$\lambda = \sigma \phi_\text{R}$ into the above equation. 

}

\section{Variables and relevant quantities}
For each variable we provide a numerical value, a reference to the paper reporting the presented value, and a reference to the equation of this cookbook where the quantity is firstly defined.
We present a table for each section with, beneath it, an explanation of the derivation of the parameters when they are not simply measured. 
As a general rule, we advise the reader to directly check the source of the numbers to have more information on the measurements or fitting procedures and the growth conditions at which the measure was performed.

The presented values are valid for the bacterium \textit{E. coli}, unless otherwise specified. For yeast and other eukaryotes, the reader can refer to the specific references cited in this cookbook (for example Refs.~\cite{metzl-raz_principles_2017, xia_proteome_2022} for the growth laws).
Note that some of the quantities introduced in the sections are extensive quantities, whose value strongly depends on the size of the system considered. 

These tables often report quantities that are closely relate to each other, for example the number of proteins in a cell~$P$ and the cellular protein concentration~$[P]$. When these values come from different references or different models, compatibility problems could arise. In writing these tables, we checked whether related quantities coming from different sources where compatible, and we did not found any strong disagreement (i.e. no mismatching order of magnitude was found). However, when designing a quantitative model even small differences can lead to numerous problems. We therefore advise the reader to check the level of agreements between the chosen parameters/quantities before using them together in a quantitative model.

\newpage

\onecolumngrid
\begin{table*}[t!bh]
\centering
\begin{tabular*}{1.0\textwidth}{lllrr}
\hline
\multicolumn{5}{c}{\textbf{Section II. Appetizer: Classic Proteome Allocation Theory}} \\
\hline
\hline
\textbf{Observable} & \textbf{Description} & \textbf{Range [units]} & \textbf{Reference} & \textbf{Defined in} \\
\hline
$\lambda$ & {\footnotesize Cell culture exponential growth rate} & 0 - 2.0 [$h^{-1}$] & \cite{scott_interdependence_2010} & \ref{box1:exp_balanced} \\
$M$ & {\footnotesize Total cell culture mass (extensive)} & $M/\# \text{cells}=$ 0.25-2 [pg $\text{cell}^{-1}$] & \cite{Basan2015} & \ref{box1:exp_balanced} \\
$V$ & {\footnotesize Total cell culture volume (extensive)} & $V/\# \text{cells}=$ 1 - 8 [$\mu m^3$] & \cite{Basan2015} & \ref{box1:exp_balanced} \\
$[P]$ & {\footnotesize Protein number concentration} & 4-5E6 [prot $\mu m^{-3}$] & \cite{balakrishnan_principles_2022} & \ref{box1:exp_balanced} \\ 
$M_p$ & {\footnotesize Total protein mass (extensive)} & $M_p/\#\text{cells}=$0.2 - 1.6 [pg $\text{cell}^{-1}$] & \cite{Basan2015} & \ref{box1:exp_balanced} \\
$\phi_\text{R}$ & {\footnotesize Ribosome and affiliated mass fraction} & 0.049-0.55 [mass fraction] & \cite{scott_interdependence_2010} & Eq.~(\ref{eq:GL_first}) \\
$\phi_\text{P}$ & {\footnotesize P sector mass fraction} & 0 - 0.50 [mass fraction] & \cite{scott_interdependence_2010} & Eq.~(\ref{eq:nu_phiC}) \\
$\nu$ (a) & {\footnotesize Nutrient quality} & 0 - 5 [$h^{-1}$] & \cite{scott_interdependence_2010} & Eq.~(\ref{eq:GL_second}) \\
$\phi_\text{U}$ & {\footnotesize U sector (unnecessary proteins) mass fraction} & 0 - 0.50 [mass fraction] & \cite{scott_interdependence_2010} & Eq.~(\ref{eq:GL_third (unnecessary)}) \\
$M_\text{R}$ & {\footnotesize Total mass of R sector proteins (extensive)} & $M_p/\#\text{cells}=$ 0 - 0.38 [pg $\text{cell}^{-1}$] & (b) & \ref{box2:computing_phi_R} \\
$r$ & {\footnotesize $M_\text{RNA}/M_p$} & 0.1 - 0.72 [mass fraction] & \cite{scott_interdependence_2010} & \ref{box2:computing_phi_R} \\
$M_\text{RNA}$ & {\footnotesize Total RNA mass (extensive)} & $M_\text{RNA}/\#\text{cells}=$ 0 - 0.5 [pg $\text{cell}^{-1}$] & \cite{Basan2015} & \ref{box2:computing_phi_R} \\
$M_\text{rRNA}$ & {\footnotesize Total ribosomal RNA mass (extensive)} & $M_\text{rRNA}/\#\text{cells}=$ 0 - 0.43  [pg cell$^{-1}$] & (c) & \ref{box2:computing_phi_R} \\
$M_\text{rp}$ & {\footnotesize Total mass of ribosomal proteins only (extensive)} & $M_\text{rp}/\#\text{cells}=$ 0 - 0.23 [pg cell$^{-1}$] & (d) & \ref{box2:computing_phi_R} \\
$\psi_\text{R}$ & {\footnotesize Ribosomal protein number fraction (\# rib. prot./$P$)} & 0 - 0.4 [number fraction] & (e) & \ref{box2:computing_phi_R} \\
$R$ & {\footnotesize Total number of ribosomes (extensive)} & $R/\#\text{cells}=$ 0 - 1.8E5 [$\text{cell}^{-1}$] & (f) & \ref{box2:computing_phi_R} \\
$P$ & {\footnotesize Total number of proteins (extensive)} & $P/\#\text{cells}=$ 3 - 40E6 [$\text{cell}^{-1}$] & (g) & \ref{box2:computing_phi_R} \\
$\varepsilon$ & {\footnotesize Codon elongation rate} & 8.4 - 17.1 [$aa$ $s^{-1}$] & \cite{Dai2016} & Eq.~(\ref{eq:maaloe}) \\
$f_\text{a}$ (h) & {\footnotesize Fraction of active ribosomes (no prot. degradation)} & 0 - 1 [fraction] & \cite{Dai2016} & Eq.~(\ref{eq:maaloe}) \\
$f_\text{i}$ (h) & {\footnotesize Fraction of inactive ribosomes (no prot. degradation)} & 0 - 1 [fraction] & \cite{Dai2016} & Eq.~(\ref{eq:maaloe}) \\
$\varepsilon_\text{R}$ & {\footnotesize Rescaled elongation rate $\varepsilon_\text{R} = \varepsilon/L_\text{R}$} & 5-10 [$h^{-1}$] & - & Eq.~(\ref{eq:GL_first_fa}) \\
$J_\text{synt}$ & {\footnotesize Biomass synthesis current $J_\text{synt} = \varepsilon_\text{R} f_\text{a} M_\text{R}$ (extensive)} & [mass per unit time] & - & Eq.~(\ref{eq:cat_syn_fluxes}) \\
$J_\text{cat}$ & {\footnotesize Biosynthesis precursor current $J_\text{cat} = \nu M_p$ (extensive)} & [mass per unit time] & - & Eq.~(\ref{eq:cat_syn_fluxes}) \\\\
\hline
\textbf{Parameter} & \textbf{Description} & \textbf{Value [units]} & \textbf{Reference} & \textbf{Defined in} \\
\hline
$\rho_\text{DM}$ & {\footnotesize Dry-mass density $\rho_\text{DM} = M/V$} & 0.28 [pg/$\mu m^3$] & (h) & \ref{box1:exp_balanced} \\
$\phi_\text{R}^\text{min}$ & {\footnotesize Offset in the 1st growth law} & 0.049 [mass fraction] & \cite{scott_interdependence_2010} & Eq.~(\ref{eq:GL_first}) \\
$\gamma_\text{0}$ & {\footnotesize Slope of the 1st growth law} & 11.02 [$h^{-1}$] & \cite{Erickson2017} & Eq.~(\ref{eq:GL_first}) \\
$\phi_\text{R}^\text{max}$ & {\footnotesize Maximum size of the R sector (2nd growth law)} & 0.55 [mass fraction] & \cite{scott_interdependence_2010} & Eq.~(\ref{eq:GL_second}) \\
$\phi_\text{U}^\text{max}$ & {\footnotesize Maximum size of the U sector $\phi_\text{U}^\text{max} = \phi_\text{R}^\text{max} - \phi_\text{R}^\text{min}$} & 0.50 [mass fraction] & \cite{scott_interdependence_2010} & Eq.~(\ref{eq:GL_third (unnecessary)}) \\
$C_{\text{r} \rightarrow \phi}$ & {\footnotesize Conversion factor $r$ to $\phi_R$} & 0.76 [number] & \cite{scott_interdependence_2010} & \ref{box2:computing_phi_R} \\
$C_{\text{r} \rightarrow \psi}$ & {\footnotesize Conversion factor $r$ to $\psi_R$} & 0.89 [number] & (i) & \ref{box2:computing_phi_R} \\
$\mu_{\text{rRNA}}^\text{ribo}$ & {\footnotesize RNA mass in a single ribosome} & 1.48E6 [Da] & \cite{scott_interdependence_2010} & \ref{box2:computing_phi_R} \\
$n_{\text{rp}}^\text{ribo}$ & {\footnotesize Number of ribosomal proteins in a single ribosome} & 54 [number] & \cite{Dai2016} & \ref{box2:computing_phi_R} \\
$\mu_{\text{p}}$ & {\footnotesize Mass of the average, typical protein} & 28 [kDa] & (j) & \ref{box2:computing_phi_R} \\
$L_\text{p}$ & {\footnotesize \# of amino acids in the typical protein} & 253 [aa] & \cite{balakrishnan_principles_2022} & \ref{box2:computing_phi_R} \\
$L_\text{rp}$ & {\footnotesize \# of amino acids in the typical ribosomal protein} & 133 [aa] & (k) & \ref{box2:computing_phi_R} \\
$\mu_\text{aa}$ & {\footnotesize Average amino acid mass} & 110 [Da] & \cite{Dai2016} & Eq.~(\ref{eq:maaloe}) \\
$L_\text{R}$ & {\footnotesize \# of amino acids in a single ribosome} & 7155 [aa] & (k) & Eq.~(\ref{eq:GL_first_fa}) \\
$\phi_\text{Q}$ & {\footnotesize Q sector (housekeeping) mass fraction} & $\approx$ 0.45 [mass fraction] & \cite{scott_interdependence_2010} & Eq.~(\ref{eq:phiQ}) \\\\
\end{tabular*}

\caption{\footnotesize{Constants and parameters introduced in Sec.~\ref{sec2:classical}, with citations for data sources and the equation of this manuscript where they are defined. The values refers to measurements in the bacteria species \textit{E. coli}.\\
    (a) Observe that these values are extremely dependent of the model chosen, see Secs.~\ref{sec2:classical}-\ref{sec7:shifts}-\ref{sec10:communities}.\\
    (b) Value obtained by multiplying~$M_\text{RNA}$ by the conversion factor~$C_{\text{r} \rightarrow \phi}$=0.76.\\
    (c) Value computed multiplying~$M_\text{RNA}$ by the conversion factor 0.86 from Ref.~\cite{scott_interdependence_2010}.\\
    (d) Value computed multiplying~$M_\text{rRNA}$ by the conversion factor 0.53 from Ref.~\cite{scott_interdependence_2010}.\\
    (e) Value computed by multiplying $r$ times~$C_{r \rightarrow \psi_\text{R}}$. Note that for the value of~$r$ we use the value of 0.45, as what Ref.~\cite{scott_interdependence_2010} reports for the first growth law.\\
    (f) value found using the total mass of ribosomal proteins~$M_\text{rp}$ and the average ribosomal protein mass~$\mu_\text{aa} L_\text{R}$, from this table. This result is compatible with the reported number density of ribosomes, BIONUMBERS ID 108603, if divided by the cellular volume.\\
    (g) Value obtained multiplying~$[P]$ by the protein volume~$V$.\\
    (h) Value obtained from a linear fit of the data presented in Ref.~\cite{Basan2015}.\\
    (i) Value computed using the formula in~\ref{box2:computing_phi_R}~$C_{\text{r} \rightarrow \psi} = \frac{\mu_\text{p}}{\mu^\text{ribo}_\text{rRNA}}\frac{M_\text{rRNA}}{M_\text{RNA}} n^\text{ribo}_\text{rp}$.\\
    (j) Value obtained via $L_\text{p} \times \mu_\text{aa}$.\\
    (k) Values obtained by filtering proteomics data set in~\cite{balakrishnan_principles_2022} for ribosomal proteins.\\
    }}
    \label{tab:params_sec_2_GL}
\end{table*}
\twocolumngrid

\newpage
\onecolumngrid
\begin{table*}[t!bh]
\begin{centering}
\begin{tabular*}{1.0\textwidth}{lllrr}
\hline
\multicolumn{5}{c}{\textbf{Section III. Main Course: On active ribosomes and a more mechanistic view of translation}} \\
\hline
\hline
\textbf{Observable} & \textbf{Description} & \textbf{Range [units]} & \textbf{Reference} & \textbf{Defined in} \\
\hline
$[m_\text{i}]$ & {\footnotesize mRNA concentration of transcript i} & 1000 - 4500 [$\mu m^{-3}$] & \cite{balakrishnan_principles_2022}  & Eq.~(\ref{eq:mech_mass_current}) \\
$L_i$ & {\footnotesize \# of amino acids in protein type i} & 21 - 1654 [aa] & \cite{Mori2021} & Eq.~(\ref{eq:mech_mass_current}) \\
$[R^\text{f}]$ & {\footnotesize Concentration of free ribosomes} & 1000 - 6000 [$\mu m^{-3}$] & \cite{balakrishnan_principles_2022} & Eq.~(\ref{eq:initiation}) \\
$\alpha^\text{TL}$ & {\footnotesize Translation initiation rate} & 0.002 - 8.6 [$s^{-1}$] & \cite{balakrishnan_principles_2022} & Eq.~(\ref{eq:mech_mass_current}) \\
$J_\text{i}^\text{TL}$  & {\footnotesize Translational (ribosomal) current per mRNA i} & 0.002 - 8.6 [prot $s^{-1}$] 
& (a) & Eq.~(\ref{eq:mech_mass_current}) \\
$\rho$ & {\footnotesize Ribosome density on an mRNA (\# rib. per mRNA/ mRNA length )} & 0.005 [\#rib/nt] & \cite{balakrishnan_principles_2022} & Eq.~(\ref{eq:ribo_density})\\\\
\hline
\textbf{Parameter} & \textbf{Description} & \textbf{Value [units]} & \textbf{Reference} & \textbf{Defined in} \\
\hline
$\varepsilon^\text{max}$ & {\footnotesize Maximum value of ribosome elongation rate} & 19 [$aa$ $s^{-1}$] & \cite{Dai2016} & Eq.~(\ref{eq:epsilon}) \\
$\alpha_\text{0}^\text{TL}$ & {\footnotesize Translation initiation rate constant} & 3.3E-7 [$s^{-1} \mu m^3$] & (b) & Eq.~(\ref{eq:initiation}) \\
$K_m$ & {\footnotesize $[m]$ at which complex-formation becomes relevant $K_m = \frac{\varepsilon}{L_\text{p} \alpha_\text{0}^\text{TL}}$} & 0.05-0.15 [$\mu$M] & \cite{calabrese_how_2024} & Eq.~(\ref{eq:J_TL}) \\
\end{tabular*}

\caption{Constants and parameters introduced in Sec.~\ref{sec3:translation}, with citations for data sources and the equation of this manuscript where they are defined.\\
(a) Assuming~$J_\text{i}^\text{TL}$ equal to~$\alpha^\text{TL}$.\\
(b) Computed from~$\frac{\alpha^\text{TL}}{[R^\text{f}]}$.\\
}
\label{tab:params_sec_3_TL}
\end{centering}
\end{table*}
\twocolumngrid

\begin{table*}[t!bh]
\centering
\begin{tabular*}{1.0\textwidth}{l l l r r}
\hline
\multicolumn{5}{c}{\textbf{Section IV. Side Dish: Linking mRNA Abundance to Growth}} \\
\hline
\hline
\textbf{Observable} & \textbf{Description} & \textbf{Range [units]} & \textbf{Reference} & \textbf{Defined in} \\
\hline
$J_\text{i}^\text{TX}$ & {\footnotesize Transcriptional (RNAP) current per gene i} & [transcripts $s^{-1}$] & - & Eq.~(\ref{eq:mech_mass_current})\\
$g_\text{i}$ & {\footnotesize Gene copy number of gene i} & 0.5-2 [$\mu m^{-3}$] & \cite{balakrishnan_principles_2022} & Eq.~(\ref{eq:transcript_dynamics}) \\
$\delta_\text{i}$ & {\footnotesize mRNA degradation rate of transcript type i} & 0.1-3 [$\text{min}^{-1}$] & \cite{balakrishnan_principles_2022} & Eq.~(\ref{eq:transcript_dynamics}) \\
$[N]$ & {\footnotesize RNAP concentration} & $\sim$1500 [$\mu m^{-3}$] & \cite{balakrishnan_principles_2022} & Eq.~(\ref{eq:transcript_dynamics}) \\
$[N^\text{f}]$ & {\footnotesize Free RNAP concentration} & 200-1500 [$\mu m^{-3}$] & \cite{balakrishnan_principles_2022} & Eq.~(\ref{eq:transcript_dynamics}) \\
$\rho_\text{i}^\text{TX}$ & {\footnotesize RNAP density on gene i (\#RNAP per gene / gene length)} & 0.0008 [RNAP/nt] & (a) & Eq.~(\ref{eq:J_i_TX}) \\
$\ell_\text{i}$ & {\footnotesize Length of a gene} & 63-4962 [nt] & \cite{balakrishnan_principles_2022} & Eq.~(\ref{eq:J_i_TX}) \\
$\alpha_\text{i}^\text{TX}$ & {\footnotesize Transcription initiation rate of gene i} & 0.01-10 [$\text{min}^{-1}$] & \cite{balakrishnan_principles_2022} & Eq.~(\ref{eq:J_i_TX}) \\
$\varepsilon^\text{TX}$ & {\footnotesize RNAP elongation rate} & 20-80 [bp/s] & \cite{Epshtein2003} & Eq.~(\ref{eq:J_i_TX}) \\
$\omega_\text{i}$ & {\footnotesize RNAP allocation $g_\text{i}/g$} & 0-1  & \cite{balakrishnan_principles_2022} & Eq.~(\ref{eq:RNAP_alloc}) \\
$f_\text{bn}$ & {\footnotesize Fraction of bound RNAP} & 0-1 & \cite{balakrishnan_principles_2022} & Eq.~(\ref{eq:fbn}) \\
$\phi_\text{N}$ & {\footnotesize RNAP mass fraction} & 0.01-0.02 & \cite{balakrishnan_principles_2022} & Eq.~(\ref{eq:lambda_phiN}) \\
$M_\text{N}$ & {\footnotesize Total mass of the N-sector proteins} & 389 [kDa] & \cite{balakrishnan_principles_2022} & Eq.~(\ref{eq:lambda_phiN}) \\
$l_\text{N}$ & {\footnotesize \# bp transcribed in RNAP genes} & 10494 [bp] & \cite{balakrishnan_principles_2022} & Eq.~(\ref{eq:lambda_phiN}) \\
$\varepsilon^\text{TX}_\text{N}$ & {\footnotesize Rescaled elongation rate $\varepsilon^\text{TX}_\text{N} = \varepsilon^\text{TX} / l_\text{N}$} & 0.29 [$\text{min}^{-1}$] & \cite{Epshtein2003,balakrishnan_principles_2022} & Eq.~(\ref{eq:lambda_phiN}) \\
$\chi_\text{i}$ & {\footnotesize Ribosome allocation $m_i/m$} & 0-1  & \cite{Erickson2017} & Eq.~(\ref{eq:chi_operative_def}) \\\\
\hline
\textbf{Parameter} & \textbf{Description} & \textbf{Value [units]} & \textbf{Reference} & \textbf{Defined in} \\
\hline
$K_\text{N}$ & {\footnotesize Michaelis-Menten-like constant for $\phi_\text{N}$. $K_\text{N} = K_m \delta / (\varepsilon^\text{TX}_\text{N}  f_\text{bn} [P])$} & 0.5E-3 & \cite{calabrese_how_2024} & Eq.~(\ref{eq:lambda_phiN}) \\
$\alpha_{0,i}^\text{TX}$ & {\footnotesize Transcription initiation rate constant } & $10^{-6}$-$10^{-2}$ [$\mu m^3$/min] & (b) & Eq.~(\ref{eq:J_i_TX}) \\\\
\end{tabular*}

\caption{Constants and parameters introduced in Sec.~\ref{sec4:transcription}, with citations for data sources and the equation of this manuscript where they are defined.\\
(a) Estimated using initiation rate from~\cite{balakrishnan_principles_2022,Baker1972} and elongation rate from~\cite{Epshtein2003}. \\
(b) Computed from $\alpha^\text{TX} / [N^\text{f}]$.\\}
\label{tab:params_sec_4_TX}
\end{table*}

\begin{table*}[!htb]
\centering
\begin{tabular*}{1.0\textwidth}{lllrr}
\hline
\multicolumn{5}{c}{\textbf{Section V. Side Dish: Protein degradation}} \\
\hline
\hline
\textbf{Observable} & \textbf{Description} & \textbf{Range [units]} & \textbf{Reference} & \textbf{Defined in} \\
\hline
$\eta$ & Global degradation rate & 0.05 - 0.009 [$h^{-1}$] & \cite{calabrese_protein_2022} & Eq.~(\ref{eq:deg:protein mass})\\
$\eta_\text{i}$ & Sector-specific degradation rate & (a) & - & Eq.~(\ref{eq:deg:protein mass})\\
$M_\text{A}$ & Total mass of free amino acids (extensive) & $M_\text{A} / \text{cell}$ 0.001 - 0.008 [pg $\text{cell}^{-1}$] & (b) & Eq.~(\ref{eq:box-A})\\
$\xi_\text{r}$ & Recycling efficiency & 0 - 1 [dimensionless] & - & Eq.~(\ref{eq:box-A})\\
$\varphi_\text{A}$ & Amino acids to potein mass~$M_\text{A} / M_p$ & 5E-3 (c) [mass fraction] & \cite{scott_emergence_2014} & Eq.~(\ref{eq:box-aa})\\
$f_\text{a}$  & {\footnotesize Fraction of active ribosomes} & 0.25-1 [fraction] & \cite{calabrese_protein_2022} & Eq.~(\ref{eq:deg:lambda}) \\
$f_\text{i}$ & {\footnotesize Fraction of inactive ribosomes} & 0-0.75 [fraction] & \cite{calabrese_protein_2022} & Eq.~(\ref{eq:deg:lambda}) \\
$J_\text{deg}$ & {\footnotesize Degradation flux} & [mass per unit time] & - & Eq.~(\ref{eq:deg:protein mass}) \\\\
\hline
\textbf{Parameter} & \textbf{Description} & \textbf{Value [units]} & \textbf{Reference} & \textbf{Defined in} \\
\hline
$\eta_0$ & Degradation rate at zero growth & 0.05 [$h^{-1}$] & \cite{Droghetti2025} & Eq.~(\ref{eq:deg:eta(lambda)})\\
$\eta_\infty$ & Degradation rate for~$\lambda \rightarrow \infty$ & 0.009 [$h^{-1}$] & \cite{Droghetti2025} & Eq.~(\ref{eq:deg:eta(lambda)})\\
$s_\eta$ & Steepness of the fit function for~$\eta(\lambda)$& 10 [$h$] & \cite{Droghetti2025} & Eq.~(\ref{eq:deg:eta(lambda)})\\\\

\end{tabular*}

\caption{Constant and parameters introduced in Sec.~\ref{sec5:degradation}, with citations for data sources and the equation of this manuscript where they are defined. The values refers to measurements in the bacteria species \textit{E. coli}.\\
    (a) To our knowledge, these quantities are not known for \textit{E. coli}.\\
    (b) Value computed from~$\psi_\text{A}$ (this table) and~$M_p$ (Tab.~\ref{tab:params_sec_2_GL}).\\
    (c) Note that, even if~$\varphi_\text{A}$ is a mass fraction, it can be greater than 1, because it's not counted in the normalization of the protein mass.\\
    }
    \label{tab:params_sec_5_deg}
\end{table*}

\begin{table*}[t!bh]
\centering
\begin{tabular*}{1.0\textwidth}{l l l r r}
\hline
\multicolumn{5}{c}{\textbf{Section VI. À la carte: Mechanistic regulation of ribosome allocation}} \\
\hline
\hline
\textbf{Observable} & \textbf{Description} & \textbf{Range [units]} & \textbf{Reference} & \textbf{Defined in} \\
\hline
${\chi}_\text{R}$ & {\footnotesize Ribosome sector allocation function} & 0.049-0.55 [number fraction] & \cite{Droghetti2025} & Eq.~(\ref{eq:6chiR_omegaR}) \\
${\omega}_\text{R}$ & {\footnotesize RNAP synthetizing ribosomal transcripts} & 0.049-0.55 [number fraction] & \cite{Droghetti2025} & Eq.~(\ref{eq:6chiR_omegaR}) \\
$[G]$ & {\footnotesize ppGpp concentration} & 0-8 [fold change] & \cite{Wu2022, chure_optimal_2023} & Eq.~(\ref{eq:6_ppGpp(epsilon)}) \\
$[G]$ & {\footnotesize ppGpp concentration} & 0-200 [$\mu$M] & \cite{marr_growth_1991} & Eq.~(\ref{eq:6_ppGpp(epsilon)}) \\
$\tau_\text{dwell}$ & {\footnotesize Ribosomal waiting time on the transcript} & 0 - $\tau_\text{trans}$ [min] (a) & \cite{Droghetti2025} & Eq.~(\ref{eq:6_epsilon(taus)}) \\\\
\hline
\textbf{Parameter} & \textbf{Description} & \textbf{Value [units]} & \textbf{Reference} & \textbf{Defined in} \\
\hline
$[G_\text{0}]$ & {\footnotesize ppGpp concentration at~$\lambda=0$} & 200 [$\mu$M] & \cite{Droghetti2025} & Eq.~(\ref{eq:6_G_0}) \\
$\tau_\text{trans}$ & {\footnotesize Translocation time} & 5.72 [min] (b) & \cite{Wu2022} & Eq.~(\ref{eq:6_epsilon(taus)}) \\\\
\end{tabular*}

\caption{Constants and parameters introduced in Sec.~\ref{sec6:mecha_regulation}, with citations for data sources and the equation of this manuscript where they are defined.\\
(a) The constraint $\tau_\text{dwell} < \tau_\text{trans}$ follows from Eq.~\ref{eq:6_epsilon(taus)} and the empirical condition $\varepsilon(\lambda=0) = 0.5 \varepsilon^\text{max}$ based on data from~\cite{Wu2022}.\\
(b) Computed as the inverse of the maximum elongation rate, i.e., $\tau_\text{trans} = (\varepsilon^\text{max})^{-1}$.}
\label{tab:params_sec_6_mecha}
\end{table*}

\begin{table*}[t!bh]
\centering
\begin{tabular*}{1.0\textwidth}{l l l r r}
\hline
\multicolumn{5}{c}{\textbf{Section VII. À la carte: Non-steady states and shifts}} \\
\hline
\hline
\textbf{Observable} & \textbf{Description} & \textbf{Range [units]} & \textbf{Reference} & \textbf{Defined in} \\
\hline
$\phi_{\text{C}_\text{i}}$ & {\footnotesize Mass fraction of catabolic protein of substrate i } & 0 - $h_\text{i}$ [mass fraction] & \cite{Erickson2017} & Eq.~(\ref{eq:FCR_J_C}) \\
$\nu_\text{i}$ & {\footnotesize Nutrient quality of substrate i} & (a1) & \cite{Erickson2017} & Eq.~(\ref{eq:FCR_J_C}) \\
$\sigma$ & {\footnotesize Translational activity} & 0-11 [$\text{h}^{-1}$] & \cite{Erickson2017} & Eq.~(\ref{eq:FCR_Jsyn}) \\
$\chi_\text{R}$ & {\footnotesize Ribosome sector allocation function} & 0.049 - 0.55 [number] (b) & \cite{Erickson2017} & Eq.~(\ref{eq:FCR_xR}) \\
$\chi_{\text{C}_\text{i}}$ & {\footnotesize Catabolic sector (of substrate i) allocation function} & 0-1 [number] (b) & \cite{Erickson2017} & Eq.~(\ref{eq:FCR_xCat}) \\\\
\hline
\textbf{Parameter} & \textbf{Description} & \textbf{Value [units]} & \textbf{Reference} & \textbf{Defined in} \\
\hline 
$h_\text{i}$& {\footnotesize Catabolic sector~$\phi_{\text{C}_\text{i}}$ maximum size} & (a2) & \cite{Erickson2017} & Eq.~(\ref{eq:FCR_xCat}) \\
$\lambda_\text{C} $& {\footnotesize Maximum growth rate achievable on carbon limited medium} & 1.17 [$h^{-1}$] & \cite{Erickson2017} & Eq.~(\ref{eq:FCR_xCat}) \\
$\tau_\omega$ & {\footnotesize Typical RNAP relocation time} & (c) & - & Eq.~(\ref{eq:7_domega_i_dt}) \\\\
\end{tabular*}

\caption{Constants and parameters introduced in Sec.~\ref{sec7:shifts}, with citations for data sources and the equation of this manuscript where they are defined.\\
(a1-a2) $h_\text{i}$, the maximal sizes of sectors specialized for each substrate, are generally not known or reported. Studies often report LacZ activity instead~\cite{you_coordination_2013, Hermsen2015}. Likewise, $\nu_\text{i}$ such that $\lambda = \sum_i \nu_i \phi_{\text{C}_i}$ are typically not reported. This challenge in modeling shift dynamics can be bypassed by defining rescaled sectors $\phi_{\text{C}_i}/h_i$ and rescaled nutrient quality $\mu_i = \nu_i h_i = \lambda / (1 - \lambda/\lambda_\text{C})$). See Ref.~\cite{Erickson2017}, supplementary section 2.3.\\
(b) In Ref.~\cite{Erickson2017}, the regulatory functions $\chi_i$ are numerical values, reflecting the fraction of biosynthesis flux allocated to a sector. In more mechanistic models like Ref.~\cite{Droghetti2025}, $\chi_i$ is defined as a transcript ratio, making its unit a number fraction.\\
(c) To our knowledge, $\tau_\omega$ has not yet been experimentally characterized.}
\label{tab:params_sec_7_shift}
\end{table*}

\begin{table*}[t!bh]
\centering
\begin{tabular*}{1.0\textwidth}{l l l r r}
\hline
\multicolumn{5}{c}{\textbf{Section VIII. Intermezzo: Density homeostasis and osmo-metabolic regulation of volume}} \\
\hline
\hline
\textbf{Observable} & \textbf{Description} & \textbf{Range [units]} & \textbf{Reference} & \textbf{Defined in} \\
\hline
$\rho_\text{DM}$ & {\footnotesize Dry mass density} & 0.06-0.12 [kg $\text{l}^{-1}$] & \cite{rollin_physical_2023} & Eq.~(\ref{eq.8_drhodt}) \\
$\lambda_V$ & {\footnotesize Volume growth rate} & 0.1-0.8 [$\text{h}^{-1}$] (a) & ~\cite{Neurohr2020} & Eq.~(\ref{eq.8_volume_growth}) \\
$\lambda_M$ & {\footnotesize Mass growth rate} & 0.1-0.8 [$\text{h}^{-1}$] (b) & \cite{Neurohr2020} & Eq.~(\ref{eq.8_mass_growth}) \\
$M$ (c) & {\footnotesize Mass (extensive)} & M/\# cells = 4.3-21.6E-15 [kg $\text{cell}^{-1}$] & (d) & Sec.~\ref{sec8:density} \\
$V$ (c) & {\footnotesize Volume (extensive)} & V/\# cells =  60-1000E-15 [l $\text{cell}^{-1}$] & \cite{Neurohr2020} & Sec.~\ref{sec8:density} \\
$n_{+}$ & {\footnotesize Cytoplasmic cation concentration} & 160 [mM] & \cite{rob_philips_ron_milo_cell_2015} & Eq.~(\ref{eq.8_electroneutrality}) \\
$n_{-}$ & {\footnotesize Cytoplasmic anion concentration} & 20 [mM] & \cite{rob_philips_ron_milo_cell_2015} & Eq.~(\ref{eq.8_electroneutrality}) \\
$c$ (e) & {\footnotesize Concentration of non-ionic impermeant osmolites} & 116.6 [mM] & \cite{rollin_physical_2023} & Eq.~(\ref{eq.8_electroneutrality}) \\
$z$ & {\footnotesize Average charge of non-ionic impermeant osmolites} & 1.2 [\#electron] & \cite{rollin_physical_2023} & Eq.~(\ref{eq.8_electroneutrality}) \\
$T$ & {\footnotesize Temperature} & 23-30 [$^\circ$C] & \cite{Neurohr2020} & Eq.~(\ref{eq.8_electroneutrality}) \\
$V_\text{ext}$ & {\footnotesize Cellular excluded volume} & 375E9 [$\text{nm}^3$] & \cite{rollin_physical_2023} & Eq.~(\ref{eq:8_Vext}) \\
$A$ & {\footnotesize Number of free amino acids} & 3.14-70E9 [$\text{cell}^{-1}$] & (f) & Eq.~(\ref{eq.8_rho_pump_leak}) \\
$z_\text{a}$ & {\footnotesize Average free amino acids charge} & 1 [\#electron] &\cite{rollin_physical_2023} & Eq.~(\ref{eq.8_rho_pump_leak}) \\\\
\hline
\textbf{Parameter} & \textbf{Description} & \textbf{Value [units]} & \textbf{Reference} & \textbf{Defined in} \\
\hline
$\rho_\text{DM}^*$ & {\footnotesize Homeostatic dry mass density} & 0.12 [kg $\text{l}^{-1}$] & \cite{rollin_physical_2023} & Eq.~(\ref{eq.8_rho*}) \\
$\theta_{M}$ & {\footnotesize Time scale of mass growth rate relaxation} & 1 [$\text{h}^{-1}$] (g) & \cite{Srivastava2025} & Eq.~(\ref{eq.8_mass_growth}) \\
$\theta_{V}$ & {\footnotesize Time scale of volume growth rate relaxation} & 1 [$\text{h}^{-1}$] (g) & \cite{Srivastava2025} & Eq.~(\ref{eq.8_volume_growth}) \\
$\mu_{M}$ & {\footnotesize Mass growth rate density-dependent steady state} & (h) & \cite{Srivastava2025} & Eq.~(\ref{eq.8_mass_growth}) \\
$\mu_{V}$ & {\footnotesize Volume growth rate density-dependent steady state} & (h) & \cite{Srivastava2025} & Eq.~(\ref{eq.8_volume_growth}) \\
$\Delta \Pi$ (i1) & {\footnotesize Osmotic pressure difference} & 10-100 [Pa] & \cite{rollin_physical_2023} & Eq.~(\ref{eq.8_osmotic}) \\
$\Delta P$ (i2) & {\footnotesize Hydrostatic pressure difference} & 10-100 [Pa] & \cite{rollin_physical_2023} & Eq.~(\ref{eq.8_osmotic}) \\
$k_\text{B}$ & {\footnotesize Boltzman constant} & 1.3807E-23 [J $\text{K}^{-1}$] & \cite{Newell2019} & Eq.~(\ref{eq.8_osmotic}) \\
$n_\text{0}$ (l) & {\footnotesize Outside ion concentration} & 150 [mM] & \cite{rob_philips_ron_milo_cell_2015} & Eq.~(\ref{eq.8_electroneutrality}) \\
$a$ & {\footnotesize Ion pumps pumping efficiency} & 0.14 [dimensionless] (m) & \cite{rollin_physical_2023} & Eq.~(\ref{eq.8_ionic}) \\\\
\end{tabular*}

\caption{Constants and parameters introduced in Sec.~\ref{sec8:density}, with citations for data sources and the equation of this manuscript where they are defined. Observe that these data are for the yeast strain \textit{S. cerevisiae}, onless otherwise specified.\\
(a) Obtained by dividing $\text{d}V/\text{d}t$ by $V$. Volume data from Ref.~\cite{Neurohr2020}.\\
(b) During balanced growth, mass and volume growth rates coincide.\\
(c) Values apply to both balanced and unbalanced growth~\cite{Neurohr2020}.\\
(d) Computed using volume and buoyant mass density reported in Ref.~\cite{Neurohr2020}.\\
(e) Derived using $z$, $n_+$, $n_-$ and Eq.~(\ref{eq.8_electroneutrality}).\\
(f) Estimated from data in Ref.~\cite{rollin_physical_2023}.\\
(g) Values correspond to one condition (growth rate 0.03~$h^{-1}$ in HeLa cells) in Ref.~\cite{Srivastava2025}.\\
(h) See Ref.~\cite{Srivastava2025} for possible forms of $\mu_\text{M}$ and $\mu_\text{V}$ for HeLa cells.\\
(i1) $\Delta \Pi$ is condition-dependent and difficult to measure; range reported in~\cite{rollin_physical_2023}.\\
(i2) Assumed equal to $\Delta \Pi$, see Eq.~(\ref{eq.8_osmotic}).\\
(l) Typical value for mammalian cells~\cite{rob_philips_ron_milo_cell_2015}.\\
(m) $a$ can be safely approximated to 0, see Ref.~\cite{rollin_physical_2023}.}
\label{tab:params_sec_8_osmo}
\end{table*}

\begin{table*}[t!bh]
\centering
\begin{tabular*}{1.025\textwidth}{l l l r r}
\hline
\multicolumn{5}{c}{\textbf{Section IX. Dessert: Crosstalk between growth and cell-cycle progression}} \\
\hline
\hline
\textbf{Observable} & \textbf{Description} & \textbf{Range [units]} & \textbf{Reference} & \textbf{Defined in} \\
\hline
$M_p$ & {\footnotesize Total mass of P sector proteins (extensive)} & M/\#cells= 0-0.8 [pg $\text{cell}^{-1}$] & (a) & Eq.~(\ref{eq:micro}) \\
$s$ (b) & {\footnotesize Size (volume, extensive)} & V/ \#cells =1-8 [$\mu m^3$] & \cite{Basan2015} & Eq.~(\ref{eq:macro}) \\
$s$ (b) & {\footnotesize Size (mass, extensive)} & M/\#cells = 0.25-2 [pg $\text{cell}^{-1}$] & \cite{Basan2015} & Eq.~(\ref{eq:macro}) \\
$X$ & {\footnotesize Division protein copy number} & (c) 0 - $X^*$ [copy number] & - & Eq.~(\ref{eq:macro}) \\
$M_X$ & {\footnotesize Division protein mass (extensive)} & (c) [mass] & - & Eq.~(\ref{eq:macro}) \\
$k_\text{X}$ & {\footnotesize Division protein size dependent production rate} & (d) [$m^{-3}\,\text{h}^{-1}$] or [$kg^{-1}\,\text{h}^{-1}$] & - & Eq.~(\ref{eq:macro}) \\
$k$ & {\footnotesize Division protein basal production rate} & (d) [$\text{h}^{-1}$] & - & Eq.~(\ref{eq:macro}) \\
$\eta_\text{X}$ & {\footnotesize Division protein degradation rate} & (d) [$\text{h}^{-1}$] & - & Eq.~(\ref{eq:macro}) \\
$\tau$ & {\footnotesize Average division time} & (e) 20 - $\infty$ [min] & \cite{scott_interdependence_2010} & Eq.~(\ref{eq:divisionstrategy}) \\
$s_\text{0}$ (b) & {\footnotesize Average size at birth (see values for volume and mass)} & [pg] or [$\mu$m$^{3}$] & - & Eq.~(\ref{eq:linearexp}) \\
$s_\text{f}$ (b) & {\footnotesize Average size at division (see values for volume and mass)} & $2s_0$ (f) [pg] or [$\mu$m$^{3}$] & - & Eq.~(\ref{eq:linearexp}) \\
$\delta s^\text{j}_\text{0}$ (g) & {\footnotesize Fluctuaction of the size at birth} &  [pg] or [$\mu$m$^{3}$] & - & Eq.~(\ref{eq:linearexp}) \\
$\delta s^\text{j}_\text{f}$ (g) & {\footnotesize Fluctuaction of the size at division} &  [pg] or [$\mu$m$^{3}$] & - & Eq.~(\ref{eq:linearexp}) \\
$\zeta_g$ & {\footnotesize Minus the slope of the adder plot ($\delta s$ VS $s_\text{0}$)} & -1 to 1 [adimensional] & - & Eq.~(\ref{eq:zetaG}) \\\\
\hline
\textbf{Parameter} & \textbf{Description} & \textbf{Value [units]} & \textbf{Reference} & \textbf{Defined in} \\
\hline
$X^*$ & {\footnotesize Condition-dependent division protein threshold to start division} & (c) [copy number] & - & Eq.~(\ref{eq:divisionstrategy}) \\
$\Delta$ (h) & {\footnotesize Adder size} & $s_0$ [pg] or [$\mu$m$^{3}$] & - & Eq.~(\ref{eq:adder}) \\\\
\end{tabular*}

\caption{Constants and parameters introduced in Sec.~\ref{sec9:cell_cycle}, with citations for data sources and the equation of this manuscript where they are defined.\\
(a) Computed by multiplying the cellular protein mass $M_p$ and the sector size $\phi_\text{P}$ (both in Tab.~\ref{tab:params_sec_2_GL}).\\
(b) The variable $s$ represents cell size, which can refer to mass or volume. Values shown are typical for \textit{E. coli} and appear also in Tab.~\ref{tab:params_sec_2_GL}.\\
(c) $X$ is a placeholder for the division protein copy number. The actual identity and typical abundance of the division trigger is unclear and context-dependent. For example, for the division protein FtsZ the copy number is in the range 5-10E3~\cite{Li2014}. \\
(d) These rates follow from the model’s ansatz for $X$ production; their values depend on the identity of $X$. For an example of the values, please refer to Ref.~\cite{ElGamel2024}.\\
(e) Condition dependent, $\infty$ is the doubling time of non-growing cells.\\
(f) On average, cells divide at approximately double the size at birth: $s_f \approx 2s_0$.\\
(g) The amplitude of the fluctuaction is usually such that the coefficient of variation (CV) for the size at birth is around 15\%. The fluctuation at division is related to the one at birth via Eq.~(\ref{eq:zetaG}).\\
(h) The adder size $\Delta = s_f - s_0$ is typically close to the average birth size $s_0$.\\
}
\label{tab:params_sec_9_cycle}
\end{table*}

\begin{table*}[t!bh]
\centering
\begin{tabular*}{1.0\textwidth}{l l l r r}
\hline
\multicolumn{5}{c}{\textbf{Section X. Digestif: Ecological implications of proteome allocation and growth laws}} \\
\hline
\hline
\textbf{Quantity} & \textbf{Description} & \textbf{Range [units]} & \textbf{Reference} & \textbf{Defined in} \\
\hline
$S^\text{(i)}$ & {\footnotesize Mass of the i-th bacteria species (extensive) } &  [mass] & - & Eq.~(\ref{eq:protein_dynamics}) \\
$d_\text{i}$ & {\footnotesize Maintenance rate of the i-th species } & 0.5 [fmol CFU$^{-1}$ d$^{-1}$] (a) & \cite{Schink2019} & Eq.~(\ref{eq:protein_dynamics}) \\
$r_\text{j}$ & {\footnotesize Mass of the j-th resource (extensive) } &  [mass] & - & Eq.~(\ref{eq:resource_dynamics}) \\
$k_\text{j}$ & {\footnotesize Supply rate of the j-th resource } &  [mass s$^{-1}$] & - & Eq.~(\ref{eq:resource_dynamics}) \\
$u^\text{i}_\text{j}$ & {\footnotesize Uptake rate of resource i by species j } &  [s$^{-1}$] & - & Eq.~(\ref{eq:resource_dynamics}) \\
$V$ & {\footnotesize Total volume of the culture (medium and cells)} & (b) [ml]& - & Eq.~\ref{eq:10.extAAleak}\\
$[M^\text{ext}_\text{A}]$ & {\footnotesize External amino acids concentration} & (c) [pg ml$^{-1}$]& - & Eq.~(\ref{eq:10.extAAleak})\\
$[M_p]$ & {\footnotesize Total protein concentration} & (c) [pg ml$^{-1}$]& - & Eq.~(\ref{eq:10.extAAleak})\\
$J_\text{leak}$ & {\footnotesize Leakage flux (extensive)} & (c) [pg s$^{-1}$]& - & Eq.~\ref{eq:10.extAAleak}\\
$[M_\text{s}]$ & {\footnotesize External nutrient concentration} & (c) [pg ml$^{-1}$]& - & Eq.~(\ref{eq:10.extAAleak})\\
$[M^\text{int}_\text{A}]$ & {\footnotesize External amino acids concentration} & (c) [pg ml$^{-1}$]& - & Eq.~(\ref{eq:10.aaINTleak})\\
$\phi_{\text{C}}$ & {\footnotesize Mass fraction of catabolic protein} & 0 - $\phi_\text{P}$ [mass fraction] & \cite{Mukherjee2024_plasticity} & Eq.~(\ref{eq:10.phiP_partition_physio}) \\
$\phi_{\text{C}^\ddagger}$ & {\footnotesize Mass fraction of core catabolic protein} & 0 - $\phi_\text{C}$ [mass fraction] & \cite{Mukherjee2024_plasticity} & Eq.~(\ref{eq:10.phiP_partition_eco}) \\
$\phi_{\text{S}}$ & {\footnotesize Mass fraction of stress response protein} & 0 - $\phi_\text{P}$ [mass fraction] & \cite{Mukherjee2024_plasticity} & Eq.~(\ref{eq:10.phiP_partition_physio}) \\
$\phi_{\text{AD}}$ & {\footnotesize Mass fraction of adaptation sector} & 0 - $\phi_\text{P}$ [mass fraction] & \cite{Mukherjee2024_plasticity} & Eq.~(\ref{eq:10.phiP_partition_eco}) \\
$\chi_{\text{C}^\ddagger}$ & {\footnotesize $\phi_{\text{C}^\ddagger}$Catabolic sector (of substrate i) allocation function} & 0-$\phi_\text{C}$ [number] & \cite{Mukherjee2024_plasticity} & Eq.~(\ref{eq:FCR_xCat}) \\\\
\hline
\textbf{Parameter} & \textbf{Description} & \textbf{Value [units]} & \textbf{Reference} & \textbf{Defined in} \\
\hline
D & {\footnotesize Influx and dilution rate} & (b) [s$^{-1}$]& - & Eq.~(\ref{eq:10.extAAleak})\\
$[M^\text{ext}_\text{A,0}]$ & {\footnotesize Amino acids influx cencentration} & (b) [pg ml$^{-1}$]& - & Eq.~(\ref{eq:10.extAAleak})\\
$[M_\text{s,0}]$ & {\footnotesize Nutrients influx cencentration} & (b) [pg ml$^{-1}$]& - & Eq.~(\ref{eq:10.extGLU})\\
\end{tabular*}

\caption{Constants and parameters introduced in Sec.~\ref{sec10:communities}, with citations for data sources and the equation of this manuscript where they are defined.\\
(a) Value for \textit{E. coli}, measured during starvation. \\
(b) The values for the considered volume, dilution rate, and influx concentrations depend on the experimental settings.\\
(c) All these quantities can reach a steady state value in chemostat settings, depending on the value of the dilution rate.\\
}
\label{tab:params_sec_10_communities}
\end{table*}

\end{document}